\documentclass[journal,draftclsnofoot,onecolumn]{IEEEtran}

\newcommand{\figref}[1]{Fig.~\ref{#1}}

\usepackage{times}
\usepackage{epsfig}
\usepackage{colortbl}
\usepackage{float}
\usepackage{subfig}
\usepackage{verbatim}

\usepackage[latin1]{inputenc}
\usepackage[T1]{fontenc}
\usepackage{lmodern}
\usepackage{amsmath}
\usepackage{amsthm}
\usepackage{amssymb}
\usepackage{bbm}

\usepackage{graphicx}
\usepackage{tikz}
\usepackage{pgfplots}
\usepackage[right]{eurosym}
\usepackage{multirow}
\usepackage{color}
\usepackage{psfrag}
\usepackage{todonotes}
\usepackage[normalem]{ulem} 
\usepackage{cancel}
\usepackage{cite} 

\usepackage{cleveref} 
\usepackage{pdfpages} 

\usetikzlibrary{shapes,decorations,patterns}
\usetikzlibrary{arrows,backgrounds,fit}
\tikzstyle{block} = [draw, align=center, rectangle, minimum height=2.5em, minimum width=2.5em]
\tikzstyle{dmc_block} = [draw, align=center, rectangle, rounded corners, minimum height=6em, minimum width=5em]
\tikzstyle{receiver_block} = [draw, align=center, rectangle, rounded corners, minimum height=2.5em, minimum width=5em]
\tikzstyle{transmitter_block} = [draw, align=center, rectangle, rounded corners, minimum height=4em, minimum width=5em]
\tikzstyle{mul} = [draw, circle, inner sep=0pt]
\tikzstyle{dot} = [draw, circle, minimum size=0.2pt,scale=0.3, fill=black,black]

\pgfdeclarepatternformonly{soft horizontal lines}{\pgfpointorigin}{\pgfqpoint{100pt}{1pt}}{\pgfqpoint{100pt}{3pt}}%
{
  \pgfsetstrokeopacity{0.3}
  \pgfsetlinewidth{0.1pt}
  \pgfpathmoveto{\pgfqpoint{0pt}{0.5pt}}
  \pgfpathlineto{\pgfqpoint{100pt}{0.5pt}}
  \pgfusepath{stroke}
}

\pgfdeclarepatternformonly{soft crosshatch}{\pgfqpoint{-1pt}{-1pt}}{\pgfqpoint{4pt}{4pt}}{\pgfqpoint{3pt}{3pt}}%
{
  \pgfsetstrokeopacity{0.3}
  \pgfsetlinewidth{0.4pt}
  \pgfpathmoveto{\pgfqpoint{3.1pt}{0pt}}
  \pgfpathlineto{\pgfqpoint{0pt}{3.1pt}}
  \pgfpathmoveto{\pgfqpoint{0pt}{0pt}}
  \pgfpathlineto{\pgfqpoint{3.1pt}{3.1pt}}
  \pgfusepath{stroke}
}

\newtheorem{theorem}{Theorem}
\newtheorem{corollary}[theorem]{Corollary}
\newtheorem{lemma}[theorem]{Lemma}

\makeatletter

\newcommand{\Rmnum}[1]{\expandafter\@slowromancap\romannumeral #1@}
\makeatother

\newfont{\bbb}{msbm10 scaled 500}

\newfont{\bb}{msbm10 scaled 1100}

\newcommand{\Ec}{{\cal E}}

\newcommand{\Tc}{{\cal T}}

\makeatletter
\newcommand*{\rom}[1]{\expandafter\@slowromancap\romannumeral #1@}
\makeatother

\definecolor{OXO-emph}{RGB}{153,0,0}

\graphicspath{{./fig/}}
\DeclareGraphicsExtensions{.pdf} 

\IEEEoverridecommandlockouts
\begin{document}

\title{Individual Secrecy for the Broadcast Channel
\thanks{This paper was presented in part at IEEE International Symposium on Information Theory, Hong Kong, Jun. 2015.}
\thanks{
Y. Chen and A. Sezgin are with the Institute of Digital Communication Systems, Ruhr University Bochum, Germany (e-mail: yanling.chen-q5g@rub.de, aydin.sezgin@rub.de).
O. O. Koyluoglu is with the Department of Electrical and Computer
Engineering, The University of Arizona, Tucson, AZ 85721, USA (e-mail:
ozan@email.arizona.edu).}
}

\author{\IEEEauthorblockN{Yanling Chen, O.~Ozan~Koyluoglu, and Aydin Sezgin}}

\maketitle


\begin{abstract}
This paper studies the problem of secure communication over broadcast channels under the \emph{individual} secrecy constraints. That is, the transmitter wants to send two independent messages to two legitimate receivers in the presence of an eavesdropper, while keeping the eavesdropper ignorant of {\it each} message (i.e., the information leakage from {\em each} message to the eavesdropper is made vanishing).  Building upon Carleial-Hellman's secrecy coding, Wyner's secrecy coding, the frameworks of superposition coding and Marton's coding together with techniques such as rate splitting and indirect decoding, achievable rate regions are developed. The proposed regions are compared with those satisfying joint secrecy and without secrecy constraints, and the individual secrecy capacity regions for special cases are characterized. In particular, capacity region for the deterministic case is established, and for the Gaussian model, a constant gap (i.e., 0.5 bits within the individual secrecy capacity region) result is obtained. Overall, when compared with the joint secrecy constraint, the results allow for trading-off secrecy level and throughput in the system.
\end{abstract}


\section{Introduction}

\subsection{Background}
The broadcast channel (BC) involves the simultaneous communication of information from one transmitter to multiple receivers. 
The broadcast nature makes the communication susceptible to eavesdropping. Therefore, it is desirable to offer a reliable communication with a certain level of security guarantee, especially to ensure that sensitive information is protected from unauthorized parties. 

The most fundamental model of the BC is the two-receiver BC with two independent messages. 
This basic model and its extensions with or without an external eavesdropper have been well studied \cite{src:Cover1972, src:Bergmans1973, src:Gallager1974, src:Ahlswede1975, src:Marton1979, src:ElGamal1981, src:Ekrem2009, src:Nair2009,Chia:Three-receiver12, src:Ekrem2013}. However, capacity regions have still remained open for the basic model (i.e., two independent private messages are dedicated to two legitimate receivers, respectively), and its extension with an external eavesdropper subject to a \emph{joint} secrecy constraint (whereby the information leakage from \emph{both} messages to the eavesdropper is made vanishing). Nevertheless, in case that the channels to all the receivers (and the eavesdropper) fulfill a certain degradation order, the capacity regions are characterized and superposition coding is shown to be optimal in both settings \cite{src:Cover1972,src:Bergmans1973,src:Gallager1974, src:Ekrem2009}.  

\begin{figure}[t]
 \centering
  \includegraphics[width=0.75\textwidth]{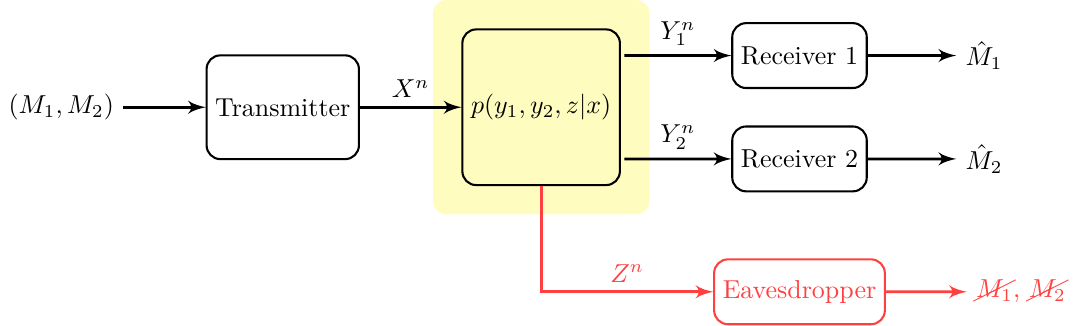}
  \caption{BC with an external eavesdropper.} \label{fig: degraded BC with an external eavesdropper}
 \end{figure}

In this paper, we primarily focus on the problem of secure communication over the BC subject to the \emph{individual} secrecy constraint. The channel model is shown in \figref{fig: degraded BC with an external eavesdropper}. Differently from the joint secrecy constraint, here one aims to minimize the information leakage from \emph{each} message to the eavesdropper. Remarkably, the joint secrecy constraint offers a higher secrecy level from the system design perspective (but unfortunately not always affordable \cite{src:ChenKS2014_ISIT}), while the individual secrecy constraint could provide an acceptable security strength from the end user's point of view with potential gains in increasing transmission rates. Therefore, the notion of individual secrecy allows for trading-off of the throughput and secrecy level.

To ensure a pre-specified secrecy level, there are popular cryptographic means as demonstrated by Shannon \cite{src:Shannon1949} that rely on the secret keys shared only between the transmitter and the intended receiver in advance of the communication. Another well-known information-theoretic means is by Wyner's secrecy coding introduced in \cite{src:Wyner1975}, where he proposed the model of the wiretap channel. The main idea is to explore the advantage of the channel of the legitimate receiver against a \emph{degraded} eavesdropper by the means of trading rate for secrecy. More specifically, sufficient randomness is added to the codeword in order to keep the eavesdropper totally ignorant of the transmitted message. Later on, a sharper result for the wiretap channel is obtained by Csisz\'{a}r and K\"{o}rner \cite{Csisz'ar:Broadcast78} by considering a general setup of transmitting the common and confidential messages over a channel where the eavesdropper's channel is not necessarily a degraded version of the channel of the legitimate receiver. 

Another secrecy coding method that has not attracted enough attention, but plays an important role in this paper for the purpose of individual secrecy, is a coding technique introduced by Carleial and Hellman in \cite{src:Carleial-Hellman1977} for a special case of the wiretap channel, where the channel to the legitimate receiver is noiseless and the eavesdropper's channel is a binary symmetric channel (BSC) with cross probability $p$. It is demonstrated that it is possible to send message (of length $n$ and divided into $n/l$ pieces each with $l$ bits) at capacity (i.e., rate $1$) over the main channel while still keeping the eavesdropper totally ignorant of \emph{each} piece of the message provided that $nh(p)\geq l.$ The main observation is that message pieces can hide each other without a need of additional randomness (under this weaker secrecy notion). In this paper, we generalize this coding idea to the broadcast channel scenarios and refer to it as Carleial-Hellman's secrecy coding. The goal remains the same, i.e., to keep the eavesdropper totally ignorant of each piece of the messages. Differently from the original proposal in \cite{src:Carleial-Hellman1977},  the channel is not restricted to be noiseless or BSC, each piece of the messages are destined at different receivers, and each piece of the messages are not necessarily of the same length. 

\subsection{Contributions}
In this paper, we consider the problem of secure communication over the broadcast channel, where the transmitter wants to send two independent messages to two legitimate receivers in the presence of an external eavesdropper. (See \figref{fig: degraded BC with an external eavesdropper}.)  In the following, we summarize the main contributions of the paper:

\begin{itemize}
\item The linear deterministic model is studied and corresponding capacity regions under different secrecy constraints are characterized. Study of this specific model provides insights into the capacity regions for the Gaussian case under different secrecy constraints, especially in the high SNR regime.

\item To investigate the fundamental limits of communication under the individual secrecy constraints, constructions building upon {Carleial-Hellman's} secrecy coding, {Wyner's} secrecy coding, superposition coding, and Marton's coding, rate splitting and indirect decoding are proposed for the general discrete memoryless broadcast channel  (DM-BC) with an external eavesdropper. 
\begin{itemize}
	\item First construction, referred to as the primitive approach, utilizes {Carleial-Hellman's} secrecy coding in the sense that it regards one message as (partial) randomness for ensuring the individual secrecy of the other. This approach is shown to be optimal if the channels to both legitimate receivers are statistically identical. 
	
	\item The primitive approach is suboptimal for the case when one legitimate receiver's channel is less noisy than the other. To further benefit from the channel advantage of the strong receiver, we propose the superposition coding scheme by taking the primitive approach as the cloud coding layer, and adding to it another satellite coding layer. Differently from the cloud layer that employs {Carleial-Hellman's} secrecy coding, we employ Wyner's secrecy coding in the satellite layer to ensure the secrecy of the additional message to the strong receiver. This approach is shown to be optimal for the case of a {\em comparable} eavesdropper (compared to the weak legitimate receiver); and the case that the weak legitimate receiver has a {\em deterministic} channel and the eavesdropper's channel is a degraded version of it. 
	
	\item Considering the general case where there may not be less noisiness order between the channels to the legitimate receivers, we devise a coding scheme by utilizing Marton's coding. The idea is to explore the advantage of rate splitting at the encoding phase (with introduction of jointly distributed satellite codewords that carry independent message pieces intended for each legitimate receiver); and recovering the individual satellite codewords at the decoding phase. As a result, a general achievable individual secrecy rate region is established, which includes regions obtained by the primitive approach and superposition coding approach as special cases. 
\end{itemize}
	
\item Following the (Marton's) coding scheme proposed and appropriately modifying its analysis for secrecy, an achievable {\em joint secrecy} rate region is established. This region is contrasted with the previous regions reported in the literature.

\item Gaussian model is studied, and a constant gap result (i.e., 0.5 bits within the individual secrecy capacity region) is obtained. In particular,  the individual secrecy capacity region is characterized for the {\em comparable} eavesdropper case (defined by satisfying $\sigma_2^2\leq \sigma_e^2\leq 2\sigma_2^2$ for the noise variances of weaker receiver and eavesdropper). To visualize the impact of different secrecy constraints on the fundamental limits, comparisons are made among the capacity regions of Gaussian-BC without secrecy constraint, and with individual and joint secrecy constraints. 
\end{itemize}

\subsection{Related Work}

The broadcast channel involves the simultaneous communication of information from one transmitter to multiple receivers. Generally speaking, the information may be independent or nested. For the general two-receiver BC with two independent messages, the capacity region is yet unknown. Nevertheless, if one receiver's channel is degraded to the other, then the capacity region is fully characterized and it is shown that superposition coding is optimal \cite{src:Cover1972,src:Bergmans1973,src:Gallager1974}. In general, the best known achievable rate region is obtained by Marton's coding in \cite{src:Marton1979}. For the BC with nested information, one instance is the two-receiver BC with one common and two private messages. The model was first introduced  by  K\"{o}rner and Marton in \cite{src:Korner-Marton1977}, and the general capacity region still remains as unknown. Nevertheless, in \cite{src:Korner-Marton1977}, the capacity region was established for the two-receiver BC with degraded message sets (i.e., when one of the private message has rate zero). 
	In \cite{src:Nair2009}, Nair and El Gamal extended the two-receiver BC with degraded message sets to the three-receiver case. In particular, they studied the specific case where one common message is sent to all three receivers, while one private message is sent to only one receiver. They proposed a new coding referred to as \emph{indirect decoding} and showed that the resulting region of this technique is strictly greater than the straightforward extension of the K\"{o}rner-Marton region for this scenario. Other studies on BC with different message degradation setups include \cite{src:Nair2009, src:Diggavi-Tse2006, src:Nari-Wang2009}, see also~\cite{ElGamal:2012} for an overview.

Due to the very broadcast nature of the communications, adversaries may overhear the transmissions, resulting in data leakage. Secure broadcasting refers to the situation  where  one  transmitter communicates with several legitimate receivers in the presence of an adversary (external eavesdropper). Inspired by the pioneering works \cite{src:Shannon1949, src:Wyner1975, Csisz'ar:Broadcast78} that studied the point-to-point secure communication, there has been a growing body of literature that investigate the problem of secure broadcasting with two or more receivers \cite{src:Ekrem2009, src:Ekrem2011MIMO, src:Ekrem2012Compound, Chia:Three-receiver12, src:Ekrem2013, src:ChenKS2014_ISIT, src:ChenKS2015_ISIT}.

The \emph{joint} secrecy capacity region for some special cases are established in\cite{src:Ekrem2009}, especially for certain degradation orders among the channels. The results of \cite{src:Ekrem2009} were extended to the Gaussian scenario in \cite{src:Ekrem2011MIMO}; and to the degraded compound multi-receiver broadcast channel in \cite{src:Ekrem2012Compound}. Moreover, \cite{src:Ekrem2013} studied the BC with two receivers and one eavesdropper, where the transmitter wants to transmit a pair of public and confidential messages to each legitimate receiver, and established the joint secrecy capacity for the degraded channels and when the confidential message to the strong receiver is absent. Nested information transmission with secrecy constraints were considered in \cite{Chia:Three-receiver12}. This work investigated the transmission of one common and one confidential message over a BC with two receivers and one eavesdropper, where the common message is to be delivered to both legitimate receivers and the eavesdropper, whilst the confidential message is to be delivered to both legitimate receivers but kept secret from the eavesdropper. A general achievable rate region is derived, and the secrecy capacity is established when the two legitimate receivers are less noisy than the eavesdropper. In some cases, the indirect decoding is shown to provide an inner bound that is strictly larger than the direct extension of Csisz\'{a}r and K\"{o}ner's approach.
Another relevant direction is the BC with privacy constraints \cite{src:Cai-Lam2000, src:Liu2008, src:GoldfeldKP2015}. The model was first introduced by Cai and Lam in \cite{src:Cai-Lam2000}, where each receiver not only should correctly decode its own message but also obtain no information about the message of the other receiver. In \cite{src:Cai-Lam2000}, the authors focused on the deterministic BC and established its capacity region. The general inner and outer bound were established later in \cite{src:Liu2008}. Recently, the authors of \cite{src:GoldfeldKP2015} considered an extension of this two-receiver BC model (i.e., BC with one common and two private messages, where each private message should satisfy a pre-specified constraint measured at the other receiver). The capacity regions are determined for semi-deterministic and physically degraded BCs and the BC with a degraded message set.  

Consider secure broadcasting when there is no common message 
or public message 
 involved. In case that only one confidential message is to be delivered, then at the eavesdropper both the \emph{joint} secrecy constraint and the \emph{individual} secrecy constraint reduce to the same. However, in case that independent confidential massages are to be delivered to multi-receivers, the two secrecy constraints can be quite different. By definition, the individual secrecy constraint is weaker than the joint one. The joint secrecy, however, is not always affordable \cite{src:ChenKS2014_ISIT}, and satisfying the individual secrecy can provide positive rates under these scenarios. Especially this secrecy notion offers an acceptable security level (that keeps each message to individually leak negligible information to eavesdropper), while potentially improving transmission efficiency. 
In \cite{src:Carleial-Hellman1977}, this notion of secrecy is analyzed for the point-to-point channel, and message pieces can be made individually secret without any degradation of channel capacity.
In \cite{Koyluoglu:Broadcast14}, we considered the problem of achieving individual secrecy over a BC with receiver side information (where each receiver has the desired message of the other receiver as side information). The individual secrecy rate region results are obtained for general models with full characterization for some special cases (e.g.: of either a strong or weak eavesdropper compared to both legitimate receivers). More detailed discussion and results on this model are presented in \cite{src:Chen2015_BCRSI}. The joint secrecy counterpart for this problem is studied in \cite{Wyrembelski:Secrecy11} and in \cite{Mansour:Capacity15}, where the latter work also considers nested information models (referred to as cognitive messages therein) under both individual and joint secrecy constraints. We remark that, in these models with side information, the readily available message of the other user can serve as secret key (in one-time pad fashion). And, this coding strategy satisfies the individual secrecy condition as the analysis for secrecy is performed per message basis (i.e., in an individual fashion), where each analysis considers the other message as secret key. 

In this work, the problem of secure broadcasting subject to the \emph{individual} secrecy constraints is analyzed. Wyner's secrecy coding continues to play an important role. Nevertheless, we find that Carleial-Hellman's secrecy coding is also essential for the individual secrecy setting. (As compared to prior works, the side information is absent at receivers in this model). Using the insights gained from the previous studies, we construct a superposition coding approach for special class of BCs (e.g., for certain less noisiness/degradation orders) and utilize Marton's coding for the general case. Overall, the results here establishes a comparison between different secrecy notions in BCs, in particular comparing BC with no secrecy constraints, BC with joint secrecy constraints with that of individual secrecy constraints.

\subsection{Notations and Organization}
In this paper, we follow the convention to denote random variables by capital letters, their realizations by the corresponding lower case letters and their images (or ranges) by calligraphic letters. In addition, we use $X^n$ to denote the sequence of variables $(X_1, \cdots, X_n),$ where $X_i$ is the $i$-th variable in the sequence, $X^{i-1}$ the sequence $(X_1, \cdots, X^{i-1})$ and $X_{i+1}^n$ the sequence $(X_{i+1}, \cdots, X_n)$. $\mathcal{R}_+$ is used to denote the set of nonnegative real numbers. $[a:b]$ is used to represent the set of natural numbers between $a$ and $b.$ We use shorthands $[a]^+=\max\{0, a\}$, and $C(x)=\frac{1}{2}\log_2(1+x)$.

The rest of the paper is organized as follows. Section~\ref{sec:model} introduces the system model, and Section~\ref{sec:deterministic} provides the results for the determistic case. Main results for the discrete memoryless model is given in Section~\ref{sec:DMC}, and for the Gaussian case in Section~\ref{sec:gaussian}. Section~\ref{sec:conclusion} concludes the paper. To enhance the flow, details are relegated to appendices.


\section{System model}\label{sec:model}

Consider a DM-BC with two legitimate receivers and one passive eavesdropper defined by $p(y_1,y_2,z|x).$  The model is shown in Fig. \ref{fig: degraded BC with an external eavesdropper}. The transmitter aims to send messages $m_1,m_2$ to receiver $1,2,$ respectively. Suppose that $x^n$ is the channel input, whilst $y_1^n$ (at receiver 1), $y_2^n$ (at receiver 2) and $z^n$ (at eavesdropper), are the channel outputs. By the \emph{discrete memoryless} nature of the channel, we have
\begin{equation}
	p(y_1^n, y_2^n, z^n|x^n) = \prod_{i=1}^{n} p(y_{1i}, y_{2i}, z_{i}|x_i). \label{def: dm}
\end{equation}

A $(2^{nR_1}, 2^{nR_2}, n)$ secrecy code for the DM-BC $p(y_1, y_2, z|x)$ consists of
\begin{itemize}
	\item Two message sets $\mathcal{M}_1$ and $\mathcal{M}_2,$ where $m_1\in \mathcal{M}_1=[1:2^{nR_1}]$ and $m_2\in \mathcal{M}_2=[1:2^{nR_2}];$
	\item  a (randomized) encoder that assigns a codeword $x^n$ to each message pair $(m_1, m_2);$ and 
	\item  two decoders, where decoder $i$ (at legitimate receiver $i$) assigns an estimate of $m_{i},$ say $\hat{m}_{i},$ or an error to each received sequence $y_{i}^n.$ 
\end{itemize}
 
The messages $M_1, M_2$ are assumed to be uniformly distributed over their corresponding message sets. Therefore, we have $R_i =\frac{1}{n}H(M_i)$, for $i=1,2$. Associated with the $(2^{nR_1}, 2^{nR_2}, n)$ secrecy code, the \emph{individual} information leakage rates are defined as $R_{L, i} =\frac{1}{n} I(M_i; Z^n)$ for $i=1,2$,
while the \emph{joint} information leakage rate is defined as $R_L = \frac{1}{n} I(M_1, M_2; Z^n).$
 
Denote the \emph{average probability of decoding error} at receiver $i$ as $P_{e,i}^n=\Pr(M_i\neq \hat{M}_i)$. The rate pair $(R_1,R_2)$ is said to be \emph{individual secrecy achievable}, if there exists a sequence of $(2^{nR_1}, 2^{nR_2}, n)$ codes such that
\begin{align} 
	  P_{e,i}^n &\leq \epsilon_n, \quad \mbox{for}\ i=1,2 \label{eq:Reliability} \\
	  R_{L,i}	&\leq	\tau_n, \quad \mbox{for}\ i=1,2 \label{eq:IndSec} \\
	  \lim\limits_{n\to\infty} \epsilon_n&= 0 \quad \mbox{and} \quad \lim\limits_{n\to\infty} \tau_n= 0. \label{eq:n to infty} 
\end{align}
Note that, \eqref{eq:IndSec} corresponds to the \emph{individual} secrecy constraints. If the coding schemes fulfill \eqref{eq:Reliability}, \eqref{eq:n to infty} and
\begin{equation}\label{eq:JointSec}
	R_L \leq \tau_n,
\end{equation}
then the rate pair $(R_1,R_2)$ is said to be achievable under \emph{joint secrecy}.  Clearly, the joint secrecy constraint \eqref{eq:JointSec} implies the individual secrecy \eqref{eq:IndSec}, and hence the jointly secret achievable rate pairs are by definition achievable as individually secret.  

 Two important classes of DM-BC are the classes of \emph{less noisy} channels and the class of \emph{degraded} channels, and will be also addressed in this paper. 
 Given a DM-BC that is defined by $p(y_1,y_2,z|x),$ formally, $Y$ is said to be \emph{less noisy} than $Z,$ if 
 \begin{equation}
 	I(U; Y)\geq I(U;Z)
 \end{equation}
 holds for any random variable $U$ such that $U\to X\to (Y, Z)$ forms a Markov chain. And, $Z$ is said to be a \emph{physically degraded} version of $Y,$ if 
 \begin{equation}
 	p(y,z|x)=p(y|x)p(z|y),
 \end{equation}
 i.e., $X\to Y\to Z$ forms a Markov chain for any input random variable $X.$ More generally, $Z$ is said to be a \emph{stochastically degraded}  (or simply \emph{degraded}) version of $Y,$  if there exists a random variable $\tilde{Y}$ such that $\tilde{Y}$ has the same conditional probability mass function as $Y$ (given X), and $X\to \tilde{Y}\to Z$ forms a Markov chain.
 
 
 \section{A special instance: linear deterministic case}\label{sec:deterministic}

Let us first take a look at the deterministic broadcast channel. In this model, the received signals at the legitimate receivers and the eavesdropper are given by 
\begin{align}
	{Y}_1&=D^{q-n_1} {X};\\
	{Y}_2&=D^{q-n_2} {X};\\
	Z&=D^{q-n_e} {X};
\end{align}
where ${X}$ is the binary input vector of length $q=\max\{n_1,n_2,n_e\}$; $D$ is the $q\times q$ down-shift matrix; $n_1, n_2$ and $n_e$ are the integer channel gains of the channels from the transmitter to receiver 1, receiver 2, and the eavesdropper, respectively. Without loss of generality, we assume that $n_1\geq n_2.$ Under this assumption, $Y_2$ is a degraded version of $Y_1$ according to the channel definition. In this case, we have the following theorem:
\begin{theorem}\label{thm:deterministic}
The individual secrecy capacity region of the linear deterministic broadcast channel with an external eavesdropper is the set of the rate pairs $(R_1, R_2)\in \mathcal{R}_{+}^2$ defined by
\begin{align}
\begin{split}\label{eqn: D_Ins}
	 R_1&\leq [n_1-n_e]^{+};\\
	 R_2&\leq [n_2-n_e]^{+};\\
	 R_1+R_2&\leq n_1.
\end{split}
\end{align}
\end{theorem}

\begin{IEEEproof}
See Appendix~\ref{sec:AppDeterministic}.
\end{IEEEproof}

	{\em Remark:} Note that in our achievability schemes, the elements of the input vector $X$ are i.i.d. $\mbox{Bern}(\frac{1}{2})$ in all scenarios. That is, $\mbox{Bern}(\frac{1}{2})$ serves as an optimal input distribution to achieve the individual secrecy capacity. Nevertheless, this universal choice is not the only optimal one. As an alternative, one can simply zero-pad those $r(k)$ bits, where random bits are set in our proposals. 

Similarly one can derive the following theorems for the linear deterministic broadcast channel: without secrecy constraint, and under the joint secrecy constraint.
\begin{theorem}
The capacity region of the linear deterministic broadcast channel is the set of the rate pairs $(R_1, R_2)\in \mathcal{R}_{+}^2$ defined by
\begin{align}
\begin{split}\label{eqn: D_NoS}
	 R_2&\leq n_2;\\
	 R_1+R_2&\leq n_1.
\end{split}
\end{align}
\end{theorem}

\begin{IEEEproof}
	Under the assumption that $n_1\geq n_2,$ \eqref{eqn: D_NoS} follows directly from the capacity region of the degraded BC \cite{src:Bergmans1973, src:Gallager1974, ElGamal:2012}.
\end{IEEEproof}

\begin{theorem}
The joint secrecy capacity region of the linear deterministic broadcast channel with an external eavesdropper is the set of the rate pairs $(R_1, R_2)\in \mathcal{R}_{+}^2$ defined by
\begin{align}\label{eqn: D_JoS}
\begin{split}
	 R_2&\leq [n_2-n_e]^{+};\\
	 R_1+R_2&\leq [n_1-n_e]^{+}.
\end{split}
\end{align}
\end{theorem}
\begin{IEEEproof}
Under the assumption that $n_1\geq n_2,$ we consider the following different scenarios:
	\begin{itemize}
		\item In case of a more noisy eavesdropper, i.e., as $q=n_1\geq n_2\geq n_e,$ \eqref{eqn: D_JoS} follows directly from the {joint secrecy} capacity region of the degraded BC \cite[Corollary 2]{src:Ekrem2009}; 
		\item In other cases (i.e., as $q=n_1\geq n_e \geq n_2$, or as $q=n_e\geq n_1 \geq n_2$),  the channel degenerates to a degraded wiretap channel as $R_2=0$ or $R_1=0$. As a direct consequence, its {joint secrecy} capacity region \eqref{eqn: D_JoS} reduces to the ones for the wiretap channel \cite{src:Wyner1975, Csisz'ar:Broadcast78}.
	\end{itemize} 
\end{IEEEproof}

\begin{figure}[!ht]
  \centering
    \subfloat[$0<n_2-n_e\leq n_e$\label{Determinstic: Ind-SC}]{
      \begin{tikzpicture}[scale=0.5]
      \begin{axis}[
      xlabel={$\mathbf{R_{1}}$},
      ylabel={$\mathbf{R_{2}}$},
      xtick=\empty,ytick=\empty,
      xmin=0,xmax=5,
      ymin=0,ymax=5,
      x=2cm,y=2cm,
      grid=major,
      extra y ticks={1.6,4},extra x ticks={0,2.6,5},
      extra y tick labels={{\Large $n_2-n_e$}, {\Large $n_2$}},
      extra x tick labels={{\Large 0}, {\Large$n_1-n_e$}, {\Large $n_1$}},
      area legend,
      legend style={cells={anchor=west},font=\large},
      ]
      \addplot[densely dotted, teal, fill=yellow!5, very thick, opacity=0.6] plot coordinates
                  { (0.05, 0.05) (0.05, 4) (0.95, 4) (4.90, 0.05) (0.05, 0.05)};
      \addplot[dashed, blue, fill=yellow!15, very thick, opacity=0.6] plot coordinates 
      { (0.05, 0.05) (0.05, 1.6) (2.6, 1.6) (2.6, 0.05) (0.05, 0.05)};          
      \addplot[red, fill=yellow!20, very thick, opacity=0.6] plot coordinates
            { (0.08, 0.08) (0.08, 1.57) (0.98, 1.57) (2.57, 0.08) (0.08, 0.08)};  
      \legend{
       	\ Without secrecy, \ Individual secrecy,\ Joint secrecy
                }
      \end{axis}
      \draw[->,color=black, semithick] (0,0) -- (11,0);
      \draw[->,color=black, semithick] (0,0) -- (0,11);
      \end{tikzpicture}
    }
    \hfill
    \subfloat[$n_2-n_e> n_e$\label{Determinstic: Ind-SC2}]{
      \begin{tikzpicture}[scale=0.5]
      \begin{axis}[
      xlabel={$\mathbf{R_{1}}$},
      ylabel={$\mathbf{R_{2}}$},
      xtick=\empty,ytick=\empty,
      xmin=0,xmax=5,
      ymin=0,ymax=5,
      x=2cm,y=2cm,
      grid=major,
      extra y ticks={1, 3,4},extra x ticks={0,4,5},
      extra y tick labels={{\Large $n_e$}, {\Large $n_2-n_e$}, {\Large $n_2$}},
      extra x tick labels={{\Large 0}, {\Large $n_1-n_e$}, {\Large $n_1$}},
      area legend,
      legend style={cells={anchor=west}, font=\large}
      ]
      \addplot[densely dotted, teal, fill=yellow!5, very thick, opacity=0.6] plot coordinates
                        { (0.05, 0.05) (0.05, 4) (0.95, 4) (4.90, 0.05) (0.05, 0.05)};
      \addplot[dashed, blue, fill=yellow!15, very thick, opacity=0.6] plot coordinates
      { (0.05, 0.05) (0.05, 3) (1.95, 3) (4, 0.95) (4, 0.05)(0.05, 0.05)};
      \addplot[red, fill=yellow!20, very thick, opacity=0.6] plot coordinates
            { (0.08, 0.08) (0.08, 2.97) (1.08, 2.97) (3.97, 0.08)(0.08, 0.08)};
      
    \legend{
            \ Without secrecy,\ Individual secrecy,\ Joint secrecy
           }
      \end{axis}
      \draw[->,color=black, semithick] (0,0) -- (11,0);
      \draw[->,color=black, semithick] (0,0) -- (0,11);
      \end{tikzpicture}
    }
    \caption{Capacity regions of deterministic BC.}
    \label{fig:deterministic BC}
  \end{figure}
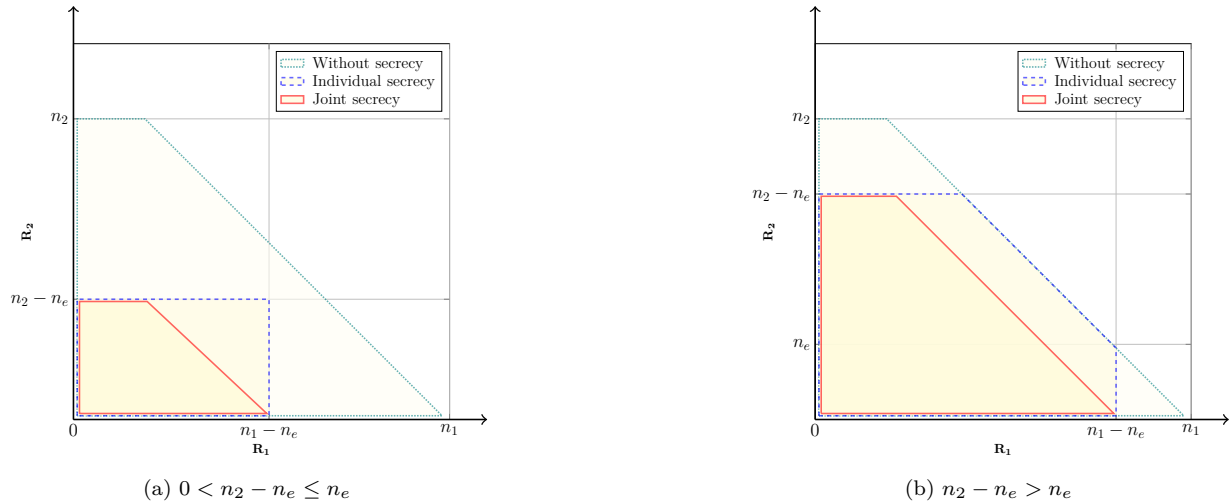

For the linear deterministic BC, we note that non-degenerate individual/joint secrecy rate regions are possible only for the case as $n_1\geq n_2\geq n_e.$ Its capacity regions under different secrecy constraints are depicted in Fig. \ref{fig:deterministic BC}.
	\begin{enumerate}
		\item Without any secrecy constraints, the capacity region is a {\em triangle with one missing corner}, where the triangle is caused by the non-negativity of the rates and the upper bound on the sum rate (since two legitimate receivers share the same transmission channel); while the missing corner is due to the fact that the transmission rate $R_2$ is upper bounded by its channel capacity (i.e., $n_2$).
		\item Under individual secrecy constraint:  
			\begin{itemize}
				\item The capacity region is a {\em rectangle} in case of $n_e\leq n_2\leq 2 n_e.$ In this case, the transmitter could send messages to both legitimate receivers up to their individual secrecy capacity ($n_1-n_e,$ $n_2-n_e$ bits, respectively) in one channel use (up to $n_1$ bits). 
				\item The capacity region is a {\em rectangle with one missing corner} in case of $n_2>2 n_e.$ In this case, the transmitter could not send secret messages to both legitimate receivers up to their individual secrecy capacity ($n_1-n_e,$ $n_2-n_e$ bits, respectively) in one channel use (up to $n_1$ bits).
			\end{itemize}
			Note that in both cases, receiver 1 could decode the message to receiver 2 (due to the degradedness of the channel). Thus $m_2$ could be regarded as side information available at receiver 1. This advantage could be explored in the transmission phase where part of $m_1$ could be secured via one-time pad \cite{src:Shannon1949} with $m_2$ while the rest via Wyner's secrecy coding \cite{src:Wyner1975, Csisz'ar:Broadcast78}. Besides, compared to the capacity region without any secrecy constraints, there is $n_e$ bits loss for the maximal transmission rates $R_1, R_2,$ respectively, due to the individual secrecy constraint.
		\item Under joint secrecy constraint, the capacity region is a {\em triangle with one missing corner}. Compared to the capacity region without any secrecy constraints, there is not only a loss of $n_e$ bits for the maximal transmission rates $R_1, R_2,$ respectively, (as under the individual secrecy constraint), but also $n_e$ bits loss for the sum rate $R_1+R_2.$ This additional loss on the sum rate $R_1+R_2$ is due to the fundamental difference between the {\em joint} secrecy \eqref{eq:IndSec} and the {\em individual} secrecy \eqref{eq:JointSec} constraints.
	\end{enumerate}
 
 
 \section{DM-BC with an external eavesdropper}\label{sec:DMC}
 
 In this section, we investigate the DM-BC with an external eavesdropper where the individual secrecy constraint is imposed. For this channel model, similar to the discussion in \cite{src:ChenKS2014_ISIT}, positive rate pairs  (i.e., $(R_1, R_2)\in \mathcal{R}_{+}^2$) are not possible, if the eavesdropper's channel is less noisy than either legitimate receiver's channel. 
 
\subsection{Primitive approach} \label{sec: primitive approach}

A primitive approach is to utilize the secrecy coding while regarding one message as (partial) randomness for ensuring the individual secrecy of the other. The idea is similar to Carleial-Hellman's secrecy coding \cite{src:Carleial-Hellman1977}, in the sense that the eavesdropper should be kept totally ignorant of each message individually. Differently from \cite{src:Carleial-Hellman1977}, here two messages are aimed at different destinations. As a direct consequence (if only such a secrecy coding is employed), the sum rate is limited by the worse channel to the legitimate receivers. 
 
\begin{theorem}\label{thm: IndS by primitive approach}
For the DM-BC with an external eavesdropper, an achievable individual secrecy rate region is given by the union of the rate pairs $(R_1, R_2)\in \mathcal{R}_{+}^2$ satisfying
 \begin{align}
 \begin{split}\label{eqn: Region_Primitive}
 		R_1+R_{2} &\leq \min\{I(U;Y_1), I(U;Y_2)\}\\
 		\max\{R_1, R_2\} & \leq \min\{I(U;Y_1), I(U;Y_2)\}-I(U;Z)
 \end{split}
 \end{align} 
 over all $p(u)p(x|u).$ 
\end{theorem}
 
\begin{IEEEproof}
See Appendix~\ref{sec:AppPrimitive}.
\end{IEEEproof}

An intuitive interpretation of the achievable region in \eqref{eqn: Region_Primitive} is as follows. The first inequality in \eqref{eqn: Region_Primitive} imposes condition on the sum rate $R_1+ R_2,$ which is due to decodability constraints at both legitimate receivers. While, the second inequality imposes condition on both individual rates, i.e., $R_1, R_2.$ This follows from the spirit of Carleial-Hellman's secrecy coding for the purpose of individual secrecy. That is, for each message, at least $I(U;Z)$ randomness is needed to keep it secret from the eavesdropper.  

Note that \eqref{eqn: Region_Primitive} can be rewritten in a compact form as follows:
\begin{equation}\label{eqn: R_sum}
	\max\{R_1+R_{2}, \max\{R_1, R_2\}+I(U;Z)\} \leq \min\{I(U;Y_1), I(U;Y_2)\}.
\end{equation}
Remarkably, in the case that the left-hand side (LHS) of \eqref{eqn: R_sum} equals to $R_{1}+R_{2},$ we have $R_{1}, R_{2}\geq I(U;Z)$ and it implies that $U^n$ codewords are fully employed to carry the  individually secured messages. (That is, each message plays also the role of randomness to ensure the secrecy of the other and no additional randomness is needed since $R_{i}=H(M_i)/n\geq I(U;Z)$). In the other case, additional randomness is needed to ensure the individual secrecy of both messages. 

This primitive approach is optimal if the channels to both legitimate receivers are statistically identical.
However, it is not optimal in general, since both rates are limited by the worse legitimate receiver. For instance, suppose that $Y_1$ is strictly less noisy than $Y_2=Z$ (i.e., $I(U;Y_1)>I(U;Y_2)=I(U;Z)$ for any $p(u,x)$.) Then, employing this primitive approach will convey no secret information to either legitimate receiver. On the other hand, positive secret rate (i.e., $R_1\in \mathcal{R}_{+}$) is clearly possible by ignoring the worse legitimate receiver, and employing Wyner's secrecy coding for the resulting wiretap channel \cite{src:Wyner1975}. 

In order to further employ the channel advantage of the strong legitimate receiver against the eavesdropper, we propose the following superposition coding approach.

\subsection{Superposition coding approach}\label{sec: DB-BC superposition}

It is well-known that superposition coding is optimal for a degraded broadcast channel where $X\to Y_1\to Y_2$ forms a Markov chain, wherein the weak receiver could decode the cloud center whilst the strong receiver could decode both the cloud center and satellite codewords~\cite{ElGamal:2012}. Such a coding scheme explores the channel advantage of the stronger legitimate receiver, so that the messages conveyed are not bounded by the worse channel. Utilizing such a superposition coding framework with embedded Carleial-Hellman's secrecy coding in the layer of cloud codeword and Wyner's secrecy coding in the layer of the satellite codeword, we have the following achievable individual secrecy rate region.

 \begin{theorem}\label{thm: IndS by superposition}
For the DM-BC with an external eavesdropper, an achievable individual secrecy rate region is given by the union of rate pairs $(R_1,R_2)\in \mathcal{R}_{+}^2$ with $R_1=R_{1s}+R_{1k},$ where $(R_{1s}, R_{1k})\in \mathcal{R}_{+}^2,$ that satisfies
 \begin{align}
 \begin{split}\label{eqn: Region_Superposition R_1s R1k R2}
		R_{k}\leq  & I(U;Y_2) \\
		R_{1s} \leq & I(V_1;Y_1|U)-I(V_1;Z|U) \\
		R_{k} +R_{1s}\leq & I(U,V_1;Y_1)-I(V_1;Z|U)		
 \end{split}
 \end{align} 
 with
 \begin{equation} \label{eqn: R_{1k2}}
 	R_k= \max\left\{R_{1k}+R_{2} , \max\{R_{1k}, R_{2}\}+I(U;Z)\right\}
 \end{equation}
 Or, equivalently in terms of $(R_1, R_2),$
 \begin{align}
\begin{split} \label{eqn: Region_Superposition R_1 R_2}
 				R_{2}  &\leq I(U;Y_2)-I(U;Z),\\
 				R_{1} &\leq  [I(V;Y_1|U)-I(V;Z|U)]^++I(U;Y_2)-I(U;Z)\\
 				\max\{R_1, R_2\} &\leq   [I(V;Y_1|U)-I(V;Z|U)]^++ I(U;Y_1)-I(U;Z)\\
 				R_{1}+R_{2} & \leq [I(V;Y_1|U)-I(V;Z|U)]^+ +\min\{I(U;Y_1), I(U;Y_2)\}
 \end{split}
 \end{align}
 over all $p(u)p(v|u)p(x|v).$ 
 \end{theorem}
 
\begin{IEEEproof}
See Appendix~\ref{sec:AppSuperposition}.
\end{IEEEproof}

Our proposed superposition coding scheme consists of two coding layers. In particular, $m_2$ and part of $m_1$ (say $m_{1k}$) are conveyed via the first layer by employing Carleial-Hellman's secrecy coding, where each message not only plays the role of being the information to be destined to a different legitimate but also the (partial) randomness for the other message to be (individually) secured from the eavesdropper. 
In the second layer, extra information is conveyed via the satellite codewords to one of the receivers (assumed receiver 1 here), in which an extra part of the message (say $m_{1s}$) is secured by employing Wyner's secrecy coding.  Applying this superposition coding with embedded different secrecy coding in two coding layers, one readily achieves the  individual secrecy rate region as provided in Theorem \ref{thm: IndS by superposition}. We note that Theorem \ref{thm: IndS by superposition} does not require any less noisiness order between the legitimate receivers. 

We note that, the first inequality (i.e., the bound on $R_{k}$) in \eqref{eqn: Region_Superposition R_1s R1k R2} is contributed by the cloud codewords in the first coding layer for $(m_{1k}, m_2)$ and the fact that the cloud codewords will be decoded at receiver 2; whilst the second inequality in \eqref{eqn: Region_Superposition R_1s R1k R2} gives the extra secret information (if any) for the receiver 1, i.e., achievable $R_{1s},$ that is carried by the satellite codewords in the second coding layer for $m_{1s}$ (just as for a classical wiretap channel); the third inequality in \eqref{eqn: Region_Superposition R_1s R1k R2} comes from the fact that receiver 1 uses indirect decoding to decode $m_1=(m_{1k}, m_{1s})$ and there is a rate loss of $I(V_1;Z|U)$ for the sake of the individual secrecy of the message. 

Such a superposition coding scheme explores not only the advantage of Carleial-Hellman's secrecy coding for the purpose of individual secrecy that is discussed in the primitive approach, but also the channel advantage of the strong receiver (since he/she may decode both the cloud and satellite codewords) to obtain extra gains in the secret rate, i.e., $R_{1s}.$ Assuming that  $Y_1$ is less noisy than $Y_2$, Theorem \ref{thm: IndS by superposition} reduces to the following.
 \begin{corollary}\label{Cor: IndS by superposition}
For the DM-BC with an external eavesdropper such that $Y_1$ is less noisy than $Y_2$, an achievable individual secrecy rate region is given by the union of rate pairs $(R_1,R_2)\in \mathcal{R}_{+}^2$ with $R_1=R_{1s}+R_{1k},$ where $(R_{1s}, R_{1k})\in \mathcal{R}_{+}^2,$ that satisfies
 \begin{align}
 \begin{split}\label{eqn: Region_Superposition Less R_1s R1k R2}
  		 R_{k} &\leq  I(U;Y_2)		\\
  		 R_{1s}	&\leq I(V;Y_1|U)-I(V;Z|U)
 \end{split}
 \end{align} 
 with $R_{k}$ as defined in \eqref{eqn: R_{1k2}}. 
 Or, equivalently in terms of $(R_1, R_2),$
 \begin{align}
\begin{split} \label{eqn: Region_Superposition Less R_1 R_2}
 				R_{2}  &\leq  I(U;Y_2)-I(U;Z),\\
 				R_{1} &\leq  [I(V;Y_1|U)-I(V;Z|U)]^++I(U;Y_2)-I(U;Z)\\
 				R_{1}+R_{2} & \leq [I(V;Y_1|U)-I(V;Z|U)]^++I(U;Y_2)
 \end{split}
 \end{align}
 over all $p(u)p(v|u)p(x|v).$ 
 \end{corollary}

\emph{Remark:} We have the following interesting observations: 
\begin{itemize}
	\item Setting $U=\emptyset,$ i.e., $R_2=R_{1k}=0,$ the region \eqref{eqn: Region_Superposition Less R_1s R1k R2} of $R_1=R_{1s}$ coincides with the secrecy capacity region of the wiretap channel \cite{src:Wyner1975, Csisz'ar:Broadcast78};
	\item If we let $Z=\emptyset,$ and $R_{1k}=0,$ the region \eqref{eqn: Region_Superposition Less R_1s R1k R2} reduces to the capacity region of the degraded broadcast channel, as established in \cite{src:Cover1972,src:Bergmans1973,src:Gallager1974}.
	\item If we let $R_{1k}=0$, then the region \eqref{eqn: Region_Superposition Less R_1s R1k R2} reduces to the joint secrecy capacity region of the degraded broadcast channel \cite{src:Ekrem2009, Bagherikaram:Secrecy09}.  The proof follows when the individual secrecy constraints (i.e., \eqref{eq:IndSec}), are replaced with joint secrecy constraints (i.e., \eqref{eq:JointSec}), for which the resulting coding scheme, as shown in \cite{src:Ekrem2009, Bagherikaram:Secrecy09}, achieves the joint secrecy capacity region for the degraded broadcast channel.
\end{itemize}

   \begin{theorem}\label{thm: IndSC_Deterministic}
  For the DM-BC with an external eavesdropper such that 
  \begin{enumerate}
  	\item $Y_1$ is less noisy than $Y_2;$
  	\item $Y_2$ is a deterministic function of $X;$ and 
  	\item $Z$ is a degraded version of $Y_2,$
  \end{enumerate}
  the individual secrecy capacity region is given by the union of rate pairs $(R_1,R_2)\in \mathcal{R}_{+}^2$ satisfying
   \begin{align}
  \begin{split} \label{eqn: InSCRegion_Deterministic R_1 R_2}
   				R_{2}  &\leq H(Y_2|Z)\\
   				R_{1} &\leq  I(X;Y_1)-I(X;Z)\\
   				R_1+R_2 &\leq I(X;Y_1)
   \end{split}
   \end{align}
   over all $p(x).$ 
   \end{theorem}
   \begin{IEEEproof}
   The achievability follows directly from Corollary \ref{Cor: IndS by superposition} by taking $U=Y_2$ and $V=X.$ (Note that in case that $Y_2$ is a deterministic function of $X,$ and $X\to Y_2\to Z$ forms a Markov chain, we have $I(X;Z|Y_2)=0,$ $H(Y_2)=I(X;Y_2)$ and $I(Y_2;Z)=I(X;Z)$.) For the converse, the first two inequalities for $R_1, R_2,$ respectively, follow directly from the classical results of wiretap channel by simply ignoring the other legitimate receiver \cite{Csisz'ar:Broadcast78}. And, the last inequality follows directly from the upper bound on the sum rate for the relaxed setting of without any secrecy constraints.
   \end{IEEEproof} 
 
  
 
 In the following, we provide an upper bound on the individual secrecy capacity region, that will be used to derive a special case secrecy capacity result in the sequel.
 
 \begin{theorem}\label{thm: DBC IndS upper bound}
 For the DM-BC with an external eavesdropper such that 
 \begin{enumerate}
 \item $Y_2$ is a degraded version of $Y_1;$ and
 \item$Y_2$ is less noisy than $Z,$ 
 \end{enumerate}
 the individual secrecy capacity region is upper bounded by the union of rate pairs $(R_1,R_2)\in \mathcal{R}_{+}^2$ satisfying
  \begin{align}
 \begin{split} \label{eqn: InSC Region_Upper Bound R_1 R_2}
  				R_{2}  &\leq I(U;Y_2)-I(U;Z)\\
  				R_{1} &\leq  I(V;Y_1|U)-I(V;Z|U)+I(U;Y_2)-I(U;Z)\\
  				R_1+R_2 \leq & 	I(V;Y_{1}|U)+I(U;Y_2)
  \end{split}
  \end{align}
  over all $p(u)p(v|u)p(x|v).$ 
  \end{theorem}
  
 \begin{IEEEproof}
 See Appendix~\ref{sec:UpperBound1}.
 \end{IEEEproof} 
 
 \begin{theorem}\label{thm: IndSC by superposition}
For the DM-BC with an external eavesdropper such that 
\begin{enumerate}
	\item $Y_2$ is a degraded version of $Y_1;$ 
	\item $Y_2$ is less noisy than $Z;$ and
	\item $I(U;Z)\leq I(U;Y_2)\leq 2 I(U;Z)$ holds for any $p(u,v,x)$,
\end{enumerate}
the individual secrecy capacity region is given by the union of rate pairs $(R_1,R_2)\in \mathcal{R}_{+}^2$ satisfying
 \begin{align}
\begin{split} \label{eqn: InSCRegion_Superposition R_1 R_2}
 				R_{2}  &\leq I(U;Y_2)-I(U;Z),\\
 				R_{1} &\leq  I(V;Y_1|U)-I(V;Z|U)+I(U;Y_2)-I(U;Z)
 \end{split}
 \end{align}
 over all $p(u)p(v|u)p(x|v).$ 
 \end{theorem}
 \begin{IEEEproof}
 The achievability follows from Corollary \ref{Cor: IndS by superposition} when the channel fulfills the conditions 1), 2) and 3). In particular, the sum rate condition becomes redundant due to condition 3). Since the derived region in this case coincides with the upper bound in Theorem \ref{thm: DBC IndS upper bound}, it gives the individual secrecy capacity region. 
 \end{IEEEproof}
 
\emph{Remark:} One can recall the linear deterministic BC with an external eavesdropper. In particular, it is an instance of Theorem \ref{thm: IndSC_Deterministic}  in case of $n_1\geq n_2\geq n_e;$ and an instance of Theorem \ref{thm: IndSC by superposition} in case of $n_e\leq n_2\leq 2 n_e$. Its individual secrecy capacity is shown in Fig. \ref{fig:deterministic BC}, and can be obtained by taking $U=Y_2=D^{n_1-n_2}X$  (it is assumed that $n_2\leq n_1$) and $V=X$ in the superposition coding as described in Theorem \ref{thm: IndS by superposition}.

\subsection{Marton's coding approach}

In the previous subsection, the superposition coding approach is shown to be optimal for some special cases if the receivers and the eavesdropper fulfill a certain degradation/less noisiness order. 

 Here, we consider the general case where there may not be degradation/less noisiness order between the legitimate receivers, and devise a coding scheme by utilizing Marton's coding, attempting to send extra secret information to both legitimate receivers. In particular, the common message extended version of Marton's coding approach allows for a transmission of a cloud center to \emph{both} receivers. In addition to this cloud center, two separate codewords can be formed via the Marton's coding. We have the following result.
 
\begin{theorem}\label{thm:Marton}
For the DM-BC with an external eavesdropper, an achievable  individual secrecy rate region is given by the union of rate pairs $(R_1,R_2)\in \mathcal{R}_{+}^2$ with $R_1=R_{1s}+R_{1k}$ and $R_2=R_{2s}+R_{2k}$, where $(R_{1k}, R_{1s}, R_{2k}, R_{2s})\in \mathcal{R}_{+}^4,$ that satisfies 
\begin{align}
\begin{split}\label{eqn: DM-BC IndS region by Marton coding}
	R_{1s} \leq & I(V_1;Y_1|U)-I(V_1;Z|U) \\
	R_{2s} \leq & I(V_2;Y_2|U)-I(V_2;Z|U)\\
	R_{k} +R_{1s}\leq & I(U,V_1;Y_1)-I(V_1;Z|U)\\
	R_{k} +R_{2s}\leq & I(U,V_2;Y_2)-I(V_2;Z|U)
\end{split}
\end{align}
with
\begin{equation} \label{eqn: R_{k}}
	R_k= \max\left\{R_{1k}+R_{2k} , \max\{R_{1k}, R_{2k}\}+I(U;Z)\right\}
\end{equation}
over all $p(u)p(v_1,v_2|u)p(x|v_1,v_2)$ subject to $I(V_1;V_2|U)+I(V_1,V_2; Z|U)\leq I(V_1;Z|U)+I(V_2;Z|U);$

Or, equivalently in terms of $(R_1, R_2),$
\allowdisplaybreaks
\begin{align}
\begin{split} \label{eqn: Region_Marton R_1 R_2}
	R_2 &\leq \left[I(V_2;Y_2|U)-I(V_2;Z|U)\right]^+ +I(U;Y_2)-I(U;Z)\\
	R_2 &\leq \sum_{i=1}^{2} \left[I(V_i;Y_i|U)-I(V_i;Z|U)\right]^+ +I(U;Y_1)-I(U;Z)\\
	R_1 &\leq \left[I(V_1;Y_1|U)-I(V_1;Z|U)\right]^+ +I(U;Y_1)-I(U;Z)\\
	R_1 &\leq \sum_{i=1}^{2} \left[I(V_i;Y_i|U)-I(V_i;Z|U)\right]^+  +I(U;Y_2)-I(U;Z)\\
	R_1+R_2 &\leq \sum_{i=1}^{2} \left[I(V_i;Y_i|U)-I(V_i;Z|U)\right]^+ +\min\{I(U; Y_1), I(U; Y_2)\}
 \end{split}
\end{align}
over all $p(u)p(v_1,v_2|u)p(x|v_1,v_2)$ subject to $I(V_1;V_2|U)+I(V_1,V_2; Z|U)\leq I(V_1;Z|U)+I(V_2;Z|U).$
\end{theorem}
 
\begin{IEEEproof}
See Appendix~\ref{sec:AppMarton}.
\end{IEEEproof}

The coding approach we develop here is built on the aforementioned primitive approach and superposition approach, but with the framework of Marton's coding. That is, we split $M_{i}$ into $M_{i}=(M_{ik}, M_{is})$, for $i=1,2.$ In particular, $(M_{1k}, M_{2k})$ are encoded into the cloud codeword $U^n$ (as in the primitive approach), where individual secrecy is guaranteed by employing Carleial-Hellman's secrecy coding; moreover, additional information $M_{1s}, M_{2s}$ are carried by individual satellite codewords $V_1^n, V_2^n,$ respectively, (as in the superposition approach for each legitimate receiver). Note that, the secrecy of $M_{is}$ for $i=1,2,$ is ensured by employing Wyner's secrecy coding. Finally, following the spirit of Marton's coding, $(V_1^n, V_2^n)$ is chosen jointly, and corresponding codeword $X^n$ is sent to the channel.

As reflected in the obtained region in \eqref{eqn: DM-BC IndS region by Marton coding}, $R_{k}$ (as defined in \eqref{eqn: R_{k}}) is contributed by applying Carleial-Hellman's secrecy coding in the cloud layer on $(M_{1k}, M_{2k})$ to obtain their individual secrecy; the first two inequalities are contributed by employing Wyner's secrecy coding in the individual satellite layer to ensure the secrecy of the extra message $M_{is}$ to each legitimate receiver $i$. The last two inequalities in \eqref{eqn: DM-BC IndS region by Marton coding} come from the fact that receiver $i,$ $i=1,2,$ uses indirect decoding to decode $m_i=(m_{ik}, m_{is})$ and there is a rate loss of $I(V_i;Z|U)$ for the sake of the individual secrecy. 

 \emph{Remark:} We report the following observations: 
 \begin{itemize}
	 \item Setting $V_1, V_2, X=U,$ i.e., $R_{1s}=R_{2s}=0,$ the region reduces to the one in \eqref{eqn: Region_Primitive} by the primitive approach. 
	 \item If we let $V_2=U$ and $X=V_1,$ i.e., $R_{2s}=0,$ the region reduces to the one in \eqref{eqn: Region_Superposition R_1s R1k R2} by the superposition approach. 
 \end{itemize}
 
 \subsection{Joint secrecy rate region}
Revising the secrecy proofs by fulfilling the joint secrecy constraints \eqref{eq:JointSec} (instead of the individual secrecy constraints \eqref{eq:IndSec} as considered in previous subsections), we obtain achievable joint secrecy rate region by utilizing the aforementioned coding approaches. We note that both the primitive approach and the superposition approach serve as underneath coding layers for the Marton's coding approach. Therefore, their resultant rate regions are also included as special cases of the region derived by the Marton's coding approach, which is given as follows. 
 
 \begin{theorem} \label{thm: JoS Marton}
 (Achievable joint secrecy rate region via Marton's coding)
 	For the DM-BC with an external eavesdropper, an achievable joint secrecy rate region  obtained by Marton's coding is the union of the rate pairs $(R_1, R_2)\in \mathcal{R}_{+}^2$ satisfying
 		\allowdisplaybreaks
 		\begin{align}\label{eqn: JoS Marton}
 		\begin{split}
 			R_1& \leq \left[I(V_1;Y_1|U)-I(V_1;Z|U)\right]^+ +I(U;Y_1)-I(U;Z)\\
 			R_2& \leq \left[I(V_2;Y_2|U)-I(V_2;Z|U)\right]^+ +I(U;Y_2)-I(U;Z)\\
 			R_1+R_2& \leq \sum_{i=1}^{2} \left[I(V_i;Y_i|U)-I(V_i;Z|U)\right]^++\min\{I(U;Y_1), I(U;Y_2)\}-I(U;Z)
 		\end{split}
 		\end{align}
 	over any  $p(u,v_1,v_2,x)=p(u)p(v_1,v_2|u)p(x|v_1,v_2)$ subject to $I(V_1, V_2;Z|U)\leq I(V_1;Z|U)+I(V_2;Z|U)-I(V_1;V_2|U).$ 
 \end{theorem}
 
 \begin{IEEEproof}
 See Appendix~\ref{sec:App Jos Marton}.
 \end{IEEEproof}
 
 \emph{Remark:} We have the following observations:
 \begin{itemize}
	\item In case that $Y_1$ is less noisy than $Y_2$, one can take $U=V_2,$ then the region \eqref{eqn: JoS Marton} reduces to
\begin{align}\label{eqn: JoS Superposition}
	 		\begin{split}
	 			R_2& \leq I(U;Y_2)-I(U;Z)\\
	 			R_1+R_2& \leq [I(V_1;Y_1|U)-I(V_1;Z|U)]^+ +I(U;Y_2)-I(U;Z).
	 		\end{split}
\end{align}
As shown in \cite{src:Ekrem2009, Bagherikaram:Secrecy09}, \eqref{eqn: JoS Superposition} is the joint secrecy capacity region of the degraded broadcast channel. Interestingly, comparing \eqref{eqn: JoS Superposition} with \eqref{eqn: InSCRegion_Superposition R_1 R_2}, the only difference is that the term $[I(V_1;Y_1|U)-I(V_1;Z|U)]^+ +I(U;Y_2)-I(U;Z)$ upper bounds the sum rate $R_1+R_2$ in \eqref{eqn: JoS Superposition} under the joint secrecy constraint, while it upper bounds $R_1$ in \eqref{eqn: InSCRegion_Superposition R_1 R_2} under the individual secrecy constraint. This implies that in case of a comparable eavesdropper (i.e., $I(U;Z)\leq I(U;Y_2)\leq 2 I(U;Z)$ holds for any $p(u,v,x)$), the strong receiver could gain in the transmission rate up to that of the weak receiver when a weaker (individual) secrecy constraint is imposed.   
	\item Compare \eqref{eqn: JoS Marton} with the individual secrecy achievable region in \eqref{eqn: Region_Marton R_1 R_2}. There is a gain of $I(U;Z)$ bits on the sum transmission rate $R_1+R_2$ as a trade for having a weaker notion of security. 
	\item \cite[Theorem 1]{src:Ekrem2013} gives an achievable rate region of the BC with two receivers and one eavesdropper, where the transmitter wants to transmit a pair of public and confidential messages to each legitimate receiver. (No secrecy constraints on the public messages, but two confidential messages are required to fulfill the joint secrecy constraint at the eavesdropper.) Setting both rates for two public messages to be zero in \cite[Theorem 1]{src:Ekrem2013}, one can obtain the following achievable joint secrecy rate region:
	\begin{align}\label{eqn: JoS by Ekrem2013}
		 		\begin{split}
		 			R_1& \leq [I(V_1;Y_1|U)-I(V_1;Z|U)]^++\min\{I(U;Y_1), I(U;Y_2)\}-I(U;Z)\\
		 			R_2& \leq [I(V_2;Y_2|U)-I(V_2;Z|U)]^++\min\{I(U;Y_1), I(U;Y_2)\}-I(U;Z)\\
		 			R_1+R_2& \leq \sum_{i=1}^{2} I(V_i;Y_i|U)+\min\{I(U;Y_1), I(U;Y_2)\} -I(V_1;V_2|U)-I(U,V_1,V_2;Z)
		 		\end{split}
	\end{align}
	 	over any  $p(u,v_1,v_2,x)=p(u)p(v_1,v_2|u)p(x|v_1,v_2)$ subject to $I(V_1, V_2;Z|U)\leq I(V_1;Z|U)+I(V_2;Z|U)-I(V_1;V_2|U)$ and $I(U;Z)\leq I(U;Y_i)$, $I(V_i;Z|U)\leq I(V_i;Y_i|U)$ for $i=1,2.$ 
	 	
	Comparing \eqref{eqn: JoS by Ekrem2013} with our joint secrecy rate region result in \eqref{eqn: JoS Marton}, we see that our upper bounds on $R_1, R_2$ are potentially greater while the upper bound on $R_1+R_2$ is potentially smaller. The reason is caused by the fact that in our achievablility scheme, indirect decoding is applied at each legitimate receiver (note that joint unique decoding works the same here without any potential rate loss); while in \cite{src:Ekrem2013}, sequential decoding is employed at both legitimate receivers (i.e., decode $U^n$ first, then $V_i^n.$ This also results in an additional constraint on $U,$ i.e., $I(U;Z)\leq I(U;Y_i)$ for $i=1,2$). Besides, the difference on the sum rate bound is due to the fact that in our joint secrecy proof, $V_1^n, V_2^n,$ are processed individually, whereas in \cite[Theorem 1]{src:Ekrem2013} $(V_1^n, V_2^n)$ as jointly.
 \end{itemize}


\section{Gaussian BC with an external eavesdropper} \label{sec:gaussian}

\begin{figure}
\centering
  \includegraphics[width=0.75\textwidth]{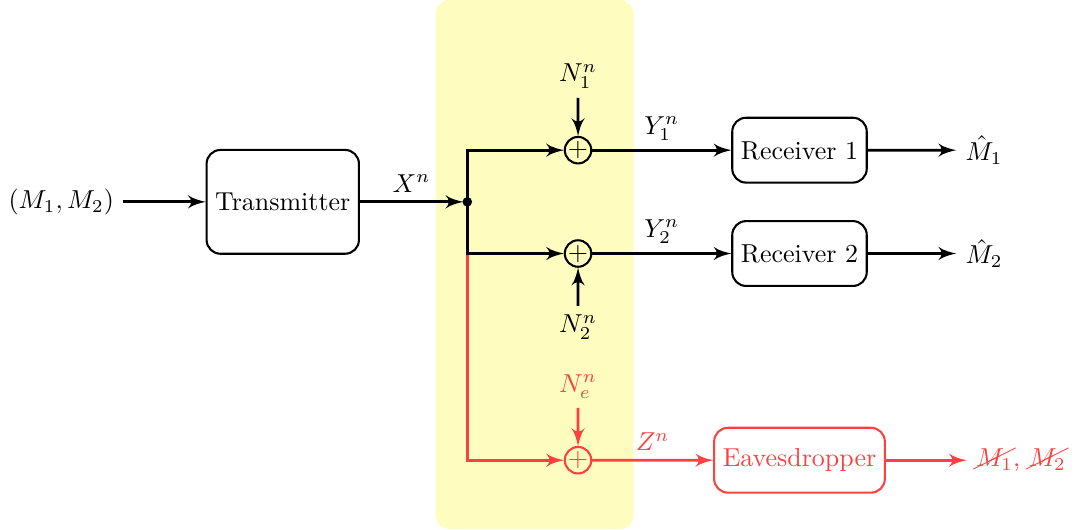}
\caption{Gaussian BC with an external eavesdropper.}
\label{fig:gaussian}
\end{figure}

The Gaussian BC with an external eavesdropper is shown in Fig. \ref{fig:gaussian}. Suppose ${X}$ is the channel input with a power constraint $P$ on it and the signals received by both receivers and the eavesdropper are  given by
	\begin{align*}
		{Y}_1&={X}+{N}_{1};\\
		{Y}_2&={X}+{N}_{2};\\
		{Z}&={X}+{N}_{e},
	\end{align*}
where ${N}_{1}, {N}_{2}$ and ${N}_{e}$ are additive white Gaussian noise (AWGN) independent of ${X}$, where ${N}_{1}\sim \mathcal{N}(0, \sigma_1^2), $  ${N}_{2}\sim \mathcal{N}(0, \sigma_2^2)$ and ${N}_{e}\sim \mathcal{N}(0, \sigma_e^2), $ respectively.

According to the noise level in the channels to both receivers and the eavesdropper, the overall channel can be regarded to be {\it stochastically} degraded. For simplicity, we only consider its corresponding {\it physically} degraded instances. The reason is that the same analysis can be easily extended to the stochastically degraded case. That is, the scenario: $\sigma_e^2\geq \sigma_2^2 \geq \sigma_1^2,$ as $X\to Y_1\to Y_2 \to Z$ forms a Markov chain, is of our interest.

In the previous section, single-letter expressions for the achievable individual secrecy rate regions and upper bounds have been proposed  for the DM-BC, which involve auxiliary variable $U$ and $V$. Applying the standard discretization procedure \cite{ElGamal:2012}, one can extend these results to the Gaussian case. However, it is in general not clear what would be the optimal choice of $(U, V).$ In this section, we are going to derive \emph{computable} inner and outer bounds on the individual secrecy capacity region of the Gaussian BC with an external eavesdropper. Interestingly, we show that our inner bound (by employing the superposition coding)  approaches the individual secrecy capacity region within a constant gap (i.e., 0.5 bits). 

\subsection{An outer bound}

\begin{theorem}\label{prop: Gaussian upper bound}  An outer bound to the individual secrecy capacity region for the Gaussian BC with an external eavesdropper (where $X\to Y_1 \to Y_2 \to Z$ forms a Markov chain) is given by the union of the rate pairs $(R_1, R_2)\in \mathcal{R}_{+}^2$ satisfying
 \begin{align}
 		R_{1}\leq & C\left(\frac{\alpha(1-\gamma) P}{\gamma\alpha P+\sigma_1^2}\right)- C\left(\frac{\alpha(1-\gamma) P}{\gamma\alpha P+\sigma_e^2}\right)+\min\left\{
 		R_2,  C\left(\frac{(1-\gamma\alpha)P}{\gamma\alpha P+\sigma_e^2}\right)\right\} 
 		\label{eqn: IndS Gaussian Outer bound on R1}\\
 		R_{2}\leq & C\left(\frac{(1-\alpha)P}{\alpha P+\sigma_2^2}\right)-C\left(\frac{(1-\alpha) P}{\alpha P+\sigma_e^2}\right)\label{eqn: IndS Gaussian Outer bound on R2},
 \end{align}
where $\alpha, \gamma\in[0,1].$ 
\end{theorem}
\begin{IEEEproof} 
First let us consider $R_2.$
\begin{align*}
nR_2&=	H(M_2)\stackrel{(a)}{\leq}I(M_2;Y_2^n)+n \lambda_2(\epsilon_n)\\
	&\stackrel{(b)}{\leq} I(M_2;Y_2^n)-I(M_2; Z^n)+n \lambda_2(\epsilon_n, \tau_n)\\
	&{=} \underbrace{h(Y_2^n)-h(Z^n)}_{nR_2^1}-(\underbrace{h(Y_2^n|M_2)-h(Z^n|M_2)}_{nR_2^2})+n \lambda_2(\epsilon_n, \tau_n),
\end{align*}
where 
	$(a)$ is due to the reliability constraint \eqref{eq:Reliability}, Fano's inequality and by taking $\lambda_2(\epsilon_n)=1/n+\epsilon_n R_2;$ 
	$(b)$ is due to the individual secrecy constraint \eqref{eq:IndSec} and by taking $\lambda_2(\epsilon_n, \tau_n)=\tau_n+\lambda_2(\epsilon_n).$

Note that according to \cite[Lemma 10 and equation (75)]{src:Leung1978}, $nR_2^1$ can be bounded by:
\begin{equation}\label{eqn: R_21}
	nR_2^1=h(Y_2^n)-h(Z^n)\leq \frac{n}{2}\log \frac{P+\sigma_2^2}{P+\sigma_e^2}.
\end{equation}
Further, due to the channel degradedness, we have for $nR_2^2:$
\begin{align*}
	nR_2^2&\geq h(Y_2^n|X^n)-h(Z^n|X^n)=\frac{n}{2}\log \frac{\sigma_2^2}{\sigma_e^2};\\
	nR_2^2 &\leq h(Y_2^n)-h(Z^n)\leq \frac{n}{2}\log \frac{P+\sigma_2^2}{P+\sigma_e^2}.
\end{align*} 
Hence, there exists an $\alpha\in [0, 1]$ such that 
\begin{equation}\label{eqn: R_22}
	nR_2^2=h(Y_2^n|M_2)-h(Z^n|M_2)=\frac{n}{2}\log \frac{\alpha P+\sigma_2^2}{\alpha P+\sigma_e^2}.
\end{equation}
Combining \eqref{eqn: R_21} and \eqref{eqn: R_22}, we have
\begin{align*}
	nR_2	&=nR_2^1-nR_2^2+n \lambda_2(\epsilon_n, \tau_n)\\
				&\leq \frac{n}{2}\log \frac{P+\sigma_2^2}{P+\sigma_e^2}-\frac{n}{2}\log \frac{\alpha P+\sigma_2^2}{\alpha P+\sigma_e^2}+n \lambda_2(\epsilon_n, \tau_n)\\
				&=\frac{n}{2}\log \frac{(P+\sigma_2^2)(\alpha P+\sigma_e^2)}{(\alpha P+\sigma_2^2)(P+\sigma_e^2)}+n \lambda_2(\epsilon_n, \tau_n).
\end{align*}
That is, 
\begin{equation}\label{eqn: Gaussian R2 outer proof}
	R_2\leq C\left(\frac{(1-\alpha)P}{\alpha P+\sigma_2^2}\right)-C\left(\frac{(1-\alpha) P}{\alpha P+\sigma_e^2}\right)+ \lambda_2(\epsilon_n, \tau_n).
\end{equation}

Now we proceed to bound $R_1.$ 
\begin{align}
	nR_1 & =	H(M_1)=H(M_1|M_2)\stackrel{(c)}{\leq}I(M_1;Y_1^n|M_2)+n \lambda_1(\epsilon_n)\nonumber\\
			&= I(M_1;Y_1^n|M_2)-I(M_1; Z^n|M_2)+I(M_1; Z^n|M_2)+n \lambda_1(\epsilon_n)\nonumber\\
			&= \underbrace{h(Y_1^n|M_2)-h(Z^n|M_2)}_{nR_1^1}-(\underbrace{h(Y_1^n|M_1, M_2)-h(Z^n|M_1, M_2)}_{nR_1^2})+\underbrace{I(M_1; Z^n|M_2)}_{nR_1^3}+n \lambda_1(\epsilon_n), \label{eqn: R_1 step 1}
\end{align}
where $(c)$ is due to the reliability constraint \eqref{eq:Reliability}, Fano's inequality and by taking $\lambda_1(\epsilon_n)=1/n+\epsilon_n R_1.$

Applying Costa's entropy power inequality (EPI) \cite[Theorem 1]{src:Costa1985} and using \eqref{eqn: R_22}, we obtain
\begin{equation}\label{eqn: upper bound nR_1^1}
	nR_1^1=h(Y_1^n|M_2)-h(Z^n|M_2)\leq \frac{n}{2}\log \frac{\alpha P+\sigma_1^2}{\alpha P+\sigma_e^2}.
\end{equation} 
(A more detailed proof of \eqref{eqn: upper bound nR_1^1} is given in Appendix \ref{Sec: Appdix proof using Costas EPI}.) 

For $nR_1^2,$ due to the channel degradedness, we have
\begin{align*}
	nR_1^2& \geq h(Y_1^n|X^n)-h(Z^n|X^n)=\frac{n}{2}\log \frac{\sigma_1^2}{\sigma_e^2};\\
	nR_1^2& \leq h(Y_1^n|M_2)-h(Z^n|M_2)\leq \frac{n}{2}\log \frac{\alpha P+\sigma_1^2}{\alpha P+\sigma_e^2}.
\end{align*} 
Hence, there exists a $\gamma\in[0,1]$ such that 
\begin{align} 
	nR_1^2	&=h(Y_1^n|M_1, M_2)-h(Z^n|M_1, M_2) \nonumber\\
			&=\frac{n}{2}\log \frac{\gamma\alpha P+\sigma_1^2}{\gamma\alpha P+\sigma_e^2}.\label{eqn: nR_1^2 value}
\end{align}
Applying the entropy power inequality (EPI) \cite{src:Shannon1948} and using \eqref{eqn: nR_1^2 value} for $h(Y_1^n|M_1, M_2)-h(Z^n|M_1, M_2)$, we can bound $h(Z^n|M_1, M_2)$ by
\begin{equation}\label{eqn: upper bound h(Z^n|M_1, M_2)}
	h(Z^n|M_1, M_2)\geq \frac{n}{2}\log 2\pi e (\gamma\alpha P+\sigma_e^2).
\end{equation}
(A more detailed proof of \eqref{eqn: upper bound h(Z^n|M_1, M_2)} is given in Appendix \ref{Sec: Appdix proof using EPI}.) 
 
For $nR_1^3,$ we observe that 
\begin{align*}
	nR_1^3	&=I(M_1; Z^n|M_2)=I(M_1, M_2;Z^n)-I(M_2;Z^n)\\
			&=I(M_2;Z^n|M_1)+I(M_1;Z^n)-I(M_2;Z^n)\\
			&\stackrel{(d)}{\leq}  nR_2+n \tau_n,
\end{align*}
where $(d)$ is due to the individual secrecy constraint \eqref{eq:IndSec}. 

Moreover, we can bound $nR_1^3$ as follows 
\begin{align*}
	nR_1^3	&=I(M_1; Z^n|M_2)=h(Z^n|M_2)-h(Z^n|M_1, M_2)\\
			&\leq h(Z^n)-h(Z^n|M_1, M_2)\\
			&\leq \frac{n}{2}\log \frac{P+\sigma_e^2}{\gamma\alpha P+\sigma_e^2}.
\end{align*}
Therefore,  we have so far 
\begin{align*}
n R_1	&=n R_1^1-n R_1^2+nR_1^3+n \lambda_1(\epsilon_n)\\
		&\leq \frac{n}{2}\log \frac{\alpha P+\sigma_1^2}{\alpha P+\sigma_e^2}-\frac{n}{2}\log \frac{\gamma\alpha P+\sigma_1^2}{\gamma\alpha P+\sigma_e^2}+
		\min\left\{
 		nR_2, \frac{n}{2}\log \frac{P+\sigma_e^2}{\gamma\alpha P+\sigma_e^2}\right\} +n \lambda_1(\tau_n, \epsilon_n) \\
		&=\frac{n}{2}\log \frac{\alpha P+\sigma_1^2}{\gamma\alpha P+\sigma_1^2} - \frac{n}{2}\log \frac{\alpha P+\sigma_e^2}{\gamma\alpha P+\sigma_e^2}+\min\left\{
		nR_2, \frac{n}{2}\log \frac{P+\sigma_e^2}{\gamma\alpha P+\sigma_e^2}\right\}+n \lambda_1(\tau_n, \epsilon_n),
\end{align*}
where $\lambda_1(\tau_n, \epsilon_n)=\tau_n+\lambda_1(\epsilon_n).$
That is,
\begin{equation} \label{eqn: Gaussian R1 outer proof}
		R_1\leq C\left(\frac{\alpha(1-\gamma) P}{\gamma\alpha P+\sigma_1^2}\right)- C\left(\frac{\alpha(1-\gamma) P}{\gamma\alpha P+\sigma_e^2}\right)+
			\min\left\{R_2,  C\left(\frac{(1-\gamma\alpha)P}{\gamma\alpha P+\sigma_e^2}\right)\right\}+ \lambda_1(\tau_n, \epsilon_n).
\end{equation}

Letting $n\to \infty, \tau_n, \epsilon_n\to 0,$ we have $\lambda_1(\tau_n, \epsilon_n), \lambda_2(\tau_n, \epsilon_n)\to 0;$ and \eqref{eqn: Gaussian R1 outer proof}, \eqref{eqn: Gaussian R2 outer proof} reduce to 
 \eqref{eqn: IndS Gaussian Outer bound on R1}, \eqref{eqn: IndS Gaussian Outer bound on R2}, respectively. This completes our proof.
\end{IEEEproof}

By Theorem \ref{prop: Gaussian upper bound}, we easily obtain a looser outer bound as described in the following corollary.

\begin{corollary}\label{Cor: Gaussian looser outer bound}  An outer bound to the individual secrecy capacity region for the Gaussian BC with an external eavesdropper (where $X\to Y_1 \to Y_2 \to Z$ forms a Markov chain) is given by the union of the rate pairs $(R_1, R_2)\in \mathcal{R}_{+}^2$ satisfying
 \begin{align}
	\begin{split}\label{eqn: IndS Gaussian looser Outer bound}
 		R_{1}\leq & C\left(\frac{\alpha P}{ \sigma_1^2}\right)- C\left(\frac{\alpha  P}{ \sigma_e^2}\right)+
 		C\left(\frac{(1-\alpha)P}{\alpha P+\sigma_2^2}\right)-C\left(\frac{(1-\alpha) P}{\alpha P+\sigma_e^2}\right)\\
 		R_{2}\leq & C\left(\frac{(1-\alpha)P}{\alpha P+\sigma_2^2}\right)-C\left(\frac{(1-\alpha) P}{\alpha P+\sigma_e^2}\right)\\
 		R_1+R_2\leq & C\left(\frac{\alpha P}{\sigma_1^2}\right)+C\left(\frac{(1-\alpha)P}{\alpha P+\sigma_2^2}\right),
	\end{split}
 \end{align}
 where $\alpha\in[0,1].$ 
\end{corollary}

\begin{IEEEproof}
See Appendix~\ref{app:CorollaryGaussianOuterBound}
\end{IEEEproof}

\subsection{An inner bound}
\begin{theorem} \label{prop: Gaussian lower bound} An inner bound of the individual secrecy capacity region for the Gaussian BC with an external eavesdropper (where $X\to Y_1 \to Y_2\to Z$ forms a Markov chain) is given by the union of the rate pairs $(R_1, R_2)\in \mathcal{R}_{+}^2$ satisfying
 \begin{align}
 \begin{split}\label{eqn: IndS Gaussian Inner bound}
	 	R_{1}\leq 
	 		& C\left(\frac{\alpha P}{\sigma_1^2}\right)-C\left(\frac{\alpha P}{\sigma_e^2}\right)+C\left(\frac{(1-\alpha)P}{\alpha P+\sigma_2^2}\right)-C\left(\frac{(1-\alpha)P}{\alpha P+\sigma_e^2}\right)\\ 
	 	 R_{2}\leq 
	 	 	& C\left(\frac{(1-\alpha)P}{\alpha P+\sigma_2^2}\right)-C\left(\frac{(1-\alpha) P}{\alpha P+\sigma_e^2}\right)\\
	 	 R_1+R_2 \leq & C\left(\frac{\alpha P}{\sigma_1^2}\right)-C\left(\frac{\alpha P}{\sigma_e^2}\right)+C\left(\frac{(1-\alpha)P}{\alpha P+\sigma_2^2}\right),	
 \end{split}
 	\end{align}
 where $\alpha\in[0,1].$
\end{theorem}
\begin{IEEEproof}
The region is obtained from Theorem~\ref{thm: IndS by superposition} by using jointly Gaussian $(U, V)$ with $U\sim \mathcal{N}(0, (1-\alpha)P)$, $V\sim \mathcal{N}(0, \alpha P)$, $X=U+V$, where $U$ and $V$ are independent and $\alpha\in[0,1]$.
\end{IEEEproof}
\subsection{Individual secrecy capacity region}
\begin{theorem}\label{Thm: Gaussian IndSec capacity}
	As $\sigma_e^2\geq \sigma_2^2\geq \sigma_1^2$ and $P\geq \sigma_e^2(\sigma_e^2-2\sigma_2^2)/\sigma_2^2,$ the individual secrecy capacity region for the Gaussian BC with an external eavesdropper is given by the union of the rate pairs $(R_1, R_2)\in \mathcal{R}_{+}^2$ satisfying
 	\begin{align*}
 		R_{1}\leq & C\left(\frac{\alpha P}{\sigma_1^2}\right)-C\left(\frac{\alpha P}{\sigma_e^2}\right)+C\left(\frac{(1-\alpha)P}{\alpha P+\sigma_2^2}\right)-C\left(\frac{(1-\alpha) P}{\alpha P+\sigma_e^2}\right)\\
 		R_{2}\leq & C\left(\frac{(1-\alpha)P}{\alpha P+\sigma_2^2}\right)-C\left(\frac{(1-\alpha) P}{\alpha P+\sigma_e^2}\right),
 	\end{align*}
 where $\alpha\in[0,1].$ 
  In particular when $\sigma_1^2\leq \sigma_2^2\leq \sigma_e^2\leq 2 \sigma_2^2,$ the above region serves as the individual secrecy capacity region for all power levels.
\end{theorem}
\begin{IEEEproof}
	Consider the inner bound \eqref{eqn: IndS Gaussian Inner bound}.  We see that when $R_2\leq C\left(\frac{(1-\alpha) P}{\alpha P+\sigma_e^2}\right)$ holds, then the sum rate bound in \eqref{eqn: IndS Gaussian Inner bound} becomes redundant. In the case that it holds for any $\alpha\in[0,1]$, the inner bound \eqref{eqn: IndS Gaussian Inner bound} coincides with the outer bound \eqref{eqn: IndS Gaussian looser Outer bound}. This happens if $\max R_2\leq C\left(\frac{(1-\alpha) P}{\alpha P+\sigma_e^2}\right),$ i.e.,
		\begin{equation}\label{eqn: CS Gaussion condition}
			C\left(\frac{(1-\alpha)P}{\alpha P+\sigma_2^2}\right)-C\left(\frac{(1-\alpha) P}{\alpha P+\sigma_e^2}\right)
			\leq C\left(\frac{(1-\alpha) P}{\alpha P+\sigma_e^2}\right). 
		\end{equation}
	Under the stated conditions $\sigma_e^2\geq \sigma_2^2\geq \sigma_1^2$ and $P\geq \sigma_e^2(\sigma_e^2-2\sigma_2^2)/\sigma_2^2$, the inequality above, \eqref{eqn: CS Gaussion condition}, holds if and only if
	$\alpha \geq  \frac{(\sigma_e^2-\sigma_2^2)^2}{P(P+\sigma_2^2)}-\frac{\sigma_2^2}{P}=$.
	(A detailed calculation is given in Appendix \ref{Sec: Appdix calculation alpha}.)
	As $\alpha\geq 0$, \eqref{eqn: CS Gaussion condition} holds regardless of the value of $\alpha$, if 
	$\frac{(\sigma_e^2-\sigma_2^2)^2}{P(P+\sigma_2^2)}-\frac{\sigma_2^2}{P}\leq 0$ which hold as $P\geq \sigma_e^2(\sigma_e^2-2\sigma_2^2)/\sigma_2^2.$ Finally, we note that this last condition always holds if $\sigma_2^2\leq \sigma_e^2\leq 2 \sigma_2^2$ as $P\geq 0$. 
\end{IEEEproof}

\emph{Remark:} Theorem \ref{Thm: Gaussian IndSec capacity} establishes the individual secrecy capacity region for all power levels for the comparable eavesdropper channel scenario (i.e., having $\sigma_2^2\leq \sigma_e^2\leq 2 \sigma_2^2$). This is the counterpart of Theorem \ref{thm: IndSC by superposition} for the Gaussian scenario: We have $I(U;Y_2)\leq 2I(U;Z)$ for any $U\sim \mathcal{N}(0, (1-\alpha)P),$ $\alpha\in [0,1]$. In this case, the superposition coding is optimal to achieve the individual secrecy capacity region.

For the scenarios where the condition in Theorem \ref{Thm: Gaussian IndSec capacity} does not hold, i.e., when $P<\sigma_e^2(\sigma_e^2-2\sigma_2^2)/\sigma_2^2$, we note that the same achievable scheme achieves the capacity region in an approximate manner (within half a bit) as established in the following result.

\begin{theorem}\label{thm:ConstantGap}
	The achievable individual secrecy rate region as described in Theorem \ref{prop: Gaussian lower bound}, i.e., the set of $(R_1, R_2)\in \mathcal{R}_{+}^2$ satisfying \eqref{eqn: IndS Gaussian Inner bound}, approaches the individual secrecy capacity region of the Gaussian BC within 0.5 bits.
\end{theorem}

\begin{IEEEproof}
See Appendix~\ref{app:ConstantGap}
\end{IEEEproof}


\subsection{Numerical results with different secrecy constraints}

In this subsection, we provide the capacity region results for the Gaussian BC without secrecy constraint and under the joint secrecy constraint, and make comparisons with our results on the individual secrecy capacity region that are derived in the previous subsection.

\begin{theorem} \cite[Theorem 5.3]{ElGamal:2012}
The capacity region of the Gaussian BC without secrecy constraint is given by the union of the rate pairs $(R_1, R_2)\in \mathcal{R}_{+}^2$ satisfying
\begin{align}
\begin{split}\label{eqn: Gaussian_NoS}
	 R_1&\leq C\left(\frac{(1-\alpha) P}{\alpha P+\sigma_2^2}\right)\\
	 R_2&\leq C\left(\frac{\alpha P}{\sigma_1^2}\right),
\end{split}
\end{align}
 where $\alpha\in[0,1].$ 
\end{theorem}

\begin{theorem}\cite[Theorem 5]{src:Ekrem2011MIMO}
The joint secrecy capacity region of the Gaussian BC with an external eavesdropper is given by the union of the rate pairs $(R_1, R_2)\in \mathcal{R}_{+}^2$ satisfying
\begin{align}\label{eqn: Gaussian_JoS}
\begin{split}
	 R_1&\leq C\left(\frac{\alpha P}{\sigma_1^2}\right)-C\left(\frac{\alpha P}{\sigma_e^2}\right)\\
	 R_2&\leq C\left(\frac{(1-\alpha) P}{\alpha P+\sigma_1^2}\right)-C\left(\frac{(1-\alpha) P}{\alpha P+\sigma_e^2}\right),
\end{split}
\end{align}
 where $\alpha\in[0,1].$ 
\end{theorem}

\begin{figure}[!ht]
  \centering
    \subfloat[$P=1,\sigma_1^2=\sigma_2^2=1/3, \sigma_e^2=1/2$ \label{Gaussian: Ind-SCA}]{
     \includegraphics[width=0.48\textwidth]{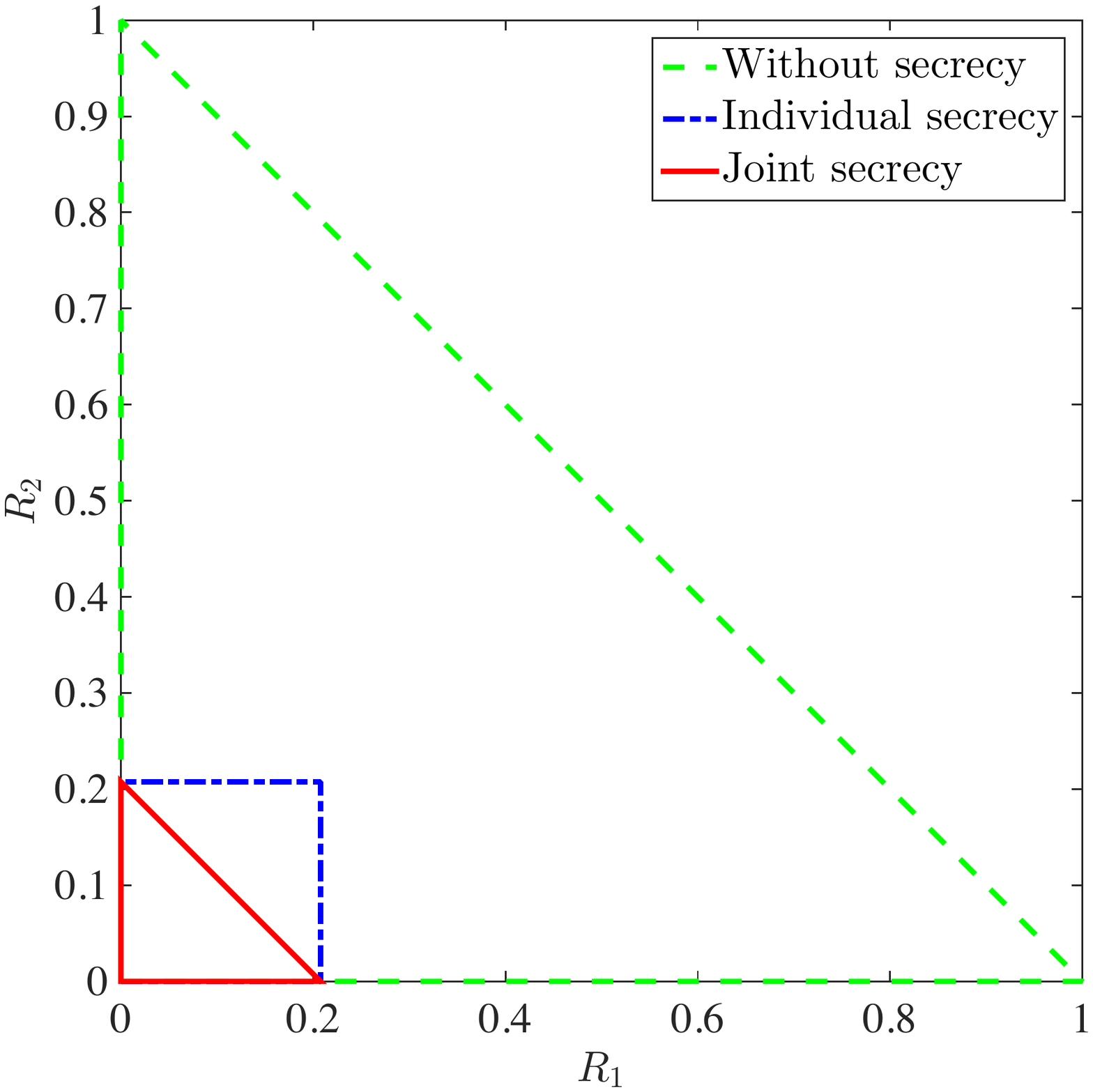}
    }
    \hfill
    \subfloat[$P=1, \sigma_1^2=3/80, \sigma_2^2=1/8, \sigma_e^2=1/2$ \label{Gaussian: Ind-SCB}]{
      \includegraphics[width=0.48\textwidth]{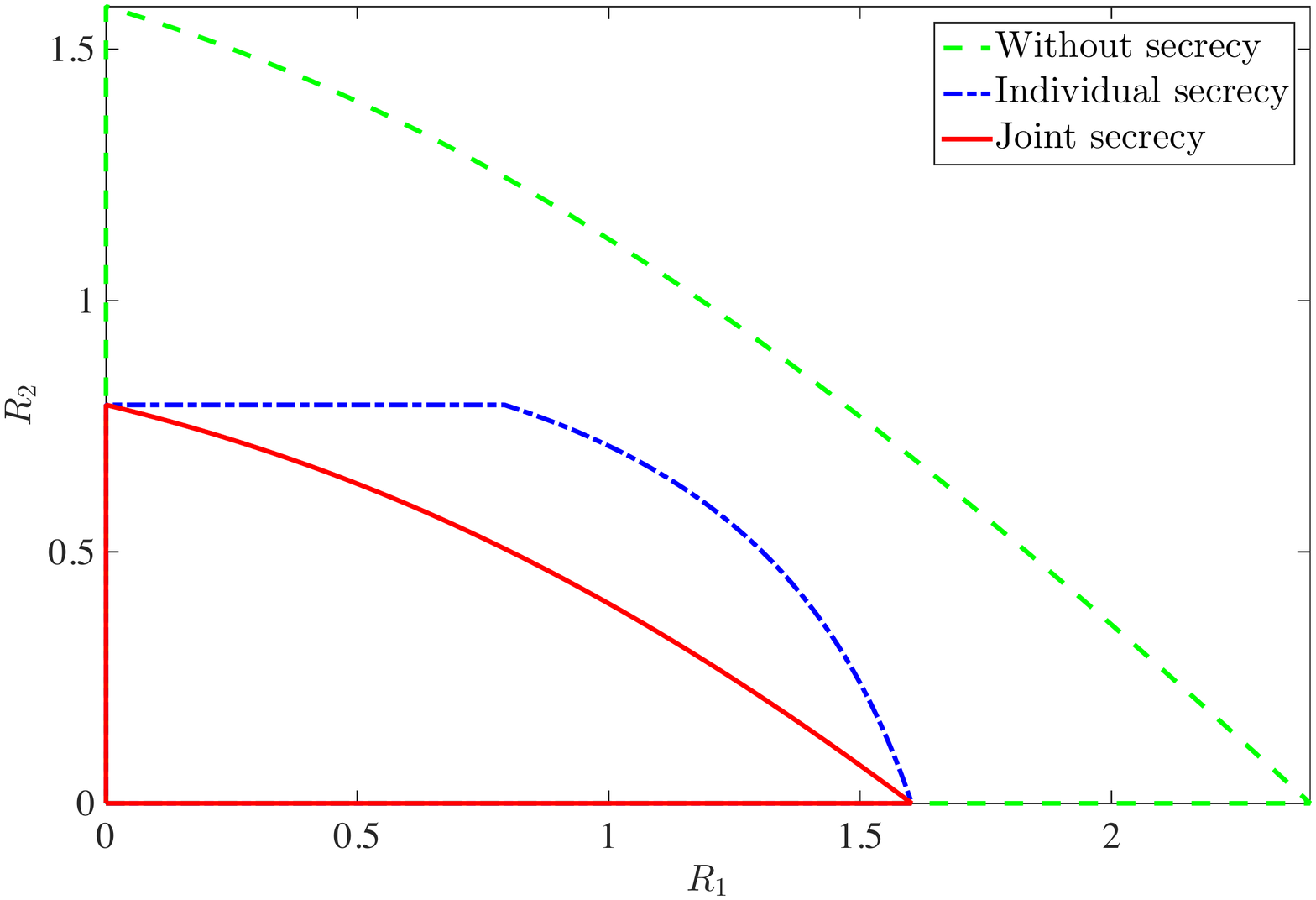}
    }\\
        \subfloat[$P=1,\sigma_1^2=1/10, \sigma_2^2=1/3, \sigma_e^2=4/3$ \label{Gaussian: Ind-SCC}]{
         \includegraphics[width=0.48\textwidth]{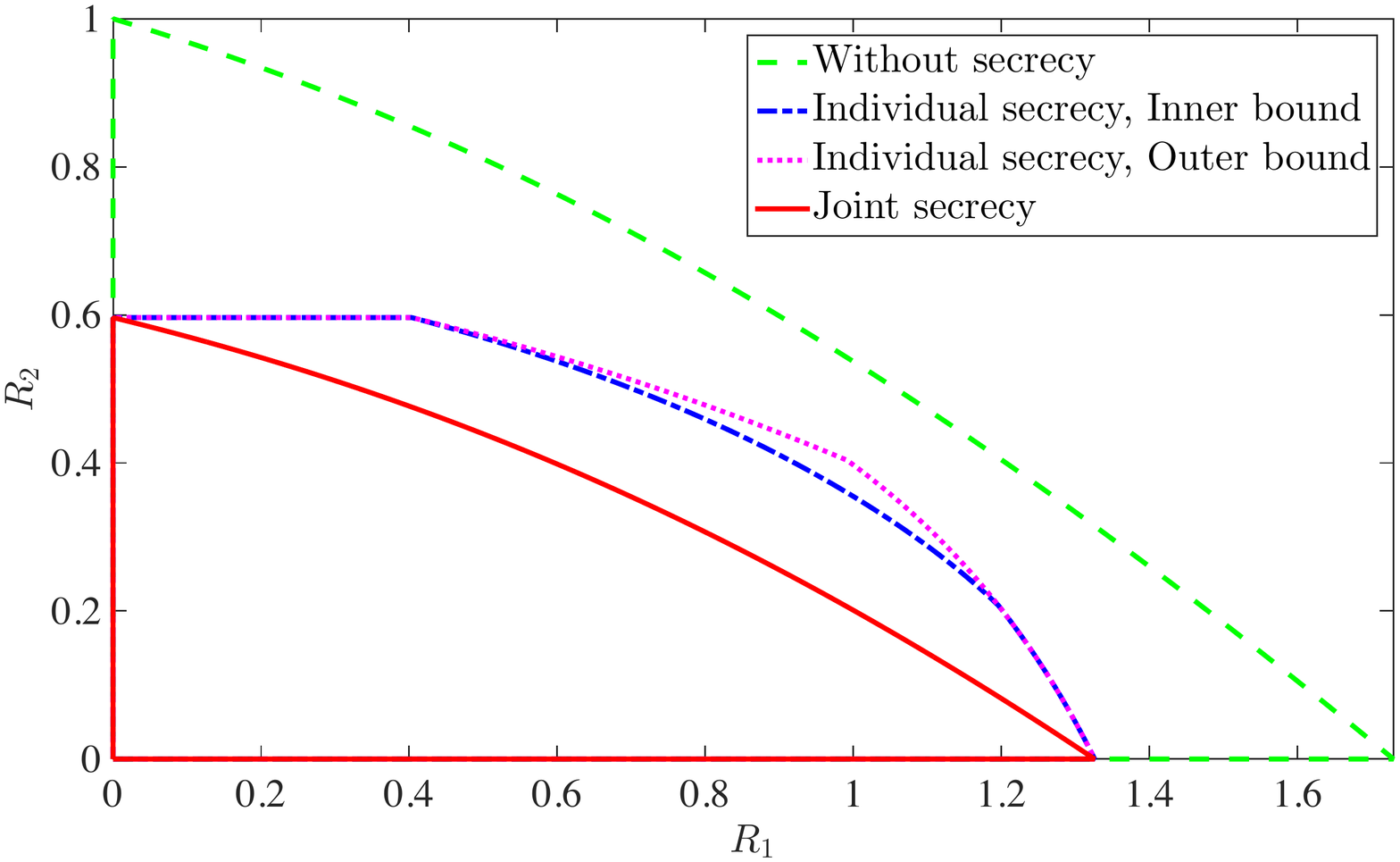}
        }
        \hfill
        \subfloat[$P=1, \sigma_1^2=1/10, \sigma_2^2=1/3, \sigma_e^2=2$ \label{Gaussian: Ind-SCD}]{
          \includegraphics[width=0.48\textwidth]{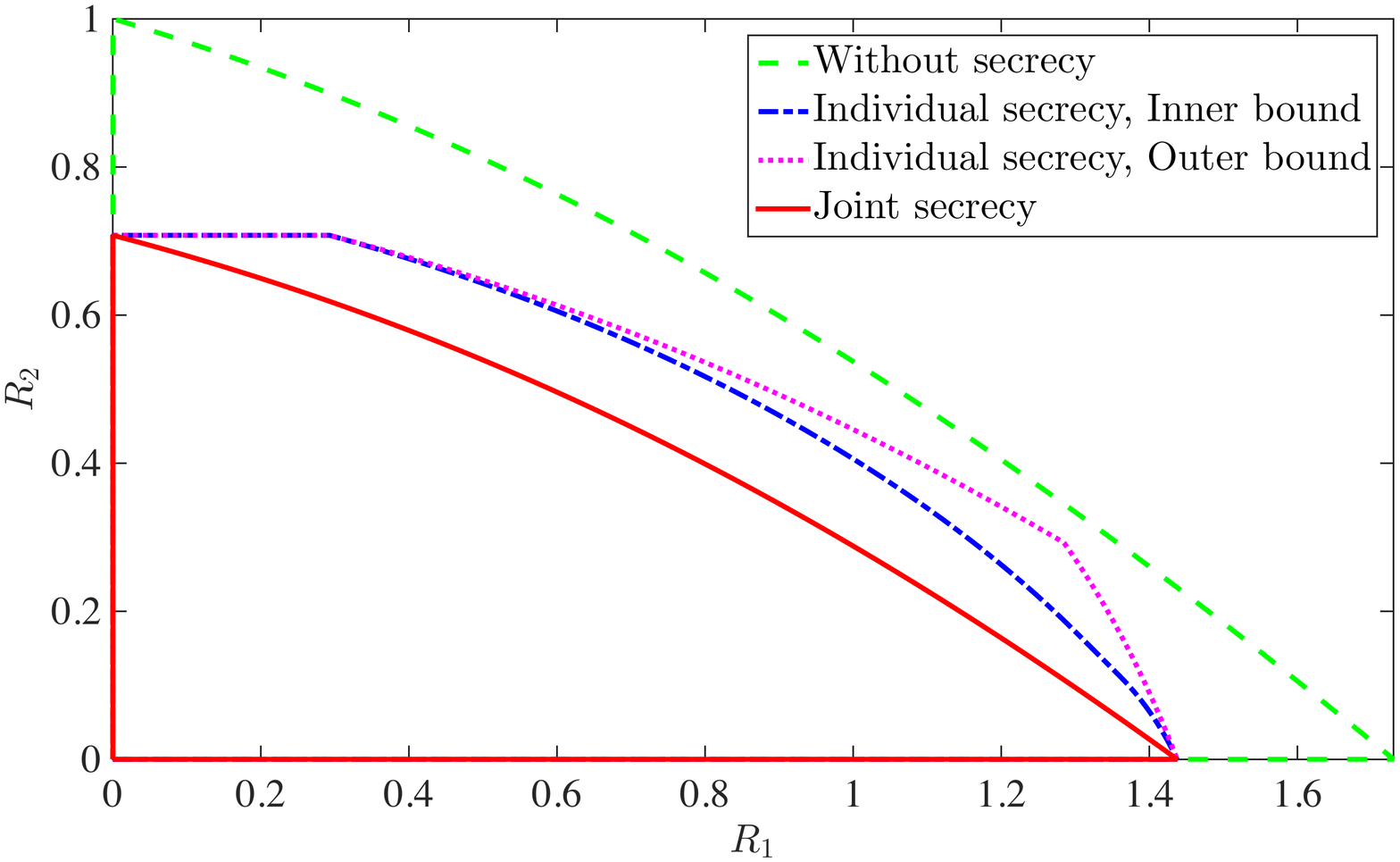}
        }
    \caption{Capacity regions of Gaussian BC, where the parameters are chosen such that the following inequalities are satisfied, in (a)-(b): $P\geq {\sigma_e^2(\sigma_e^2-2\sigma_2^2) }/{\sigma_2^2},$  further in (a): $\sigma_1^2=\sigma_2^2$ and (b): $\sigma_1^2<\sigma_2^2$; in (c)-(d): $P< {\sigma_e^2(\sigma_e^2-2\sigma_2^2) }/{\sigma_2^2},$ further in (c): $2\sigma_2^2\leq \sigma_e^2\leq P+2\sigma_2^2$ and in (d): $\sigma_e^2\geq P+2\sigma_2^2$, respectively.}
    \label{fig: Gaussian BC}
  \end{figure}

For the Gaussian BC, its capacity regions (or bounds) under different secrecy constraints are depicted in Fig. \ref{fig: Gaussian BC} (non-trivial case of $\sigma_1^2\leq \sigma_2^2\leq \sigma_e^2$ is assumed as detailed earlier). The capacity region without secrecy constraint is enclosed by (green) dashed lines;  the joint secrecy capacity region is enclosed by (red) solid lines; whilst the individual secrecy capacity region or its inner bound are enclosed by (blue) dash-dotted lines, and the outer bound by (magenta) dotted lines. We observe the followings. 

In \figref{Gaussian: Ind-SCA} and \figref{Gaussian: Ind-SCB}, we plot capacity regions under different secrecy constraints for some special cases that satisfy the condition requested in Theorem \ref{Thm: Gaussian IndSec capacity}, in which, we have for any joint secrecy achievable $(R_1, R_2),$ that $(R_1+R_2, R_2)$ is achievable with individual secrecy. 
More specifically,
\figref{Gaussian: Ind-SCA} depicts a special case where both legitimate receivers experience the same noise level. In this case, both the capacity regions without secrecy constraint and under joint secrecy constraint, are right angled isosceles triangles; while the capacity region under the individual secrecy constraint is a square, area of which doubles that of the joint secrecy case.
 \figref{Gaussian: Ind-SCB} depicts a more general case where both legitimate receivers experience different noise levels. The maximum marginal transmission rates (say $R_1^*, R_2^*,$ respectively) to both receivers are the same under either joint or individual secrecy constraints, which are strictly smaller than the ones for the scenario without any secrecy constraints. However, a distinct behavior for the individual secrecy capacity region is that, if the weak receiver operates at its maximum transmission rate, then the strong receiver can be still active (unlike the scenarios without secrecy constraint and under joint secrecy constraint). This can be visualized by the top-left part of the blue dash-dotted curve, as a straight line. Especially, for this special case as depicted in \figref{Gaussian: Ind-SCB}, we have $(R_2^*, R_2^*)$ pair as individually secret.
We note that, form Theorem \ref{Thm: Gaussian IndSec capacity}, superposition coding is optimal and $I(U;Y_2)-I(U;Z)\leq I(U;Z)$ holds for any $U\sim \mathcal{N}(0, (1-\alpha)P),$ $\alpha\in [0,1].$ As receiver 2 operates at rate $R_2\leq R_U$ with $R_U=I(U;Y_2)-I(U;Z),$ the information (say $m_{1k}$) up to $R_U$ could be carried  to receiver 1 via the cloud codeword $U$  while maintaining the individual secrecy of $m_2$ and $m_{1k}.$ Additional secret information to receiver 1 will be conveyed by the satellite codeword $V,$ similar to the joint secrecy scenario.

In \figref{Gaussian: Ind-SCC} and \figref{Gaussian: Ind-SCD}, the parameters are chosen such that $P< {\sigma_e^2(\sigma_e^2-2\sigma_2^2) }/{\sigma_2^2}$, which does not satisfy the condition given in Theorem \ref{Thm: Gaussian IndSec capacity}. Therefore, we use the inner bound (as given in Theorem \ref{prop: Gaussian lower bound}) and the outer bound (as given in Corollary \ref{Cor: Gaussian looser outer bound}). 
		More specifically, \figref{Gaussian: Ind-SCC} depicts a case where $2\sigma_2^2\leq \sigma_e^2\leq P+2\sigma_2^2$ (i.e., satisfying \eqref{eqn: value of alpha sub2}). In this case, there exists an $\alpha_0=\frac{(\sigma_e^2-\sigma_2^2)^2}{P(P+\sigma_2^2)}-\frac{\sigma_2^2}{P},$ such that a gap between the inner and outer bound occurs as $0<\alpha\leq \alpha_0.$ For the chosen parameter set, we have $\alpha_0=5/12.$ That is, the inner bound is tight for $\alpha=0$ and $\alpha\in [5/12, 1],$ which corresponds to the region where $R_2=R_2^*$  (here $R_2^*=0.7075$) and $R_2\in [0, 0.2075]$, respectively. 
\figref{Gaussian: Ind-SCD} depicts a case where $\sigma_e^2\geq P+2\sigma_2^2$ (i.e., satisfying \eqref{eqn: value of alpha sub1}). In this case, the inner and outer bound coincide at $\alpha=0$ but not for $0<\alpha<1.$ This indicates that the top-left part of the (blue) dash-dotted curve, as a straight line, is tight for the individual secrecy. 
		Differently from the scenarios in \figref{Gaussian: Ind-SCA}  and \figref{Gaussian: Ind-SCB},  we notice that in \figref{Gaussian: Ind-SCC}  and \figref{Gaussian: Ind-SCD}, $(R_2^*, R_2^*)$ pair is \emph{not} individually secret. The underlying reason is that $I(U;Y_2)-I(U;Z)\leq I(U;Z)$ does not hold for any $U\sim \mathcal{N}(0, (1-\alpha)P),$ $\alpha\in [0,1]$ in this case.  In particular, in \figref{Gaussian: Ind-SCD}, as $\alpha=0$ (i.e., all the power is assigned to the $U$ codeword), we have $R_2^*=I(U;Y_2)-I(U;Z)=0.7075$ and $I(U;Z)=0.2925.$ As receiver 2 operates $m_2$ at rate $R_2^*=0.7075,$ the maximal information $m_1$ that could be carried  to receiver 1 via the codeword $U$ will be bounded by $I(U;Z)=0.2925$ (according to \eqref{eqn: IndS Gaussian Inner bound} and \eqref{eqn: IndS Gaussian looser Outer bound}) while maintaining the individual secrecy of $m_2$ and $m_{1}.$


\section{Conclusion}\label{sec:conclusion}
In this paper, we studied the problem of secure communication over degraded broadcast channel under the individual secrecy constraint. Compared to the joint secrecy constraint, this relaxed setting allows for higher secure communication rates at the expense of having a weaker notion of security. As a general result, we derived several achievable rate regions and characterized the {individual secrecy} capacity region for some special cases. In addition, we also investigated the linear deterministic model and the Gaussian model. For the linear deterministic model, the capacity regions are fully characterized for the cases without secrecy constraint, under joint and individual secrecy constraint; while for the Gaussian model, a constant gap (i.e., 0.5 bits within the individual secrecy capacity region) result is obtained. Comparisons are made among the capacity regions for both models with different secrecy constraints (under no/individual/joint secrecy cases).



\appendices

\section{Proof of Theorem~\ref{thm:deterministic}} \label{sec:AppDeterministic}

The converse can be shown as follows: The first two inequalities for $R_1, R_2,$ respectively, follow from the classical results of wiretap channel by simply ignoring the other legitimate receiver. And, the last inequality follows directly from the upper bound on the sum rate for the relaxed case of without any secrecy constraints. 

The achievability can be shown by considering different scenarios, which is classified according to the relation between the channel gains $n_1, n_2, n_e$, and the relation between the rates $R_1, R_2$. Under the assumption that $n_1\geq n_2$, for both cases of $q=n_1\geq n_e\geq n_2$ and $n_e\geq n_1\geq n_2$, the individual secrecy capacity region reduces to the one for the wiretap channel \cite{src:Wyner1975, Csisz'ar:Broadcast78} and the achievability follows therein. Here we only need to consider the rest case $q=n_1\geq n_2\geq n_e$. The detailed achievability proof is given as follows. 

\begin{itemize}
\item If $R_2\leq n_e,$ we have two scenarios: 		
\begin{enumerate}
\item $R_1\leq  R_2.$ For this scenario, \eqref{eqn: D_Ins} reduces to the following:
				\begin{equation*}
					 		R_1\leq R_2\leq \min\{n_2-n_e, n_e\}.
				\end{equation*}
			For its achievability, given $m_1, m_2$ with $m_1=[m_1(1), \cdots, m_1(R_1)]$ and $m_2=[m_2(1), \cdots, m_2(R_2)]$, we send $X=[x(1), x(2), \cdots, x(n_1)]^T$ such that 
				\begin{equation*}
					x(k)=\left\{
							\begin{array}{lll}
								m_1(k)\oplus m_2(k) & &  1\leq k\leq R_1\\
								r(k) &	& R_1< k\leq n_e \\
								m_2(k-n_e) & & n_e< k\leq n_e+R_2\\
								r(k) &	&  n_e+R_2< k\leq n_1
							\end{array}
						\right. 
				\end{equation*}
				where $r(k)$ is randomly chosen from $\{0, 1\}.$ The construction of $X$ is illustrated in Fig. \ref{fig: X^n LDBC 1<a}. 
			
				\begin{figure}[h]
				\centering
				{
					\begin{tabular}{rcl}
							$m_1:$ & & 
									$\begin{tikzpicture}
										\node[minimum height=1.6em, anchor=base, fill=blue!25] {$m_{1}(1), \cdots, m_1(R_1)$}; 
									\end{tikzpicture}$
									\\
							$m_2:$ & & 
									$\begin{tikzpicture}
										\node[minimum height=1.6em, anchor=base, fill=teal!25] {$m_{2}(1), \cdots, m_2(R_1), \cdots, m_2(R_2)$}; 
									\end{tikzpicture}$
									\\
							$X^T:$  &&
									$\underbrace{
									\underbrace{
									\underbrace{
									\overbrace{
									\begin{tikzpicture}
										\node[minimum height=1.6em, anchor=base, fill=red!25] {$m_{1}(k)\oplus m_{2}(k)$};
									\end{tikzpicture}
									}^{R_{1}}
									\begin{tikzpicture}
										\node[minimum height=1.6em, anchor=base, fill=yellow!25] {$r(k)$};
									\end{tikzpicture}
									}_{n_e}
									\overbrace{
									\begin{tikzpicture}
										\node[minimum height=1.6em, anchor=base, fill=teal!25] {$\quad m_{2}(k-n_e)\quad $};
									\end{tikzpicture}
									}^{R_{2}}
									}_{\leq n_2}
									\begin{tikzpicture}
										\node[minimum height=1.6em, anchor=base, fill=yellow!25] {$r(k)$};
									\end{tikzpicture}
									}_{n_1}	$												
						\end{tabular}
						}
						\caption{Codeword $X$ for a) $R_1\leq R_2\leq n_e.$}
						\label{fig: X^n LDBC 1<a}											
						\end{figure}
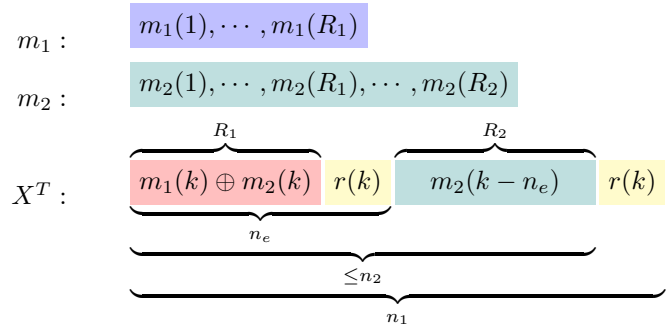
			
		\item $R_1\geq R_2.$ For this scenario, \eqref{eqn: D_Ins} reduces to the following:
					\begin{equation*}
								 R_2\leq R_1\leq n_1-n_e;\quad	R_2 \leq \min\{n_e, n_2-n_e\}.
					\end{equation*}
			For its achievability, given $m_1, m_2$ with $m_1=[m_1(1), \cdots, m_1(R_1)]$ and $m_2=[m_2(1), \cdots, m_2(R_2)]$, we send $X=[x(1), x(2), \cdots, x(n_1)]^T$ such that 
							\begin{equation*}
								x(k)=\left\{
										\begin{array}{lll}
											m_1(k)\oplus m_2(k) & &  1\leq k\leq R_2\\
											r(k) &	& R_2< k\leq n_e\\
											m_2(k-n_e) & & n_e< k\leq n_e+R_2\\
											m_1(k-n_e-R_2) & &  n_e+R_2< k\leq n_e+R_1\\
											r(k) &	& n_e+R_1< k\leq n_1
										\end{array}
									\right. 
							\end{equation*} 
			where $r(k)$ is randomly chosen from $\{0, 1\}.$ The construction of $X$ is illustrated in Fig. \ref{fig: X^n LDBC 2}. 
				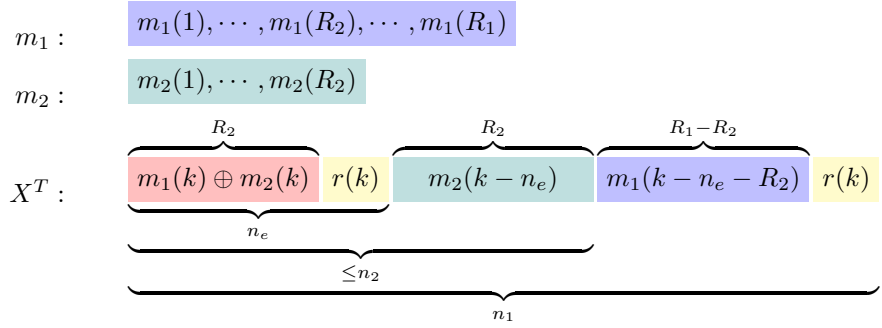
\begin{figure}[h]
								\centering
								{
									\begin{tabular}{rcl}
										$m_1:$ & & 
											$\begin{tikzpicture}
												\node[minimum height=1.6em, anchor=base, fill=blue!25] {$m_{1}(1), \cdots, m_1(R_2), \cdots, m_1(R_1)$}; 
											\end{tikzpicture}$
												\\
										$m_2:$ & & 
											$\begin{tikzpicture}
												\node[minimum height=1.6em, anchor=base, fill=teal!25] {$m_{2}(1), \cdots, m_2(R_2)$}; 
											\end{tikzpicture}$		
												\\
										$X^T:$ & &
												$\underbrace{
												\underbrace{
												\underbrace{
												\overbrace{
												\begin{tikzpicture}
													\node[minimum height=1.65em, anchor=base, fill=red!25] {$m_{1}(k)\oplus m_{2}(k)$};
												\end{tikzpicture}
												}^{R_2}
												\begin{tikzpicture}
													\node[minimum height=1.6em, anchor=base, fill=yellow!25] {$r(k)$};
												\end{tikzpicture}
												}_{n_e}
												\overbrace{
												\begin{tikzpicture}
													\node[minimum height=1.6em, anchor=base, fill=teal!25] {$\quad m_{2}(k-n_e)\quad$};
												\end{tikzpicture}
												}^{R_{2}}
												}_{\leq n_2}
												\overbrace{
												\begin{tikzpicture}
													\node[minimum height=1.6em, anchor=base, fill=blue!25] {$m_{1}(k-n_e-R_2)$};
												\end{tikzpicture}
												}^{R_1-R_2}
												\begin{tikzpicture}
													\node[minimum height=1.6em, anchor=base, fill=yellow!25] {$r(k)$};
												\end{tikzpicture}
												}_{n_1}$
												\end{tabular}
												}													
											\caption{Codeword $X$ for b) $R_2\leq n_e$ and $R_2\leq R_1.$}
											\label{fig: X^n LDBC 2}											
											\end{figure}
\end{enumerate}			
\item If $R_2\geq n_e,$ we also have two scenarios: 
			
\begin{enumerate}
\item $R_1\leq n_e.$ For this scenario, \eqref{eqn: D_Ins} reduces to the following:
						\begin{equation*}
							R_1\leq n_e\leq R_2\leq n_2-n_e.
						\end{equation*}
					Note that this scenario is possible only when $n_2\geq 2n_e.$ For its achievability, 
					given $m_1, m_2$ with $m_1=[m_1(1), \cdots, m_1(R_1)]$ and $m_2=[m_2(1), \cdots, m_2(R_2)]$, we send $X=[x(1), x(2), \cdots, x(n_1)]^T$ such that 
						\begin{equation*}
							x(k)=\left\{
									\begin{array}{lll}
										m_1(k)\oplus m_2(k) & &  1\leq k\leq R_1\\
										r(k) &	& R_1< k\leq n_e\\
										m_2(k-n_e) & & n_e+1\leq k\leq n_e+R_2\\
											r(k) &	& n_e+R_2< k\leq n_1
									\end{array}
								\right. 
						\end{equation*}
						where $r(k)$ is randomly chosen from $\{0, 1\}.$ The construction of $X$ is illustrated in Fig. \ref{fig: X^n LDBC 3}. 

				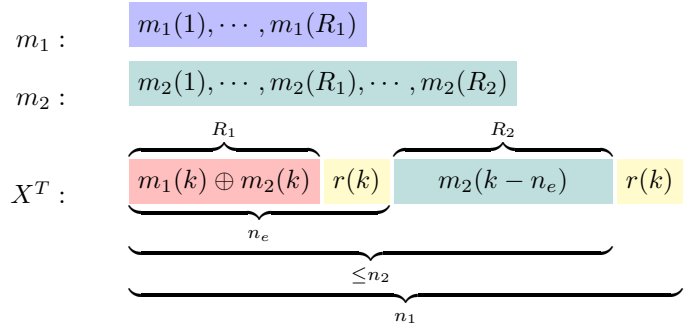
\begin{figure}[h]
								\centering
								{
						\begin{tabular}{rcl}
							$m_1:$ & & 
								$\begin{tikzpicture}
									\node[minimum height=1.6em, anchor=base, fill=blue!25] {$m_{1}(1), \cdots, m_1(R_1)$}; 
								\end{tikzpicture}$
									\\
							$m_2:$ & & 
								$\begin{tikzpicture}
									\node[minimum height=1.6em, anchor=base, fill=teal!25] {$m_{2}(1), \cdots, m_2(R_1), \cdots, m_2(R_2)$}; 
								\end{tikzpicture}$
									\\
							$X^T:$ & &
									$\underbrace{
									\underbrace{
									\underbrace{
									\overbrace{
									\begin{tikzpicture}
										\node[minimum height=1.6em, anchor=base, fill=red!25] {$m_{1}(k)\oplus m_{2}(k)$};
									\end{tikzpicture}
									}^{R_{1}}
									\begin{tikzpicture}
										\node[minimum height=1.6em, anchor=base, fill=yellow!25] {$r(k)$};
									\end{tikzpicture}
									}_{n_e}
									\overbrace{
									\begin{tikzpicture}
										\node[minimum height=1.6em, anchor=base, fill=teal!25] {$\ \quad m_{2}(k-n_e)\quad \ $};
									\end{tikzpicture}
									}^{R_{2}}
									}_{\leq n_2}
									\begin{tikzpicture}
										\node[minimum height=1.6em, anchor=base, fill=yellow!25] {$r(k)$};
									\end{tikzpicture}
									}_{n_1}	$												
								\end{tabular}
								}
								\caption{Codeword $X$ for c) $R_1\leq n_e\leq R_2.$}
								\label{fig: X^n LDBC 3}											
								\end{figure}
						
					\item $R_1\geq n_e.$ For this scenario, \eqref{eqn: D_Ins} reduces to the following: 
											\begin{equation*}
												n_e\leq R_1\leq n_1-n_e; \quad n_e\leq R_2\leq n_2-n_e.
											\end{equation*}
						For its achievability, given $m_1, m_2$ with $m_1=[m_1(1), \cdots, m_1(R_1)]$ and $m_2=[m_2(1), \cdots, m_2(R_2)]$, we send $X=[x(1), x(2), \cdots, x(n_1)]^T$ such that 
											\begin{equation*}
												x(k)=\left\{
														\begin{array}{lll}
															m_1(k)\oplus m_2(k) & &  1\leq k\leq n_e\\
															m_2(k-n_e) &	& n_e< k\leq n_e+R_2\\
															m_1(k-R_2) & & n_e+R_2< k\leq R_1+R_2\\
															r(k)& & R_1+R_2< k\leq n_1
														\end{array}
													\right. 
											\end{equation*}
											where $r(k)$ is randomly chosen from $\{0, 1\}.$ The construction of $X$ is illustrated in Fig. \ref{fig: X^n LDBC 4}.  
					
				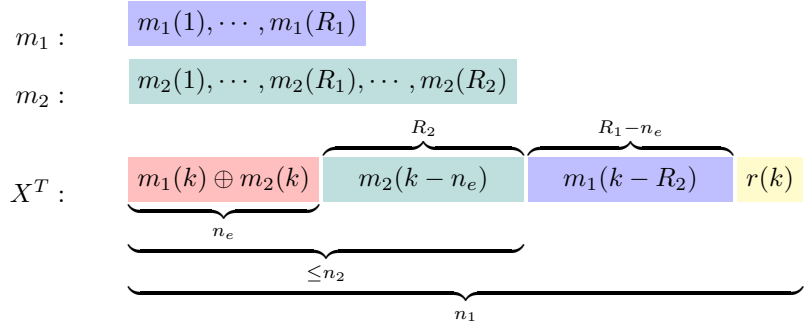
\begin{figure}[h]
								\centering
								{
						\begin{tabular}{rcl}
												$m_1:$ & & 
													$\begin{tikzpicture}
														\node[minimum height=1.6em, anchor=base, fill=blue!25] {$m_{1}(1), \cdots, m_1(R_1)$}; 
													\end{tikzpicture}$
														\\
												$m_2:$ & & 
													$\begin{tikzpicture}
														\node[minimum height=1.6em, anchor=base, fill=teal!25] {$m_{2}(1), \cdots, m_2(R_1), \cdots, m_2(R_2)$}; 
													\end{tikzpicture}$
														\\
												$X^T:$ & &
														$\underbrace{
														\underbrace{
														\underbrace{
														\begin{tikzpicture}
															\node[minimum height=1.6em, anchor=base, fill=red!25] {$m_{1}(k)\oplus m_{2}(k)$};
														\end{tikzpicture}
														}_{n_{e}}
														\overbrace{
														\begin{tikzpicture}
															\node[minimum height=1.6em, anchor=base, fill=teal!25] {$\quad m_{2}(k-n_e)\quad$};
														\end{tikzpicture}
														}^{R_{2}}
														}_{\leq n_2}
														\overbrace{
														\begin{tikzpicture}
															\node[minimum height=1.6em, anchor=base, fill=blue!25] {$\quad m_{1}(k-R_2)\quad$};
														\end{tikzpicture}
															}^{R_{1}-n_e}
														\begin{tikzpicture}
															\node[minimum height=1.6em, anchor=base, fill=yellow!25] {$r(k)$};
														\end{tikzpicture}
														}_{n_1}	$	
														\end{tabular}
														}												
													\caption{Codeword $X$ for d) $R_1\geq n_e$ and $R_2\geq n_e.$}
													\label{fig: X^n LDBC 4}											
													\end{figure}
												\end{enumerate}

\end{itemize}

	Note that in all scenarios, receiver 1 gets the first $n_1$ bits of $X;$ receiver 2 gets the first $n_2$ bits of $X;$ while the eavesdropper gets the first $n_e$ bits of $X.$ Receiver 2 can obtain the desired message $m_2$; and receiver 1 obtains the message $m_2$ first and then could decode its desired message $m_1$ with the help of $m_2;$ whilst the eavesdropper gets only $m_1(k)\oplus m_2(k)$ for $1\leq k\leq \min\{n_e, R_1, R_2\}$ and some other random bits, which gives no information on $m_1,$ $m_2$ individually.


\section{Proof of Theorem~\ref{thm: IndS by primitive approach}}\label{sec:AppPrimitive}
In the following, we provide the detailed achievability proof for a given $p(u,x)$.

{\em Codebook generation:} Fix $p(u).$ Randomly generate $2^{n(R_{1}+R_{2}+R_{r})}$ i.i.d sequences $u^{n}(m_{1}, m_{2}, m_{r}),$ with $(m_1,m_2, m_{r})\in[1:2^{nR_1}]\times[1:2^{nR_{2}}]\times[1:2^{nR_{r}}],$ according to $p(u).$  
	
\begin{figure}[h]
	\vspace{-2mm}
	\centering
	\begin{tabular}{ll}
			$u^n(m_{1}, m_{2}, m_{r}):$ & $\overbrace{
						\begin{tikzpicture}
							\node[minimum height=1.6em, anchor=base, minimum width=6em, fill=teal!25] {$m_{1}$}; 
						\end{tikzpicture}
							}^{nR_1}
						\overbrace{
						\begin{tikzpicture}
							\node[minimum height=1.6em, anchor=base, minimum width=4em, fill=red!25] {$m_{2}$}; 
						\end{tikzpicture}
							}^{nR_{2}}
						\overbrace{
								\begin{tikzpicture}
									\node[minimum height=1.6em, anchor=base, fill=red!25] {$\,m_{r}\,$}; 
								\end{tikzpicture}
								}^{nR_{r}}$	
	\end{tabular}
	\caption{Encoding}\label{fig: DM-DBC Primitive encoding}
		\end{figure}
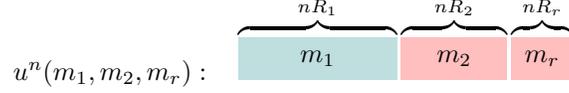

{\em Encoding:} To send messages $(m_1, m_2),$ randomly choose $m_{r}\in [1: 2^{nR_{r}}]$ and find $u^n(m_1,m_2, m_{r}).$ Given $u^n(m_1,m_2, m_{r}),$ generate $x^n$ according to $p(x|u),$ and transmit it to the channel. The choice of $u^n$ is illustrated in Fig. \ref{fig: DM-DBC Primitive encoding}.

{\em Decoding:} 
Receiver 2, upon receiving $y_2^n,$ finds $u^n(\hat{m}_{1}, \hat{m}_{2}, \hat{m}_{r})$ such that $(u^n(\hat{m}_{1}, \hat{m}_{2}, \hat{m}_{r}), y_2^n)$ is jointly typical.  
Receiver 1, upon receiving $y_1^n,$ finds $u^n(\tilde{m}_{1}, \tilde{m}_{2}, \tilde{m}_{r})$ such that $(u^n(\tilde{m}_{1}, \tilde{m}_{2},\tilde{m}_{r}), y_1^n)$ is jointly typical. 

{\em Analysis of the error probability of decoding:} 
Assume that $(M_1,M_2)=(m_1,m_2)$ is sent. 

First we consider $P_{e,2}$ at receiver 2. A decoding error happens iff one or both of the following events occur: 
\begin{align*}
	\Ec_{21} =& \{(u^n(m_{1}, m_{2}, m_{r}), y_2^n)\notin \Tc_{\epsilon}^{(n)}\},\\
	\Ec_{22} =& \{(u^n(\hat{m}_{1}, \hat{m}_{2}, \hat{m}_{r}), y_2^n)\in \Tc_{\epsilon}^{(n)}  \mbox{ for some } \hat{m}_{2}\neq m_{2}\}.	
\end{align*}
Thus, $P_{e,2}$ can be upper bounded as 
	\begin{equation*}
		P_{e,2}\leq \Pr(\Ec_{21})+\Pr(\Ec_{22}).
	\end{equation*} 
By the LLN, $\Pr(\Ec_{21})$ tends to zero as $n\rightarrow \infty.$ For $\Pr(\Ec_{22})$, since $u^n(\hat{m}_{1}, \hat{m}_{2},\hat{m}_{r})$ is independent of $(u^n(m_{1},m_{2}, m_{r}), \allowbreak y_2^n)$ for $\hat{m}_{2}\neq m_{2},$ by the packing lemma \cite{ElGamal:2012}, $\Pr(\Ec_{22})$ tends to zero as $n\rightarrow \infty$ if 
\begin{equation}\label{eqn: ReCon on R_{1}+R_{2}+R_{r} at Rx1}
	R_{1}+R_{2}+R_{r}\leq I(U;Y_2)-\delta_n(\epsilon_n).
\end{equation}

Similarly, at receiver 1, the average probability of decoding error $P_{e,2},$ can be made arbitrarily small as $n\rightarrow \infty$ if 
\begin{equation}\label{eqn: ReCon on R_{1}+R_{2}+R_{r} at Rx2}
	R_{1}+R_{2}+R_{r}\leq I(U;Y_1)-\delta_n(\epsilon_n).
\end{equation}

{\em Analysis of individual secrecy:} 
For the individual secrecy \eqref{eq:IndSec}, i.e., $R_{L,i}\leq \tau_n,$ for $i=1,2,$ it is equivalent to show that $H(M_i|Z^n)\geq H(M_i)-n\tau_n=nR_i-n\tau_n.$  First we consider $H(M_2|Z^n).$ 
\allowdisplaybreaks
	\begin{align*}
	 H(M_2|Z^n)
	 		= 	& H(M_{2}, Z^n)-H(Z^n)\\ 
			=	& H(U^n, M_{2}, Z^n)-H(U^n|M_2, Z^n)-H(Z^n)\\
			= 	& H(U^n)+H(Z^n|U^n)-H(U^n|M_2, Z^n)-H(Z^n)\\
			\stackrel{(a)}{=}	& n[R_1+R_2+R_r]+nH(Z|U)-H(U^n|M_2, Z^n)-H(Z^n)\\
			\stackrel{(b)}{\geq} & n[R_1+R_2+R_r]-nI(U;Z)-H(U^n|M_2, Z^n)\\
			\stackrel{(c)}{\geq} & n[R_1+R_2+R_r]-nI(U;Z)-n[R_1+R_r-I(U;Z)]-n\tau_n\\
			=& nR_2-n\tau_n\\
			=& H(M_2)-n\tau_n,
	\end{align*}
where 
	$(a)$ follows from the codebook construction that $H(U^n)=n[R_1+R_{2k}+R_{r}]$ and the discrete memoryless of the channel; 
	$(b)$ is due to the fact that $H(Z^n)=\sum\limits_{i=1}^{n}H(Z_i|Z^{i-1})\leq \sum\limits_{i=1}^{n}H(Z_i)=nH(Z);$ and
	$(c)$ follows from \cite[Lemma 1]{Chia:Three-receiver12} that $H(U^n|M_2, Z^n)\leq n[R_1+R_r-I(U;Z)]+n\tau_n$
if taking 
		\begin{equation}\label{eqn: SeCon on R_{1}+R_r for Rx2}
			R_{1}+R_r\geq I(U;Z)+\delta_n(\tau_n).
		\end{equation}

A similar proof can be applied to show that $H(M_1|Z^n)\geq H(M_1)-n\tau_n$ if taking  	
		\begin{equation}\label{eqn: SeCon on R_{2}+R_r for Rx1}
			R_{2}+R_r\geq I(U;Z)+\delta_n(\tau_n).
		\end{equation}	
	  
{\em Achievable individual secrecy rate region:}  The resulting region has the following constraints: 
the non-negativity for rates, i.e., $R_{1}, R_{2}, R_{r} \geq  0$, 
the conditions for a reliable communication, i.e., \eqref{eqn: ReCon on R_{1}+R_{2}+R_{r} at Rx1}, \eqref{eqn: ReCon on R_{1}+R_{2}+R_{r} at Rx2}, and 
the conditions for individual secrecy, i.e., \eqref{eqn: SeCon on R_{1}+R_r for Rx2}, \eqref{eqn: SeCon on R_{2}+R_r for Rx1}.
Eliminating $R_{r}$ here by applying Fourier-Motzkin procedure \cite{ElGamal:2012}, we get the desired rate region as given in \eqref{eqn: Region_Primitive}.


\section{Proof of Theorem~\ref{thm: IndS by superposition}}\label{sec:AppSuperposition}

For a given input probability distribution $p(u,v,x)$, let $I_1=I(V;Y_1|U)-I(V;Z|U).$  
If $I_1\leq 0$, the claimed region reduces to \eqref{eqn: Region_Primitive}, which is achievable by taking the primitive approach as described in Section \ref{sec: primitive approach}, or more specifically, by employing Carleial-Hellman's secrecy coding.
In the following, we provide the detailed achievability proof for the remaining case, i.e., if $I_1>0$ for a given $p(u,v,x)$.

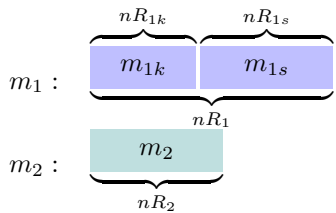
\begin{figure}[h]
\centering
\begin{tabular}{ll}
				$m_1:$ & 
				$\underbrace{
						\overbrace{
						\begin{tikzpicture}
							\node[minimum height=1.6em, anchor=base, minimum width=4em, fill=blue!25] {$m_{1k}$}; 
						\end{tikzpicture}
							}^{nR_{1k}}
						\overbrace{					
						\begin{tikzpicture}
							\node[minimum height=1.6em, anchor=base, minimum width=5em, fill=blue!25] {$m_{1s}$};
						\end{tikzpicture} 
						}^{nR_{1s}}
						}_{nR_1}$	
						\\
				$m_2:$ &  $\underbrace{
										\begin{tikzpicture}
											\node[minimum height=1.6em, minimum width=5em, anchor=base, fill=teal!25] {$m_{2}$}; 
										\end{tikzpicture}
											}_{nR_2}$
\end{tabular}\caption{Rate splitting}\label{fig: DM-DBC rate splitting}
\end{figure}

{\em Rate splitting:} As illustrated in Fig. \ref{fig: DM-DBC rate splitting}, we split $M_1$ into $(M_{1k}, M_{1s}).$ In particular, $M_{1k}, M_{1s}$ are of entropy $nR_{1k}$ and $nR_{1s},$ respectively; and $M_2$ is of entropy $nR_2.$ That is,
	\begin{align}
		R_1		&=R_{1k}+R_{1s}. 	\label{eqn: Rate Relation R_1}
	\end{align}

{\em Codebook generation:} Fix $p(u), p(v|u).$ First, randomly generate $2^{n(R_2+R_{1k}+R_{r})}$ i.i.d. sequences $u^{n}(m_2, m_{1k}, \allowbreak m_{r}),$ with $(m_2, m_{1k}, m_{r})\in[1:2^{nR_2}]\times[1:2^{nR_{1k}}]\times[1:2^{nR_{r}}],$ according to $p(u).$ Secondly, for each $u^n(m_2, m_{1k}, m_{r}),$ randomly generate i.i.d. sequences $v^n(m_2, m_{1k}, m_{r}, m_{1s}, m_{1r})$ with $(m_{1s}, m_{1r})\in [1: 2^{nR_{1s}}]\times [1: 2^{nR_{1r}}],$ according to $p(v|u).$  
	
\begin{figure}[h]
	\vspace{-2mm}
	\centering
	\begin{tabular}{ll}
			$u^n(m_2, m_{1k}, m_{r}):$ & $\overbrace{
						\begin{tikzpicture}
							\node[minimum height=1.6em, anchor=base, minimum width=5em, fill=teal!25] {$m_{2}$}; 
						\end{tikzpicture}
							}^{nR_2}
						\overbrace{
						\begin{tikzpicture}
							\node[minimum height=1.6em, anchor=base, minimum width=4em, fill=blue!25] {$m_{1k}$}; 
						\end{tikzpicture}
							}^{nR_{1k}}
						\overbrace{
								\begin{tikzpicture}
									\node[minimum height=1.6em, anchor=base, fill=red!25] {$\,m_{r}\,$}; 
								\end{tikzpicture}
								}^{nR_{r}}$	
							\\
			$v^n(m_2, m_{1k}, m_{r}, m_{1s}, m_{1r}):$ & $\underbrace{
											\begin{tikzpicture}
												\node[minimum height=1.6em, anchor=base, minimum width=5em, fill=blue!25] {$m_{1s}$}; 
											\end{tikzpicture}
											}_{nR_{1s}}
											\underbrace{
											\begin{tikzpicture}
												\node[minimum height=1.6em, anchor=base, minimum width=3em, fill=red!25] {$m_{1r}$}; 
											\end{tikzpicture}
											}_{nR_{1r}}$
	\end{tabular}
	\caption{Encoding}\label{fig: DM-DBC encoding}
		\end{figure}
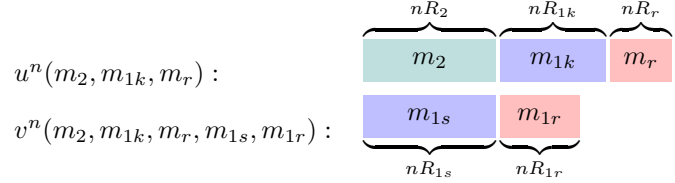

{\em Encoding:} To send messages $(m_1, m_2)$ with $m_1=(m_{1k}, m_{1s}),$ randomly choose $m_{r}\in [1: 2^{nR_{r}}]$ and find $u^n(m_2,m_{1k},m_{r}).$  Given $u^n(m_2,m_{1k}, m_{r}),$ randomly choose $m_{1r}\in [1: 2^{nR_{1r}}]$, further find the corresponding $v^n(m_2, m_{1k}, m_{r}, m_{1s}, m_{1r}).$ Generate $x^n$ according to $p(x|v),$ and transmit it to the channel.
 The choice of $u^n, v^n$ is illustrated in Fig. \ref{fig: DM-DBC encoding}.

{\em Decoding:} 
Receiver 2, upon receiving $y_2^n,$ finds  $u^n(\hat{m}_2, \hat{m}_{1k}, \hat{m}_{r})$ such that $(u^n(\hat{m}_2, \hat{m}_{1k}, \hat{m}_{r}), y_2^n)$ is jointly typical.  

Receiver 1, upon receiving $y_1^n,$ finds a unique  tuple  $(\tilde{m}_2, \tilde{m}_{1k}, \tilde{m}_{r}, \tilde{m}_{1s})$ $u^n(\tilde{m}_2, \tilde{m}_{1k}, \allowbreak\tilde{m}_{r})$ such that \\
$(u^n(\tilde{m}_2, \tilde{m}_{1k},\tilde{m}_{r}), v^n(\tilde{m}_2, \tilde{m}_{1k}, \tilde{m}_{r}, \tilde{m}_{1s}, \tilde{m}_{1r}), y_1^n)$ is jointly typical for some $\tilde{m}_{1r}$. 
Finally, decode $\tilde{m}_{1}=(\tilde{m}_{1k}, \tilde{m}_{1s}).$ 

{\em Analysis of the error probability of decoding:} 
Assume that $(M_1,M_2)=(m_1,m_2)$ with $m_1=(m_{1k}, m_{1s})$ is sent. 
 
First we consider $P_{e,2}$ at receiver 2. A decoding error happens iff one or both of the following events occur: 
\begin{align*}
	\Ec_{21} =& \{(u^n(m_{2}, m_{1k}, m_{r}), y_2^n)\notin \Tc_{\epsilon}^{(n)}\},\\
	\Ec_{22} =& \{(u^n(\hat{m}_{2}, \hat{m}_{1k}, \hat{m}_{r}), y_2^n)\in \Tc_{\epsilon}^{(n)}  \mbox{ for some } \hat{m}_{2}\neq m_{2}\}.	
\end{align*}
Thus, $P_{e,2}$ can be upper bounded as 
	\begin{equation*}
		P_{e,2}\leq \Pr(\Ec_{21})+\Pr(\Ec_{22}).
	\end{equation*} 
By the LLN, $\Pr(\Ec_{21})$ tends to zero as $n\rightarrow \infty.$ For $\Pr(\Ec_{22})$, since $u^n(\hat{m}_{2}, \hat{m}_{1k},\hat{m}_{r})$ is independent of $(u^n(m_{2},m_{1k}, m_{r}), \allowbreak y_2^n)$ for $\hat{m}_{2}\neq m_{2},$ by the packing lemma \cite{ElGamal:2012}, $\Pr(\Ec_{22})$ tends to zero as $n\rightarrow \infty$ if 
\begin{equation}\label{eqn: ReCon1 on R_2+R_{1k}+R_{r}}
	R_2+R_{1k}+R_{r}\leq I(U;Y_2)-\delta_n(\epsilon_n).
\end{equation}    

At receiver 1, the decoder makes an error iff one or more of the following events occur: 
\begin{align*}
	\Ec_{11} =& \{(u^n (m_{2}, m_{1k}, m_{r}), v^n(m_{2},m_{1k}, m_{r}, m_{1s}, m_{1r}), y_1^n)\notin \Tc_{\epsilon}^{(n)}\},\\
	\Ec_{12} =& \{(u^n(\tilde{m}_{2}, \tilde{m}_{1k}, \tilde{m}_{r}), v^n(\tilde{m}_{2}, \tilde{m}_{1k}, \tilde{m}_{r}, \tilde{m}_{1s}, \tilde{m}_{1r}),  y_1^n)\in \Tc_{\epsilon}^{(n)}  \mbox{ for some } (\tilde{m}_{2}, \tilde{m}_{1k}, \tilde{m}_{r})\neq (m_{2}, m_{1k}, m_{r})\},\\
	\Ec_{13} =& \{(u^n (m_{2}, m_{1k}, m_{r}), v^n(m_{2},m_{1k}, m_{r}, \tilde{m}_{1s}, \tilde{m}_{1r}), y_1^n)\in \Tc_{\epsilon}^{(n)}  \mbox{ for some } \tilde{m}_{1s}\neq m_{1s}\}.
\end{align*} 
So $P_{e,1}$ can be upper bounded by 
	\begin{equation*}
		P_{e,1}\leq 
		\Pr(\Ec_{11})+\Pr(\Ec_{12})+\Pr(\Ec_{13}).
	\end{equation*} 
By the LLN,  $\Pr(\Ec_{11})$ tends to zero as $n\rightarrow \infty.$ For $\Pr(\Ec_{12})$, since $u^n(\tilde{m}_{2}, \tilde{m}_{1k},  \tilde{m}_{r})$ is independent of $(u^n(m_{2}, m_{1k}, m_{r}), y_1^n)$ for $(\tilde{m}_{2}, \tilde{m}_{1k})\neq (m_{2}, m_{1k}),$ by the packing lemma \cite{ElGamal:2012}, $\Pr(\Ec_{12})$ tends to zero as $n\rightarrow \infty$ if
	\begin{equation}\label{eqn: ReCon2 on R_2+R_{1k}+R_{r}+R_{1s}+R_{1r}}
		R_2+R_{1k}+R_{r}+R_{1s}+R_{1r}\leq I(U, V_1;Y_1)-\delta_n(\epsilon_n).
	\end{equation}
For $\Pr(\Ec_{13})$, note that if $(\tilde{m}_{1s}, \tilde{m}_{1r})\neq (m_{1s}, m_{1r}),$ then for a given $u^n(m_{2}, m_{1k}, m_{r})$, $v^n(m_{2},  m_{1k}, m_{r}, \tilde{m}_{1s}, \tilde{m}_{1r})$ is independent of $(v^n(m_{2}, m_{1k},  m_{r}, m_{1s}, m_{1r}), y_1^n).$  By the packing lemma \cite{ElGamal:2012}, $\Pr(\Ec_{13})$ tends to zero as $n\rightarrow \infty$ if 
\begin{equation}\label{eqn: ReCon on R_{1s}+R_{1r}}
	R_{1s}+R_{1r}\leq I(V;Y_1|U)-\delta_n(\epsilon_n).
\end{equation}

{\em Analysis of individual secrecy:} 
For the individual secrecy \eqref{eq:IndSec}, i.e., $R_{L,i}\leq \tau_n,$ for $i=1,2,$ we show in the following its equivalent form that $H(M_i|Z^n)\geq nR_i-n\tau_n.$  

First consider $H(M_2|Z^n).$ We have
	\begin{align*}
	 H(M_2|Z^n)= & H(M_{2}, Z^n)-H(Z^n)\\ 
			=	& H(U^n, M_{2}, Z^n)-H(U^n|M_2, Z^n)-H(Z^n)\\
			= 	& H(U^n)+H(Z^n|U^n)-H(U^n|M_2, Z^n)-H(Z^n)\\
			\stackrel{(a)}{\geq} & H(U^n)+H(Z^n|U^n)-n[R_{1k}+R_{r}-I(U;Z)]-H(Z^n)-n\tau_n/2\\
			\stackrel{(b)}{=} & n[R_2+R_{1k}+R_{r}]-n[R_{1k}+R_{r}-I(U;Z)]-I(U^n;Z^n)-n\tau_n/2\\
			=& nR_2+nI(U;Z)-I(U^n;Z^n)-n\tau_n/2\\
			\stackrel{(c)}{\geq}& nR_2-n\tau_n
	\end{align*}
where 	$(a)$ follows from \cite[Lemma 1]{Chia:Three-receiver12} that $H(U^n|M_{2}, Z^n)\leq n[R_{1k}+R_{r}-I(U;Z)]+n\tau_n/2,$ if taking 
			\begin{equation}\label{eqn: IndS Cond on R_{1k}+R_{r}}
				R_{1k}+R_{r}\geq I(U;Z)+\delta_n(\tau_n);
			\end{equation}		
		$(b)$ follows from the codebook construction that $H(U^n)=n[R_2+R_{1k}+R_{r}];$ and
		$(c)$ is due to the fact that $I(U^n;Z^n)\leq nI(U;Z)+n\tau_n/2,$ the proof of which is given as follows. 
		\begin{align*}
			I(U^n;Z^n)
					=& H(Z^n)-H(Z^n|U^n)\\
					=& H(Z^n)-H(Z^n|U^n, V^n)-I(V^n;Z^n|U^n)\\
					\stackrel{(d)}{=} & H(Z^n)-nH(Z|U, V)-H(V^n|U^n)+H(V^n|U^n, Z^n)\\
					\stackrel{(e)}{\leq} & H(Z^n)-nH(Z|U, V)-H(V^n|U^n)+n[R_{1s}+R_{1r}-I(V;Z|U)]+n\tau_n/2\\
					\stackrel{(f)}{\leq } & nH(Z)-nH(Z|U, V)-n[R_{1s}+R_{1r}]+n[R_{1s}+R_{1r}-I(V;Z|U)]+n\tau_n/2\\
					=& nI(U;Z)+n\tau_n/2,
		\end{align*}
where $(d)$ is due to the discrete memoryless of the channel; $(e)$ follows from \cite[Lemma 1]{Chia:Three-receiver12} that $H(V^n|U^n, Z^n)\leq n[R_{1s}+R_{1r}-I(V;Z|U)]+n\tau_n/2,$ if taking 
					\begin{equation}\label{eqn: IndS Cond on R_{1s}+R_{1r}}
						R_{1s}+R_{1r}\geq I(V;Z|U)+\delta_n(\tau_n);
					\end{equation} 
		$(f)$ follows from the fact that $H(Z^n)=\sum_{i=1}^{n}H(Z_i|Z^{i-1})\leq \sum_{i=1}^{n}H(Z_i)=nH(Z)$ and by the codebook construction $H(V^n|U^n)=n[R_{1s}+R_{1r}].$
		
For $H(M_1|Z^n),$ we have 
	\begin{align*}
	H(M_1| Z^n)	= 
				& H(M_{1k},M_{1s}|Z^n) \\
			= 	& H(M_{2},M_{1k}, M_{r}, M_{1s}|Z^n)-H(M_{2}, M_{r}|M_{1k}, M_{1s}, Z^n) \\
		 	=	& H(U^n, M_{1s}|Z^n)-H(U^n|M_{1k}, M_{1s}, Z^n) \\
		 	\stackrel{(g)}{\geq} & H(U^n|Z^n)+H(M_{1s}|U^n,Z^n)-H(U^n|M_{1k}, Z^n) \\
		 	=	&  H(U^n|Z^n)+H(V^n|U^n, Z^n)-H(V^n|M_{1s},U^n, Z^n)-H(U^n|M_{1k}, Z^n) \\
		 	\stackrel{(h)}{\geq}  &  H(U^n, V^n|Z^n)-n[R_{1r}-I(V;Z|U)]-n[R_{2}+R_{r}-I(U;Z)]-n\tau_n\\
		 	=& H(U^n, V^n)-I(U^n, V^n;Z^n)-n[R_{1r}+R_{2}+R_{r}]+nI(U,V;Z)-n\tau_n\\
		 	\stackrel{(i)}{\geq} & n[R_2+R_{1k}+R_{r}+R_{1s}+R_{1r}]-I(U^n, V^n;Z^n)-n[R_{1r}+R_{2}+R_{r}]+nI(U,V;Z)-n\tau_n\\
		 	= & nR_1-I(U^n, V^n;Z^n)+nI(U,V;Z)-n\tau_n\\
		 	\stackrel{(j)}{\geq}& nR_1-n\tau_n, 
			\end{align*}
where 
	$(g)$ is due to the fact that conditioning reduces entropy;
	$(h)$ follows from \cite[Lemma 1]{Chia:Three-receiver12} that by taking 
			\begin{equation}\label{eqn: IndS Cond on R_{2}+R_{r}}
				R_{2}+R_{r}\geq I(U;Z)+\delta_n(\tau_n),
			\end{equation}
	we have $H(U^n|M_{1k}, Z^n)\leq n[R_{2}+R_{r}-I(U;Z)]+n\tau_n/2;$ and by taking 
			\begin{equation}\label{eqn: IndS Cond on R_{1r}}
				R_{1r}\geq I(V;Z|U)+\delta_n(\tau_n),
			\end{equation}
	we have $H(V^n|M_{1s},U^n, Z^n)\leq n[R_{1r}-I(V;Z|U)]+n\tau_n/2;$ 
	$(i)$ is by the codebook construction that $H(U^n, V^n)=n[R_2+R_{1k}+R_{r}+R_{1s}+R_{1r}];$
	$(j)$ is due to the fact that $I(U^n, V^n;Z^n)\leq nI(U,V;Z),$ the proof of which is given as follows:
		\begin{align*}
			I(U^n, V^n;Z^n) =& H(Z^n)-H(Z^n|U^n, V^n)\\
							\stackrel{(k)}{=} & H(Z^n)-nH(Z|U,V)\\
							\stackrel{(l)}{\leq} & nH(Z)-nH(Z|U,V)\\
							=& nI(U,V;Z),
		\end{align*}
where 
	$(k)$ is due to the discrete memoryless of the channel; and $(l)$ follows from the fact that $H(Z^n)=\sum_{i=1}^{n}H(Z_i|Z^{i-1})\leq \sum_{i=1}^{n}H(Z_i)=nH(Z).$ 
	 
{\em Achievable rate region:} The resulting region has the following constraints:
the non-negativity for rates, i.e., $R_{1k}, R_{1s}, R_{r}, R_{1r} \geq  0$,
the rate relations imposed by rate splitting as specified in \eqref{eqn: Rate Relation R_1},
the conditions for a reliable communication, i.e., \eqref{eqn: ReCon1 on R_2+R_{1k}+R_{r}}, \eqref{eqn: ReCon2 on R_2+R_{1k}+R_{r}+R_{1s}+R_{1r}}, \eqref{eqn: ReCon on R_{1s}+R_{1r}}, and
the conditions for individual secrecy of the messages at the eavesdropper, i.e., \eqref{eqn: IndS Cond on R_{1k}+R_{r}}, \eqref{eqn: IndS Cond on R_{1s}+R_{1r}}, \eqref{eqn: IndS Cond on R_{2}+R_{r}}, \eqref{eqn: IndS Cond on R_{1r}}.
Eliminating $R_{1r}, R_{r}$ by applying Fourier-Motzkin procedure \cite{ElGamal:2012}, we get the desired rate region as defined in \eqref{eqn: Region_Superposition R_1s R1k R2}; further eliminating $R_{1s}, R_{1k},$ we obtain \eqref{eqn: Region_Superposition R_1 R_2}.


\section{Proof of Theorem \ref{thm: DBC IndS upper bound}}\label{sec:UpperBound1}

Consider a DM-BC with an external eavesdropper such that $Y_2$ is a degraded version of $Y_1$ and $Y_2$ is less noisy than $Z$. For a reliable communication under individual secrecy constraint, we have
\allowdisplaybreaks
\begin{align}
	nR_2= H(M_2)&=I(M_2;Y_2^n)+H(M_2|Y_2^n) \nonumber\\
		\stackrel{(a)}{\leq}& I(M_2;Y_2^n)-I(M_2;Z^n)+n\lambda_2(\epsilon_n, \tau_n)\label{eqn: CProof R_2 Step 1}
\end{align}
where $(a)$ is due to the reliability constraint \eqref{eq:Reliability} and individual secrecy constraint \eqref{eq:IndSec} and by taking $\lambda_2(\epsilon_n, \tau_n)=\tau_n+1/n+\epsilon_n R_2.$

Moreover, we have
\begin{align}
	nR_1= 	& H(M_1)=H(M_1|M_2) \nonumber\\
		=	& I(M_1;Y_1^n|M_2)+H(M_1|M_2, Y_1^n) \nonumber\\
		\stackrel{(c)}{\leq}& \underbrace{I(M_1;Y_1^n|M_2)-I(M_1;Z^n|M_2)}_{nR_1^s}+\underbrace{I(M_1;Z^n|M_2)}_{nR_1^k}+n\lambda_1(\epsilon_n)  \label{eqn: CProof R_1Step 1}
\end{align}
where $(c)$ is due to the reliability constraint \eqref{eq:Reliability} and Fano's inequality, the fact that $H(M_1|M_2, Y_1^n)\leq H(M_1|Y_1^n),$ and by taking $\lambda_1(\epsilon_n)=1/n+\epsilon_n R_1.$

Note that for $nR_1^k$ in \eqref{eqn: CProof R_1Step 1}, we have
\begin{align}
	nR_1^k= & I(M_1;Z^n|M_2) \nonumber\\
		= & I(M_1; Y_2^n|M_2)-I(M_1; Y_2^n|M_2) +I(M_1;Z^n|M_2)\nonumber\\
		=& I(M_1,M_2; Y_2^n)-I(M_2;Y_2^n) -I(M_1; Y_2^n|M_2) +I(M_1;Z^n|M_2)\nonumber\\
		\stackrel{(d)}{\leq} & I(M_1,M_2; Y_2^n)-I(M_1,M_2;Z^n)-I(M_1; Y_2^n|M_2) +I(M_1;Z^n|M_2) +n\lambda_2(\epsilon_n, \tau_n)\nonumber\\
		=& I(M_2;Y_2^n)-I(M_2;Z^n)+n\lambda_2(\epsilon_n, \tau_n)\label{eqn: CProof R_1Step R_1k}
\end{align}
where $(d)$ follows that $I(M_2;Y_2^n) \geq I(M_1,M_2; Z^n)-n\lambda_2(\epsilon_n, \tau_n),$ which proof is provided as follows:
\begin{align*}
	I(M_2;Y_2^n)= 
		& H(M_2)-H(M_2|Y_2 ^n)
		\stackrel{(f)}{\geq}  H(M_2)-n\lambda_2(\epsilon_n)\\
		=& H(M_2|M_1)-n\lambda_2(\epsilon_n)
		\geq  I(M_2;Z^n|M_1)-n\lambda_2(\epsilon_n)\\
		=& I(M_1,M_2; Z^n)-I(M_1;Z^n)-n\lambda_2(\epsilon_n)\\
		\stackrel{(g)}{\geq} & I(M_1,M_2; Z^n)-n\lambda_2(\epsilon_n, \tau_n)
\end{align*}
where $(f)$ is due to the reliability constraint\eqref{eq:Reliability} and Fano's inequality and by taking $\lambda_2(\epsilon_n)=1/n+\epsilon_n R_2$; and $(g)$ is due to the individual secrecy constraint \eqref{eq:IndSec} and by taking $\lambda_2(\epsilon_n, \tau_n)=\tau_n+\lambda_2(\epsilon_n).$ 

For $I(M_2; Y_2^n)-I(M_2;Z^n)$ in \eqref{eqn: CProof R_2 Step 1} and \eqref{eqn: CProof R_1Step R_1k}, we have
\begin{align}		
 I(M_2; &  Y_2^n)-I(M_2;Z^n)
	= 	 \sum_{i=1}^n\left[ I(M_2;Y_{2i}|Y_2^{i-1})-I(M_2;Z_{i}|Z_{i+1}^n)\right]\nonumber\\
		\stackrel{(h)}{=}	& \sum_{i=1}^n \left[I(M_2;Y_{2i}|Y_2^{i-1})-I(M_2;Z_{i}|Z_{i+1}^n)\right]
			+\sum_{i=1}^n \left[I(Z_{i+1}^n;Y_{2i}|M_2,Y_2^{i-1})-I(Y_2^{i-1};Z_{i}|M_2,Z_{i+1}^n)\right]\nonumber\\
		= 	& \sum_{i=1}^n \left[I(M_2, Z_{i+1}^n;Y_{2i}|Y_2^{i-1})-I(M_2, Y_2^{i-1};Z_{i}|Z_{i+1}^n)\right]\nonumber\\
		=	& \sum_{i=1}^n \left[I(M_2;Y_{2i}|Y_2^{i-1}, Z_{i+1}^n)-I(M_2;Z_{i}|Y_2^{i-1}, Z_{i+1}^n)\right]
			+\sum_{i=1}^n \left[I(Z_{i+1}^n;Y_{2i}|Y_2^{i-1})-I(Y_2^{i-1};Z_{i}|Z_{i+1}^n)\right]\nonumber\\
		\stackrel{(h)}{=}	& \sum_{i=1}^n \left[I(M_2;Y_{2i}|Y_2^{i-1}, Z_{i+1}^n)-I(M_2;Z_{i}|Y_2^{i-1}, Z_{i+1}^n)\right]\nonumber\\
		=&\sum_{i=1}^n \left[I(M_2, Y_{2}^{i-1}, Z_{i+1}^n;Y_{2i})-I(M_2,Y_{2}^{i-1},Z_{i+1}^n;Z_{i})\right]
		-\sum_{i=1}^n \left[I(Y_{2}^{i-1}, Z_{i+1}^n;Y_{2i})-I(Y_{2}^{i-1}, Z_{i+1}^n;Z_{i})\right]\nonumber\\\
		\stackrel{(i)}{\leq} 	&\sum_{i=1}^n \left[I(M_2, Y_{2}^{i-1}, Z_{i+1}^n;Y_{2i})-I(M_2,Y_{2}^{i-1}, Z_{i+1}^n;Z_{i})\right]\nonumber\\
		=&\sum_{i=1}^n \left[I(M_2, Y_1^{i-1}, Y_{2}^{i-1}, Z_{i+1}^n;Y_{2i})-I(M_2, Y_1^{i-1},Y_{2}^{i-1}, Z_{i+1}^n;Z_{i})\right]\nonumber\\
		& -\sum_{i=1}^n \left[I(Y_1^{i-1};Y_{2i}|M_2, Y_{2}^{i-1}, Z_{i+1}^n)-I(Y_1^{i-1};Z_{i}|M_2, Y_{2}^{i-1}, Z_{i+1}^n)\right]\nonumber\\
		\stackrel{(i)}{\leq} 	&\sum_{i=1}^n \left[I(M_2, Y_1^{i-1}, Y_{2}^{i-1}, Z_{i+1}^n;Y_{2i})-I(M_2, Y_1^{i-1}, Y_{2}^{i-1}, Z_{i+1}^n;Z_{i})\right]\nonumber\\
		\stackrel{(j)}{=}	&\sum_{i=1}^n \left[I(U_i;Y_{2i})-I(U_i;Z_{i})\right], \label{eqn: CProof R_2 R_1k}
\end{align}
where $(h)$ is due to the Csisz\'{a}r sum identity; $(i)$ is due to the fact that the channel to legitimate receiver 2 is less noisy than the one to the eavesdropper; and $(j)$ is by setting $U_i=(M_2, Y_{1}^{i-1}, Y_{2}^{i-1}, Z_{i+1}^n).$

Replacing \eqref{eqn: CProof R_2 R_1k} in \eqref{eqn: CProof R_2 Step 1} and \eqref{eqn: CProof R_1Step R_1k}, respectively, we obtain
\begin{align}
	n R_2 \leq & \sum_{i=1}^n I(U_i;Y_{2i})-I(U_i;Z_{i})+n\lambda_2(\epsilon_n, \tau_n); \label{eqn: CProof R_2 Step Done}\\
	n R_1^k \leq & \sum_{i=1}^n I(U_i;Y_{2i})-I(U_i;Z_{i})+n\lambda_2(\epsilon_n, \tau_n). \label{eqn: CProof R_1k Step Done}
\end{align}

Similarly we bound $R_1^s$ in \eqref{eqn: CProof R_1Step 1} as follows:
\begin{align}
	nR_1^s 	=&I(M_1;Y_1^n|M_2)-I(M_1;Z^n|M_2) \nonumber\\
			\stackrel{(k)}{=}&\sum_{i=1}^n 
				\left[I(M_1;Y_{1i}|M_2,Y_1^{i-1}, Z_{i+1}^n)-I(M_1;Z_{i}|M_2,Y_1^{i-1}, Z_{i+1}^n)\right] \nonumber\\
			{=}& \sum_{i=1}^n 
				\left[I(M_1, Y_2^{i-1};Y_{1i}|M_2,Y_1^{i-1}, Z_{i+1}^n)-I(M_1, Y_2^{i-1};Z_{i}|M_2,Y_1^{i-1}, Z_{i+1}^n)\right] \nonumber\\
			& -\sum_{i=1}^n 
				\left[I(Y_2^{i-1};Y_{1i}|M_1, M_2,Y_1^{i-1}, Z_{i+1}^n)-I(Y_2^{i-1};Z_{i}|M_1, M_2,Y_1^{i-1}, Z_{i+1}^n)\right]\nonumber\\
			{\leq}& \sum_{i=1}^n 
				\left[I(M_1, Y_2^{i-1};Y_{1i}|M_2,Y_1^{i-1}, Z_{i+1}^n)-I(M_1, Y_2^{i-1};Z_{i}|M_2,Y_1^{i-1}, Z_{i+1}^n)\right]\nonumber\\
			=& \sum_{i=1}^n 
				\left[I(M_1;Y_{1i}|M_2,Y_1^{i-1}, Y_2^{i-1}, Z_{i+1}^n)-I(M_1;Z_{i}|M_2,Y_1^{i-1}, Y_2^{i-1}, Z_{i+1}^n)\right]\nonumber\\
			& +\sum_{i=1}^n 
				\left[I(Y_2^{i-1};Y_{1i}|M_2,Y_1^{i-1}, Z_{i+1}^n)-I(Y_2^{i-1};Z_{i}| M_2,Y_1^{i-1}, Z_{i+1}^n)\right]\nonumber\\
			\stackrel{(l)}{=}& \sum_{i=1}^n 
				\left[I(M_1;Y_{1i}|M_2,Y_1^{i-1}, Y_2^{i-1}, Z_{i+1}^n)-I(M_1;Z_{i}|M_2,Y_1^{i-1}, Y_2^{i-1}, Z_{i+1}^n)\right]\nonumber\\
			\stackrel{(m)}{\leq}&\sum_{i=1}^n 
				\left[I(V_i;Y_{1i}|U_i)-I(V_i;Z_{i}|U_i)\right] \label{eqn: CProof R_1s Step Done}
\end{align}
where $(k)$ is obtained by applying the Csisz\'{a}r sum identity twice; 
$(l)$ is due to the channel degradedness that implies the Markov chains $Y_2^{i-1}\to (M_2,Y_1^{i-1}, Z_{i+1}^n)\to (Y_{1i}, Z_{i})$; and $(m)$ follows by the fact $U_i=(M_2,Y_1^{i-1}, Y_{2}^{i-1}, Z_{i+1}^n)$ and further setting $V_i=(M_1, U_i).$

Replacing $R_1^k$ and $R_1^s$ in \eqref{eqn: CProof R_1Step 1} by \eqref{eqn: CProof R_1k Step Done} and \eqref{eqn: CProof R_1s Step Done}, respectively, we obtain
\begin{align}
	nR_1\leq & nR_1^k+nR_1^s+n\lambda_1(\epsilon_n) \nonumber \\
			\stackrel{(n)}{\leq} & \sum_{i=1}^n \left[I(V_i;Y_{1i}|U_i)-I(V_i;Z_{i}|U_i) \right]+\sum_{i=1}^n \left[I(U_i;Y_{2i})-I(U_i;Z_{i})\right]+n\lambda(\epsilon_n, \tau_n) \label{eqn: CProof R_1 Step Done}
\end{align}
where $(n)$ is by taking $\lambda(\epsilon_n, \tau_n)=\lambda_1(\epsilon_n)+\lambda_2(\epsilon_n, \tau_n).$

Now we proceed to bound $R_1+R_2.$
\begin{align}
	n(R_1+R_2)= & H(M_1|M_2)+H(M_2)\nonumber\\
		=& I(M_1;Y_1^n|M_2)+I(M_2;Y_2^n)+H(M_1|M_2,Y_1^n)+H(M_2|Y_2^n)\nonumber\\
		\stackrel{(o)}{\leq}& I(M_1;Y_1^n|M_2)+I(M_2;Y_2^n)+n\lambda(\epsilon_n)\nonumber\\
		= & I(M_1;Y_1^n|M_2)-I(M_1;Z^n|M_2)+I(M_2;Y_2^n)-I(M_2;Z^n)+I(M_1,M_2;Z^n)+n\lambda(\epsilon_n)\nonumber\\
		\stackrel{(p)}{\leq} & \sum_{i=1}^n \left[I(V_i;Y_{1i}|U_i)-I(V_i;Z_{i}|U_i)\right]+\sum_{i=1}^n \left[I(U_i;Y_{2i})-I(U_i;Z_{i})\right]+\sum_{i=1}^n I(M_1, M_2;Z_{i}|Z_{i+1}^n)+n\lambda(\epsilon_n)\nonumber\\
		\stackrel{(q)}{\leq} &\sum_{i=1}^n \left[I(V_i;Y_{1i}|U_i)-I(V_i;Z_{i}|U_i)\right]+\sum_{i=1}^n \left[I(U_i;Y_{2i})-I(U_i;Z_{i})\right]+\sum_{i=1}^n I(U_i, V_i;Z_{i})+n\lambda(\epsilon_n)\nonumber\\
		= & \sum_{i=1}^n \left[I(V_i;Y_{1i}|U_i)+I(U_i;Y_{2i})\right]+n\lambda(\epsilon_n) \label{eqn: CProof R_1+R_2 Step Done}
\end{align}
where $(o)$ is due to the reliability constraint \eqref{eq:Reliability} and Fano's inequality, the fact that $H(M_1|M_2, Y_1^n)\leq H(M_1|Y_1^n)$ and by taking $\lambda(\epsilon_n)=2/n+\epsilon_n (R_1+R_2);$ $(p)$ is due to \eqref{eqn: CProof R_2 R_1k} and \eqref{eqn: CProof R_1s Step Done}; and $(q)$ is due to the definition of $U_i$ and $V_i,$ i.e., $U_i=(M_2,Y_1^{i-1}, Y_{2}^{i-1}, Z_{i+1}^n)$ and $V_i=(M_1, U_i).$

Introducing a time-sharing random variable $Q$ which is uniform over $1, 2\cdots, n$ and taking $U=(U_Q, Q), V=V_Q, Y_1=Y_{1, Q}, Y_2=Y_{2, Q}, Z=Z_{Q},$ we proceed on \eqref{eqn: CProof R_2 Step Done}, \eqref{eqn: CProof R_1 Step Done} and \eqref{eqn: CProof R_1+R_2 Step Done} as follows:
\begin{align*}
	R_2 \leq & I(U;Y_{2})-I(U;Z)+\lambda_2(\epsilon_n, \tau_n) \\
	R_1 \leq & I(V;Y_{1}|U)-I(V;Z|U)+I(U;Y_{2})-I(U;Z)+\lambda(\epsilon_n, \tau_n)\\
	R_1+R_2 \leq & 	I(V;Y_{1}|U)+I(U;Y_2)+\lambda(\epsilon_n)					
\end{align*}
Taking the limit as $n\to \infty$ such that $\lambda_2(\epsilon_n, \tau_n), \lambda(\epsilon_n, \tau_n), \lambda(\epsilon_n)\to 0,$ we conclude our proof of the upper bound.


\section{Proof of Theorem~\ref{thm:Marton}}\label{sec:AppMarton}

For a given input probability distribution $p(u,v_1,v_2,x)$, let $I_1=I(V;Y_1|U)-I(V;Z|U)$ and $I_2=I(V;Y_2|U)-I(V;Z|U)$. If $I_1, I_2\leq 0$, the claimed region \eqref{eqn: Region_Marton R_1 R_2} reduces to \eqref{eqn: Region_Primitive}, which is achievable by taking the primitive approach as described in Section \ref{sec: primitive approach}. We assume $I_1>0$. Now, if $I_2\leq 0$, the claimed region \eqref{eqn: Region_Marton R_1 R_2} reduces to \eqref{eqn: Region_Superposition R_1 R_2}, which is achievable by employing the superposition approach as described in Section \ref{sec: DB-BC superposition}. A similar proof applies to the case of $I_1\leq 0$ and $I_2>0.$ In the following, we provide the detailed achievability proof for the remaining case, i.e., if $I_1>0$ and $I_2>0$ for a given $p(u,v_1,v_2,x)$.

\begin{figure}[h]
\centering
\begin{tabular}{rcl}
				$m_1:$ &	& 
						$
						\overbrace{
						\begin{tikzpicture}
							\node[minimum height=1.6em, minimum width=4em, anchor=base, fill=blue!25] {$m_{1k}$}; 
						\end{tikzpicture}
							}^{nR_{1k}}
						\overbrace{					
						\begin{tikzpicture}
							\node[minimum height=1.6em, minimum width=3em, anchor=base, fill=blue!25] {$m_{1s}$};
						\end{tikzpicture} 
						}^{nR_{1s}}
						$\\
				$m_2:$ &	&  
								$
									\underbrace{
										\begin{tikzpicture}
											\node[minimum height=1.6em, minimum width=3em, anchor=base, fill=teal!25] {$m_{2k}$}; 
										\end{tikzpicture}
											}_{nR_{2k}}
										\underbrace{
										\begin{tikzpicture}
											\node[minimum height=1.6em, anchor=base, fill=teal!25] {$m_{2s}$};
										\end{tikzpicture}
												}_{nR_{2s}}
									$
			\end{tabular}
			\caption{Marton's coding: Rate splitting.}
			\label{fig: BC Marton RS}	
			\end{figure}
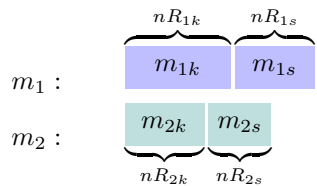

{\em Rate splitting:} As illustrated in Fig. \ref{fig: BC Marton RS}, we represent $M_1, M_2$ by $M_1=(M_{1k}, M_{1s})$ and $M_2=(M_{2k}, M_{2s})$ with $M_{1k}, M_{2k}$ of entropy $nR_{1k}, nR_{2k},$ respectively; while $M_{1s}, M_{2s}$ of entropy $nR_{1s}, nR_{2s},$ respectively. Therefore, we have
\begin{align}
	R_1& =R_{1k} +R_{1s}; \label{eqn: Marton R_1}\\
	R_2& =R_{2k} +R_{2s}. \label{eqn: Marton R_2}
\end{align}

{\em Codebook generation:} Fix $p(u), p(v_1,v_2|u).$ 

First, randomly generate $2^{n(R_{1k}+R_{2k}+R_{r})}$ i.i.d. sequences $u^{n}(m_{2k}, m_{1k}, m_{r}),$ with $(m_{2k}, m_{1k}, m_{r})\in[1:2^{nR_{2k}}]\times[1:2^{nR_{1k}}]\times[1:2^{nR_{r}}],$ according to $p(u).$

For each fixed $u^n(m_{2k}, m_{1k}, m_{r}), $ randomly generate $2^{n(R_{1s}+R_{1r}+R_{1c})}$ i.i.d. sequences $v_1^n(m_{2k}, m_{1k}, m_{r}, m_{1s}, \allowbreak m_{1r}, m_{1c})$ with $(m_{1s}, m_{1r}, m_{1c})\in [1: 2^{nR_{1s}}]\times [1: 2^{nR_{1r}}]\times [1: 2^{nR_{1c}}]$, according to $p(v_1|u)$; and similarly generate $2^{n(R_{2s}+R_{2r}+R_{2c})}$ i.i.d. sequences $v_2^n(m_{2k}, m_{1k}, m_{r}, m_{2s}, m_{2r}, m_{2c})$ with $(m_{2s}, m_{2r}, m_{2c})\in [1: 2^{nR_{2s}}]\times [1: 2^{nR_{2r}}]\times [1: 2^{nR_{2c}}],$ according to $p(v_2|u)$. 

{\em Encoding:} To send messages $(m_1, m_2),$ with $m_1=(m_{1k}, m_{1s}),$ $m_2=(m_{2k}, m_{2s}),$ randomly choose $m_{r}\in [1: 2^{nR_r}]$ and find $u^n(m_{2k}, m_{1k}, m_{r}).$

Given $u^n(m_{2k}, m_{1k}, m_{r}),$ randomly choose $(m_{1r}, m_{2r})\in [1: 2^{nR_{1r}}] \times [1: 2^{nR_{2r}}],$ and pick $(m_{1c}, m_{2c})$ such that $v_1^n(m_{2k}, m_{1k}, m_{r}, m_{1s}, m_{1r}, m_{1c})$ and $v_1^n(m_{2k}, m_{1k}, m_{r}, m_{2s}, m_{2r}, m_{2c})$ are jointly typical. (If there is more than one such jointly typical pair, choose one of them uniformly at random.)
This is possible with high probability, if 
\begin{equation}\label{eqn: Marton cond on cover index}
		R_{1c}+R_{2c}>I(V_1;V_2|U)
\end{equation}
(refer to \cite{src:ElGamal1981} for the proof). 

Finally, for the chosen jointly typical pair $(v_1^n, v_2^n)$, generate a codeword $x^n$ at random according to $p(x|v_1, v_2)$ and transmit it.

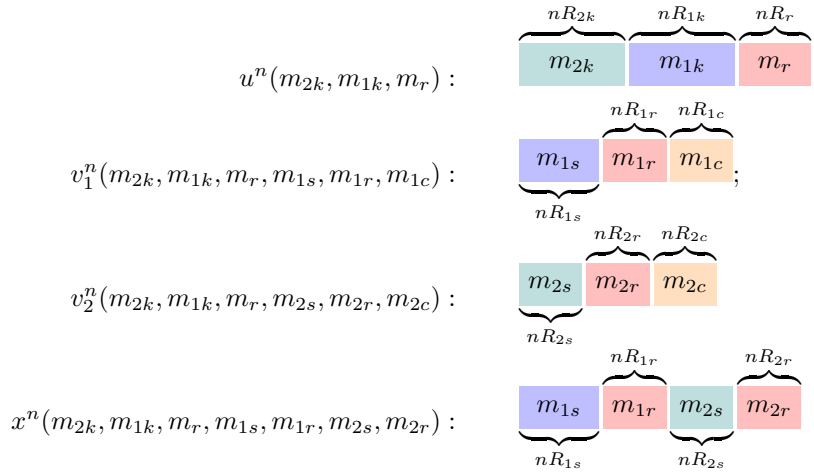
\begin{figure}[h]
\centering
	\begin{tabular}{rcl}
			$u^n(m_{2k}, m_{1k}, m_{r}):$ &	& 
						$
						\overbrace{
						\begin{tikzpicture}
							\node[minimum height=1.6em, minimum width=4em, anchor=base, fill=teal!25] {$m_{2k}$}; 
						\end{tikzpicture}
							}^{nR_{2k}}
						\overbrace{
						\begin{tikzpicture}
							\node[minimum height=1.6em, minimum width=4em, anchor=base, fill=blue!25] {$m_{1k}$}; 
						\end{tikzpicture}
							}^{nR_{1k}}
						\overbrace{
						\begin{tikzpicture}
							\node[minimum height=1.6em, anchor=base, fill=red!25] {$\ m_{r}\ $}; 
						\end{tikzpicture}
						}^{nR_r}
						$\\
			$v_1^n(m_{2k}, m_{1k}, m_{r}, m_{1s}, m_{1r}, m_{1c}):$ &	& 
														$
														\underbrace{
														\begin{tikzpicture}
															\node[minimum height=1.6em, minimum width=3em, anchor=base, fill=blue!25] {$m_{1s}$}; 
														\end{tikzpicture}
														}_{nR_{1s}}
														\overbrace{
														\begin{tikzpicture}
															\node[minimum height=1.6em, minimum width=2em, anchor=base, fill=red!25] {$m_{1r}$}; 
														\end{tikzpicture}
														}^{nR_{1r}}
						 								\overbrace{
														\begin{tikzpicture}
															\node[minimum height=1.6em, minimum width=2em, anchor=base, fill=orange!25] {$m_{1c}$}; 
														\end{tikzpicture}
														}^{nR_{1c}}; 
													$\\
			$v_2^n(m_{2k}, m_{1k}, m_{r}, m_{2s}, m_{2r}, m_{2c}):$ &	& 
																	$
																	\underbrace{
																	\begin{tikzpicture}
																		\node[minimum height=1.6em, anchor=base, fill=teal!25] {$m_{2s}$};
																	\end{tikzpicture}
																	}_{nR_{2s}}
																	\overbrace{
																	\begin{tikzpicture}
																		\node[minimum height=1.6em, minimum width=2em, anchor=base, fill=red!25] {$m_{2r}$}; 
																	\end{tikzpicture}
																	}^{nR_{2r}}
									 								\overbrace{
																	\begin{tikzpicture}
																		\node[minimum height=1.6em, minimum width=2em, anchor=base, fill=orange!25] {$m_{2c}$}; 
																	\end{tikzpicture}
																	}^{nR_{2c}} 
																$\\
			$x^n(m_{2k}, m_{1k}, m_{r}, m_{1s}, m_{1r}, m_{2s}, m_{2r}):$ &	& 
																	$\underbrace{
																	\begin{tikzpicture}
																	\node[minimum height=1.6em, minimum width=3em, anchor=base, fill=blue!25] {$m_{1s}$}; 
																	\end{tikzpicture}
																	}_{nR_{1s}}
																	\overbrace{
																	\begin{tikzpicture}
																	\node[minimum height=1.6em, minimum width=2em, anchor=base, fill=red!25] {$m_{1r}$}; 							\end{tikzpicture}}^{nR_{1r}}
																	\underbrace{
																	\begin{tikzpicture}
																		\node[minimum height=1.6em, anchor=base, fill=teal!25] {$m_{2s}$}; 
																	\end{tikzpicture}
																	}_{nR_{2s}}
																	\overbrace{
																	\begin{tikzpicture}
																		\node[minimum height=1.6em, minimum width=2em, anchor=base, fill=red!25] {$m_{2r}$}; 
																	\end{tikzpicture}
																	}^{nR_{2r}}
																	$
			\end{tabular}
		\caption{Marton's coding: Encoding.}
		\label{fig: BC Marton encoding}
		\end{figure}

The choice of $u^n, v_1^n, v_2^n, x^n$ for given $(m_1, m_2)$ is illustrated in Fig. \ref{fig: BC Marton encoding}. 

{\em Decoding:} Receiver 1, upon receiving $y_1^n,$ finds a unique tuple $(\hat{m}_{2k}, \hat{m}_{1k}, \hat{m}_{r}, \hat{m}_{1s})$ such that $(u^n(\hat{m}_{2k}, \hat{m}_{1k}, \hat{m}_{r}), \allowbreak  v_1^n(\hat{m}_{2k}, \hat{m}_{1k}, \hat{m}_{r}, \hat{m}_{1s}, \hat{m}_{1r}, \hat{m}_{1c})$ is jointly typical with $y_1^n$ for some $(\hat{m}_{1r}, \hat{m}_{1c}).$ And, receiver 2, upon receiving $y_2^n,$ finds a unique tuple $(\tilde{m}_{2k},  \tilde{m}_{1k}, \allowbreak\tilde{m}_{r}, \tilde{m}_{2s})$ such that $(u^n(\tilde{m}_{2k}, \tilde{m}_{1k}, \tilde{m}_{r}), v_2^n(\tilde{m}_{2k},  \tilde{m}_{1k}, \allowbreak\tilde{m}_{r}, \tilde{m}_{2s}, \tilde{m}_{2r}, \tilde{m}_{2c}) )$ is jointly typical with $y_2^n$ for some $(\tilde{m}_{2r}, \tilde{m}_{2c}).$  

{\em Analysis of the error probability of decoding:}  Assume that $m_1=(m_{1k}, m_{1s}),$ $m_2=(m_{2k}, m_{2s})$ is sent. 

For $P_{e,1}$, a decoding error happens if receiver 1's estimate is $(u^n(\hat{m}_{2k}, \hat{m}_{1k}, \hat{m}_{r}), v_1^n(\hat{m}_{2k}, \hat{m}_{1k}, \hat{m}_{r}, \hat{m}_{1s}, \hat{m}_{1r}, \allowbreak\hat{m}_{1c}))$ with $(\hat{m}_{2k}, \hat{m}_{1k}, \hat{m}_{r}, \hat{m}_{1s})\neq (m_{2k}, m_{1k}, m_{r}, m_{1s}).$ In more details, the error event can be partitioned into the followings:
	\begin{enumerate}
		\item Error event corresponds to $(\hat{m}_{2k}, \hat{m}_{1k}, \hat{m}_{r})\neq (m_{2k}, m_{1k}, m_{r}).$ Note that this event occurs with arbitrarily small probability (e.g.: $\epsilon_n/2$) if 
			\begin{equation}\label{eqn: cond Marton Rx1 pe1}
					R_{1k}+R_{2k}+R_{r}+R_{1s}+R_{1r}+R_{1c} \leq  I(U,V_1;Y_1)-\delta_n(\epsilon_n).
			\end{equation}
		\item Error event corresponds to $(\hat{m}_{2k}, \hat{m}_{1k}, \hat{m}_{r})= (m_{2k}, m_{1k}, m_{r})$ but $\hat{m}_{1s}\neq m_{1s}.$ Note that this event occurs with arbitrarily small probability (e.g.: $\epsilon_n/2$) if 
			\begin{equation}\label{eqn: cond Marton Rx1 pe2}
					R_{1s}+R_{1r}+R_{1c}\leq  I(V_1;Y_1|U)-\delta_n(\epsilon_n).
			\end{equation}		
	\end{enumerate}

Similar analysis can be done at the receiver 2, from which the decoding error probability $P_{e,2}$ can be made arbitrarily small (e.g.: $\epsilon_n$) if
	\begin{align}
			R_{1k}+R_{2k}+R_{r}+R_{2s}+R_{2r}+R_{2c} & \leq  I(U, V_2;Y_2)-\delta_n(\epsilon_n); \label{eqn: cond Marton Rx2 pe1}\\
			R_{2s}+R_{2r}+R_{2c} &\leq I(V_2;Y_2|U)-\delta_n(\epsilon_n).\label{eqn: cond Marton Rx2 pe2}
	\end{align} 

{\em Analysis of individual secrecy:} For the individual secrecy 
\eqref{eq:IndSec}, i.e., $R_{L,i}\leq \tau_n,$ for $i=1,2,$ it suffices to show that $H(M_1|Z^n)+H(M_2|Z^n)\geq H(M_1)+H(M_2)-n\tau_n=n(R_1+R_2)-n\tau_n.$

First consider $H(M_1|Z^n),$ we have 
	\begin{align}
	H(M_1| Z^n)	= & H(M_{1k},M_{1s}|Z^n) \nonumber\\
			= 	& H(M_{1k},M_{2k}, M_r, M_{1s}|Z^n)-H(M_{2k}, M_r|M_{1k}, M_{1s}, Z^n) \nonumber\\
		 	=	& H(U^n, M_{1s}|Z^n)-H(U^n|M_{1k}, M_{1s}, Z^n) \nonumber\\
		 	\stackrel{(a)}{\geq} & H(U^n|Z^n)+H(M_{1s}|U^n,Z^n)-H(U^n|M_{1k}, Z^n) \nonumber\\
		 	\stackrel{(b)}{\geq} & H(U^n|Z^n)+H(M_{1s}|U^n,Z^n)-n[R_{2k}+R_{r}-I(U;Z)]-n\tau_n/6 \nonumber\\
		 	=	&  H(U^n|Z^n)+H(V_1^n, M_{1s}|U^n, Z^n)-H(V_1^n|M_{1s},U^n, Z^n)-n[R_{2k}+R_{r}-I(U;Z)]-n\tau_n/6 \nonumber\\
		 	\stackrel{(b)}{\geq}& H(U^n|Z^n)+H(V_1^n|U^n, Z^n)-n[R_{1r}+R_{1c}-I(V_1;Z|U)]-n[R_{2k}+R_{r}-I(U;Z)]-n\tau_n/3 \nonumber\\
		 	= &  H(U^n, V_1^n|Z^n)-n[R_{1r}+R_{1c}-I(V_1;Z|U)]-n[R_{2k}+R_{r}-I(U;Z)]-n\tau_n/3 \label{eqn: IndS Rx1}
			\end{align}
where 
	$(a)$ is due to the fact that conditioning reduces entropy;
	$(b)$ follows from \cite[Lemma 1]{Chia:Three-receiver12} that we have 
	\begin{itemize}
		\item $H(U^n|M_{1k}, Z^n)\leq n[R_{2k}+R_{r}-I(U;Z)]+n\tau_n/6$ if taking 
					\begin{equation}\label{eqn: Marton IndS Cond on R_{2k}+R_{r}}
						R_{2k}+R_{r}\geq I(U;Z)+\delta_n(\tau_n);
					\end{equation}
		\item $H(V_1^n|M_{1s},U^n, Z^n)\leq n[R_{1r}+R_{1c}-I(V_1;Z|U)]+n\tau_n/6$ if taking
					\begin{equation}\label{eqn: Marton IndS Cond on R_{1r}+R_{1c}}
						R_{1r}+R_{1c}\geq I(V_1;Z|U)+\delta_n(\tau_n).
					\end{equation}
	\end{itemize}

Similarly, we could show that 
	\begin{align}
	H(M_2|Z^n)	= & H(M_{2k},M_{2s}|Z^n) \nonumber\\
				\geq & H(U^n, V_2^n|Z^n)-n[R_{2r}+R_{2c}-I(V_2;Z|U)]-n[R_{1k}+R_{r}-I(U;Z)]-n\tau_n/3 \label{eqn: IndS Rx2}
	\end{align}
if taking
			\begin{align}
				R_{1k}+R_{r}  & \geq I(U;Z)+\delta_n(\tau_n); 		\label{eqn: Marton IndS Cond on R_{1k}+R_{r}}\\			
				R_{2r}+R_{2c} & \geq I(V_2;Z|U)+\delta_n(\tau_n). 	\label{eqn: Marton IndS Cond on R_{2r}+R_{2c}}
			\end{align}
			
Note that 
\begin{align}
	H&(U^n, V_1^n|Z^n)+H(U^n, V_2^n|Z^n)\nonumber\\
		& =2H(U^n|Z^n)+	H(V_1^n|U^n, Z^n)+H(V_2^n|U^n, Z^n)\nonumber\\
		& \geq 2H(U^n|Z^n)+	H(V_1^n, V_2^n|U^n, Z^n)\nonumber\\
		& = 2H(U^n)-2I(U^n;Z^n)+H(V_1^n, V_2^n, Z^n|U^n)-H(Z^n|U^n)\nonumber\\
		& = 2H(U^n)-2I(U^n;Z^n)+H(V_1^n, V_2^n|U^n)+H(Z^n|U^n, V_1^n, V_2^n)-H(Z^n|U^n)\nonumber\\
		& \stackrel{(d)}{\geq} 2H(U^n)-2I(U^n;Z^n)+H(V_1^n, V_2^n|U^n)-nI(V_1, V_2;Z|U)\nonumber\\
		& \stackrel{(e)}{\geq} 2n[R_{2k}+R_{1k}+R_{r}]+n[R_{1s}+R_{1r}+R_{2s}+R_{2r}]-2I(U^n;Z^n)-nI(V_1, V_2;Z|U)\nonumber\\
		& \stackrel{(f)}{\geq} 2n[R_{2k}+R_{1k}+R_{r}]+n[R_{1s}+R_{1r}+R_{2s}+R_{2r}]-2nI(U;Z)-nI(V_1, V_2;Z|U)-n\tau_n/3 \label{Indsproof: int step}
\end{align}
where $(d)$ follows from the fact that $H(Z^n|U^n, V_1^n, V_2^n)=nH(Z|U, V_1, V_2)$ due to the discrete memoryless of the channel and  $H(Z^n|U^n)=\sum_{i=1}^n H(Z_i|U^n, Z^{i-1})\leq \sum_{i=1}^n H(Z_i|U_i)=nH(Z|U);$  $(e)$ follows from the codebook construction that $H(U^n)=n[R_{2k}+R_{1k}+R_{r}]$ and $H(V_1^n, V_2^n|U^n)\geq n[R_{1s}+R_{1r}+R_{2s}+R_{2r}];$ and $(f)$ is due to the fact that  $I(U^n;Z^n)\leq nI(U;Z)+n\tau_n/6,$ the proof of which is given as follows:
			\begin{align*}
				I(U^n;Z^n)
						= & H(Z^n)-H(Z^n|U^n)\\
						=& H(Z^n)-H(Z^n|U^n, V_1^n, V_2^n)-I(V_1^n, V_2^n;Z^n|U^n)\\
						\stackrel{(g)}{=} & H(Z^n)-nH(Z|U, V_1, V_2)-H(V_1^n, V_2^n|U^n)+H(V_1^n, V_2^n|U^n, Z^n)\\
						\stackrel{(h)}{\leq} & H(Z^n)-nH(Z|U,V_1, V_2)-H(V_1^n, V_2^n|U^n)+H(V_1^n|U^n, Z^n)+H(V_2^n|U^n, Z^n)\\
						\stackrel{(i)}{\leq } & H(Z^n)-nH(Z|U,V_1, V_2)-H(V_1^n, V_2^n|U^n)\\
						&+n[R_{1s}+R_{1r}+R_{1c}-I(V_1;Z|U)]+n[R_{2s}+R_{2r}+R_{2c}-I(V_2;Z|U)]+n\tau_n/6\\
						\stackrel{(j)}{\leq } & nH(Z)-nH(Z|U,V_1, V_2)-n[R_{1s}+R_{1r}+R_{2s}+R_{2r}]\\
						&  +n[R_{1s}+R_{1r}+R_{1c}-I(V_1;Z|U)]+n[R_{2s}+R_{2r}+R_{2c}-I(V_2;Z|U)]+n\tau_n/6\\
						=& nI(U;Z)+n[R_{1c}+R_{2c}+I(V_1;V_2;Z|U)-I(V_1;Z|U)-I(V_2;Z|U)]+n\tau_n/6\\
						\stackrel{(k)}{\leq } & nI(U;Z)+n\tau_n/6
			\end{align*}
	where $(g)$ is due to the discrete memoryless of the channel; 
 	$(h)$ follows the fact that $H(A,B|C)\leq H(A|C)+H(B|C);$
 	$(i)$  follows from \cite[Lemma 1]{Chia:Three-receiver12} that we have 
 		\begin{itemize}
 			\item $H(V_1^n|U^n, Z^n)\leq n[R_{1s}+R_{1r}+R_{1c}-I(V_1;Z|U)]+n\tau_n/12$ if taking
 						\begin{equation}\label{eqn: Marton IndS Cond on R_{1s}+R_{1r}+R_{1c}}
 							R_{1s}+R_{1r}+R_{1c}\geq I(V_1;Z|U)+\delta_n(\tau_n).
 						\end{equation}
 					(Note that \eqref{eqn: Marton IndS Cond on R_{1s}+R_{1r}+R_{1c}} holds if \eqref{eqn: Marton IndS Cond on R_{1r}+R_{1c}} holds.)
 		    \item $H(V_2^n|U^n, Z^n)\leq n[R_{2s}+R_{2r}+R_{2c}-I(V_2;Z|U)]+n\tau_n/12$ if taking
 			 						\begin{equation}\label{eqn: Marton IndS Cond on R_{2s}+R_{2r}+R_{2c}}
 			 							R_{2s}+R_{2r}+R_{2c}\geq I(V_2;Z|U)+\delta_n(\tau_n).
 			 						\end{equation}
 			 		(Note that \eqref{eqn: Marton IndS Cond on R_{2s}+R_{2r}+R_{2c}} holds if \eqref{eqn: Marton IndS Cond on R_{2r}+R_{2c}} holds.)
 		\end{itemize}
		$(j)$ follows from the fact that $H(Z^n)=\sum_{i=1}^{n}H(Z_i|Z^{i-1})\leq \sum_{i=1}^{n}H(Z_i)=nH(Z)$ and by the codebook construction $H(V_1^n, V_2^n|U^n)\geq n[R_{1s}+R_{1r}+R_{2s}+R_{2r}];$ 
   		and $(k)$ is by taking
   			\begin{equation}\label{eqn: Marton IndS Cond on R_{1c}+R_{2c}}
   					R_{1c}+R_{2c}\leq I(V_1;Z|U)+I(V_2;Z|U)-I(V_1, V_2;Z|U).
   			\end{equation}

Combining \eqref{eqn: IndS Rx1} and \eqref{eqn: IndS Rx2}, we obtain
	\begin{align*}
	 	H(M_1| Z^n)+H(M_2|Z^n) & \stackrel{(l)}{\geq} H(U^n, V_1^n|Z^n)-n[R_{1r}+R_{1c}-I(V_1;Z|U)]-n[R_{2k}+R_{r}-I(U;Z)]-n\tau_n/3 \\
	 			& +H(U^n, V_2^n|Z^n)-n[R_{2r}+R_{2c}-I(V_2;Z|U)]-n[R_{1k}+R_{r}-I(U;Z)]-n\tau_n/3\\
	 			&\stackrel{(m)}{\geq} n[R_1+R_2]-n[R_{1c}+R_{2c}]+n[I(V_1;Z|U)+I(V_2;Z|U)-I(V_1, V_2;Z|U)]-n\tau_n\\
	 			&\stackrel{(n)}{\geq} n[R_1+R_2]-n\tau_n,
	\end{align*}
where $(l)$ is due to \eqref{eqn: IndS Rx1} and \eqref{eqn: IndS Rx2}; $(m)$ is according to \eqref{Indsproof: int step} and the fact that $R_1=R_{1k}+R_{1s}$ and $R_2=R_{2k}+R_{2s}$ as defined in \eqref{eqn: Marton R_1} and \eqref{eqn: Marton R_2}, respectively; and $(n)$ is due to \eqref{eqn: Marton IndS Cond on R_{1c}+R_{2c}}.

{\em Achievable rate region:} We summarize the rate requirements in order to guarantee a reliable communication to both legitimate receivers and satisfy the individual secrecy constraints at the eavesdropper as follows:
\begin{itemize}
		\item the non-negativity for rates, i.e., 
			\begin{equation*}
				R_{1k}, R_{2k}, R_{1s}, R_{2s}, R_{r}, R_{1r}, R_{2r}, R_{1c}, R_{2c} \geq  0;
			\end{equation*}
		\item the rate relations imposed by rate splitting as specified in \eqref{eqn: Marton R_1} and \eqref{eqn: Marton R_2}, i.e.,
			\begin{align*}
						R_1& =R_{1k} +R_{1s}; \\
						R_2& =R_{2k} +R_{2s}.
			\end{align*}
		\item the conditions for a reliable communication to both legitimate receivers, i.e., \eqref{eqn: Marton cond on cover index}, \eqref{eqn: cond Marton Rx1 pe1}, \eqref{eqn: cond Marton Rx1 pe2}, \eqref{eqn: cond Marton Rx2 pe1}, \eqref{eqn: cond Marton Rx2 pe2}:
			\begin{align}
							R_{1c}+R_{2c}				&>I(V_1;V_2|U)												\label{eqn: FM 1}\\
							R_{1k}+R_{2k}+R_{r}+R_{1s}+R_{1r}+R_{1c} 	&\leq  I(U, V_1;Y_1)	\label{eqn: FM 2}\\
							R_{1s}+R_{1r}+R_{1c}	&\leq I(V_1;Y_1|U)										  \label{eqn: FM 4}\\
							R_{1k}+R_{2k}+R_{r}+R_{2s}+R_{2r}+R_{2c} 	& \leq I(U, V_2;Y_2)	\label{eqn: FM 3}\\
							R_{2s}+R_{2r}+R_{2c} 	&\leq I(V_2;Y_2|U)										  \label{eqn: FM 5}
			\end{align}
		\item the conditions for individual secrecy of the messages at the eavesdropper, i.e., \eqref{eqn: Marton IndS Cond on R_{2k}+R_{r}}, \eqref{eqn: Marton IndS Cond on R_{1r}+R_{1c}}, \eqref{eqn: Marton IndS Cond on R_{1k}+R_{r}}, \eqref{eqn: Marton IndS Cond on R_{2r}+R_{2c}}, \eqref{eqn: Marton IndS Cond on R_{1c}+R_{2c}}:
					\begin{align}
						R_{2k}+R_{r}  &\geq I(U;Z)				\label{eqn: FM 6_1}\\
						R_{1r}+R_{1c} &\geq I(V_1;Z|U)		\label{eqn: FM 7}\\
						R_{1k}+R_{r}  & \geq I(U;Z)				\label{eqn: FM 6_2}\\			
						R_{2r}+R_{2c} & \geq I(V_2;Z|U)		\label{eqn: FM 8}\\
						R_{1c}+R_{2c}& \leq I(V_1;Z|U)+I(V_2;Z|U)-I(V_1, V_2;Z|U)	\label{eqn: FM 10}
					\end{align}
	Note that \eqref{eqn: FM 6_1} and \eqref{eqn: FM 6_2} can be replaced by the following inequality
					\begin{equation}
						\min\{R_{1k}, R_{2k}\}+R_{r} \geq I(U;Z).  \label{eqn: FM 6}
					\end{equation}
\end{itemize}

Eliminating $R_r, R_{1r}, R_{2r}, R_{1c}, R_{2c}$ by applying Fourier-Motzkin procedure \cite{ElGamal:2012}, we obtain the region of $(R_1, R_2)=(R_{1k}+R_{1s}, R_{2k}+R_{2s})$ in terms of $(R_{1k}, R_{1s}, R_{2k}, R_{2s})$ as given in \eqref{eqn: DM-BC IndS region by Marton coding} in Theorem~\ref{thm:Marton}.  Note that a sketch of this Fourier-Motzkin procedure is provided in Appendix \ref{App: FM}. Further eliminate $R_{1k}, R_{1s}, R_{2k}, R_{2s},$ one can derive the same region in terms of $(R_1, R_2)$ as given in \eqref{eqn: Region_Marton R_1 R_2} in Theorem~\ref{thm:Marton}.


\section{Fourier-Motzkin Elimination for Theorem~\ref{thm:Marton}}\label{App: FM}
Here we briefly outline the Fourier-Motzkin procedure in the proof of Theorem~\ref{thm:Marton}. 

\begin{itemize}
	\item To eliminate $R_{r},$ we consider the non-negativity of the rate $R_{r}$ and the inequalities \eqref{eqn: FM 2}, \eqref{eqn: FM 3} and \eqref{eqn: FM 6} which involve $R_{r}.$ We end up with
		\begin{align}
			R_{k}+R_{1s}+R_{1r}+R_{1c} 			& \leq I(U, V_1;Y_1)				\label{eqn: FM 11r}\\
			R_{k}+R_{2s}+R_{2r}+R_{2c} 			& \leq I(U, V_2;Y_2)				\label{eqn: FM 12r},
		\end{align}
		where $R_{k}=\max\left\{R_{1k}+R_{2k} , \max\{R_{1k}, R_{2k}\}+I(U;Z)\right\}.$
\item To eliminate $R_{1r},$ we consider the non-negativity of the rate $R_{1r}$ and the inequalities \eqref{eqn: FM 4}, \eqref{eqn: FM 7} and \eqref{eqn: FM 11r} which involve $R_{1r}.$ We end up with
		\allowdisplaybreaks
		\begin{align}
			R_{1s}+R_{1c} 			& \leq I(V_1;Y_1|U)				\label{eqn: FM 15}\\
			R_{k}+R_{1s}+R_{1c} & \leq I(U, V_1;Y_1)				\label{eqn: FM 16r}\\
			R_{1s}			& \leq I(V_1;Y_1|U)-I(V_1;Z|U)				\label{eqn: FM 18}\\
			R_{k}+R_{1s} 			& \leq I(U, V_1;Y_1)-I(V_1;Z|U)				\label{eqn: FM 19r} 
		\end{align}
\item To eliminate $R_{2r},$ we consider the non-negativity of the rate $R_{2r}$ and the inequalities \eqref{eqn: FM 5}, \eqref{eqn: FM 8} and \eqref{eqn: FM 12r} which involve $R_{2r}.$ We end up with
		\allowdisplaybreaks
		\begin{align}
			R_{2s}+R_{2c} 			& \leq I(V_2;Y_2|U)				\label{eqn: FM 24}\\
			R_{k}+R_{2s}+R_{2c} 		& \leq I(U, V_2;Y_2)				\label{eqn: FM 25r}\\
			R_{2s}			& \leq I(V_2;Y_2|U)-I(V_2;Z|U)				\label{eqn: FM 27}\\
			R_{k}+R_{2s} 	& \leq I(U, V_2;Y_2)-I(V_2;Z|U)				\label{eqn: FM 28r} 
		\end{align}
\item To eliminate $R_{1c},$ we consider the non-negativity of the rate $R_{1c}$ and the inequalities \eqref{eqn: FM 1}, \eqref{eqn: FM 10}, \eqref{eqn: FM 15} and \eqref{eqn: FM 16r} which involve $R_{1c}.$ We end up with the following inequalities after canceling the redundant ones. 
		\allowdisplaybreaks
		\begin{align}
			I(V_1;V_2|U)  &\leq I(V_1;Z|U)+I(V_2;Z|U)-I(V_1,V_2;Z|U)	\label{eqn: FM 49} \\
			R_{2c} 			& \leq 	I(V_1;Z|U)+I(V_2;Z|U)-I(V_1,V_2;Z|U)	\label{eqn: FM 39-43}\\
			R_{2c}-R_{1s}  & \geq I(V_1;V_2|U)-I(V_1;Y_1|U)  \label{eqn: FM 50} \\
			R_{2c}-R_{k}-R_{1s}&\geq I(V_1;V_2|U)-I(U, V_1;Y_1)  \label{eqn: FM51r} 
		\end{align}
\item To eliminate $R_{2c},$ we consider the non-negativity of the rate $R_{2c}$ and the inequalities \eqref{eqn: FM 24}, \eqref{eqn: FM 25r},  \eqref{eqn: FM 39-43}, \eqref{eqn: FM 50} and \eqref{eqn: FM51r} which involve $R_{2c}.$ All the resulting inequalities are redundant (i.e., they all can be derived by combinations of other existing inequalities).  Thus no new inequalities are introduced. 
\end{itemize}

So far, we have for $R_{1s}, R_{2s}$  the inequalities \eqref{eqn: FM 18} and \eqref{eqn: FM 27}, respectively; and for their combinations with $R_{1k}, R_{2k}$ (implied by $R_{k}$) the inequalities \eqref{eqn: FM 19r} and \eqref{eqn: FM 28r}. Additionally, the inequality \eqref{eqn: FM 49} need to be fulfilled by the choices of $(U, V_1, V_2)$. 
This yields the desired region in terms of $(R_{1k}, R_{1s}, R_{2k}, R_{2s})$ as given in \eqref{eqn: DM-BC IndS region by Marton coding} in Theorem~\ref{thm:Marton}.

\section{Proof of Theorem~\ref{thm: JoS Marton}}\label{sec:App Jos Marton}

In this appendix, we establish the rate region as given in Theorem \ref{thm: JoS Marton} under the \emph{joint} secrecy constraint. To this end, we utilize the same encoding and decoding schemes as described in Appendix \ref{sec:AppMarton}. As a direct consequence, the reliability proof (i.e., analysis of the error probability of decoding) remains the same. However, we need to revise the secrecy analysis under the joint secrecy constraint.  That is, the achievability scheme needs to fulfill the joint secrecy constraint \eqref{eq:JointSec}, unlike the analysis given in Appendix \ref{sec:AppMarton}, in which the individual secrecy constraint \eqref{eq:IndSec} is satisfied.

{\em Analysis of joint secrecy:} 
For the joint secrecy \eqref{eq:JointSec}, i.e., $R_{L}\leq \tau_n,$ it is equivalent to show that $H(M_1, M_2|Z^n)\geq H(M_1, M_2)-n\tau_n=n(R_1+R_2)-n\tau_n.$ 
\allowdisplaybreaks
	\begin{align*}
	H(M_1, M_2| Z^n)	
			=   	& H(M_{1k},M_{2k},M_{1s},M_{2s}|Z^n)\\
			=   	& H(M_{1k},M_{2k},M_{1s},M_{2s},M_{r}|Z^n) - H(M_r|M_{1k},M_{2k},M_{1s},M_{2s}, Z^n)\\
			=   	& H(U^n,M_{1s},M_{2s}|Z^n) - H(U^n|M_{1k},M_{2k},M_{1s},M_{2s}, Z^n)\\
			\geq	& H(U^n, M_{1s},M_{2s}|Z^n) - H(U^n|M_{1k},M_{2k},Z^n)\\
		 	=		& H(U^n, V_1^n, V_2^n|Z^n) -H(M_{1r}, M_{1c},M_{2r}, M_{2c}|U^n,M_{1s},M_{2s}, Z^n) -H(U^n|M_{1k}, M_{2k}, Z^n)\\
		 	\stackrel{(a)}{\geq}	& H(U^n, V_1^n, V_2^n|Z^n) -H(U^n|M_{1k}, M_{2k}, Z^n) \\
		 									& -H(M_{1r}, M_{1c}|U^n,M_{1s},M_{2s}, Z^n)-H(M_{2r}, M_{2c}|U^n,M_{1s},M_{2s}, Z^n)\\
		 	\stackrel{(b)}{\geq}	& H(U^n, V_1^n, V_2^n|Z^n) -H(U^n|M_{1k}, M_{2k}, Z^n)\\
		 									& -H(M_{1r}, M_{1c}|U^n,M_{1s}, Z^n)-H(M_{2r}, M_{2c}|U^n,M_{2s}, Z^n) \\
		 	=		& H(U,V_1^n,V_2^n)-H(Z^n)+H(Z^n|U^n, V_1^n, V_2^n)-H(U^n|M_{1k}, M_{2k}, Z^n)\\
		 			&-H(V_1^n|U^n,M_{1s}, Z^n)-H(V_2^n|U^n,M_{2s}, Z^n) \nonumber\\
		 	\stackrel{(c)}{\geq} & n[R_{1k}+R_{2k}+R_{r}+R_{1s}+R_{2s}+R_{1r}+R_{2r}]-nI(U, V_1, V_2; Z)-n[R_r-I(U;Z)]\\
		 								& -n[R_{1r}+R_{1c}-I(V_1;Z|U)]-n[R_{2r}+R_{2c}-I(V_2;Z|U)]-n\tau_n\\
		 	\stackrel{(d)}{\geq} & n[R_1+R_2]-n\tau_n
			\end{align*}
where 
	$(a)$ follows from the fact that $H(A,B|C)\leq H(A|C)+H(B|C);$
	$(b)$ is due to the fact that conditioning reduces entropy;
	$(c)$ follows from that 
	\begin{enumerate}
		\item $H(U^n)=n[R_{2k}+R_{1k}+R_{r}]$ and $H(V_1^n, V_2^n|U^n)\geq n[R_{1s}+R_{1r}+R_{2s}+R_{2r}]$ by the codebook construction; 
		\item $H(Z^n)=\sum_{i=1}^n H(Z_i|Z^{i-1})\leq \sum_{i=1}^n H(Z_i)=nH(Z|U);$ 
		\item $H(Z^n|U^n, V_1^n, V_2^n)=nH(Z|U, V_1, V_2)$ due to the discrete memoryless of the channel;
		\item applying \cite[Lemma 1]{Chia:Three-receiver12}, we have that
			\begin{itemize}
				\item 	 $H(U^n|M_{1k}, M_{2k}, Z^n)\leq n[R_{r}-I(U;Z)]+n\tau_n/3$ if taking 
							\begin{equation}\label{eqn: Marton JoS Cond on R_{r}}
								R_{r}\geq I(U;Z)+\delta_n(\tau_n);
							\end{equation}
				\item  $H(V_1^n|M_{1s},U^n, Z^n)\leq n[R_{1r}+R_{1c}-I(V_1;Z|U)]+n\tau_n/3$ if taking \eqref{eqn: Marton IndS Cond on R_{1r}+R_{1c}}, i.e.,
							\begin{equation*}
								R_{1r}+R_{1c}\geq I(V_1;Z|U)+\delta_n(\tau_n);
							\end{equation*}
				\item $H(V_2^n|M_{2s},U^n, Z^n)\leq n[R_{2r}+R_{2c}-I(V_1;Z|U)]+n\tau_n/3$ if taking \eqref{eqn: Marton IndS Cond on R_{2r}+R_{2c}}, i.e.,
									\begin{equation*}
										R_{2r}+R_{2c}\geq I(V_2;Z|U)+\delta_n(\tau_n);
									\end{equation*}
			\end{itemize}
	\end{enumerate}
$(d)$ is by taking \eqref{eqn: Marton IndS Cond on R_{1c}+R_{2c}}, i.e.,
	\begin{equation*}
		R_{1c}+R_{2c} \leq I(V_1;Z|U)+I(V_2;Z|U)-I(V_1,V_2;Z|U).
	\end{equation*}
We note that a stronger constraint \eqref{eqn: Marton JoS Cond on R_{r}} is imposed on $R_r$ in order to guarantee the joint secrecy, instead of the \eqref{eqn: Marton IndS Cond on R_{2k}+R_{r}} and \eqref{eqn: Marton IndS Cond on R_{1k}+R_{r}} for the case of individual secrecy.

{\em Achievable rate region:}  
The resulting region has the following constraints:
the non-negativity for rates, i.e., $R_{1k}, R_{2k}, R_{1s}, R_{2s}, R_{r}, R_{1r}, R_{2r}, R_{1c}, R_{2c} \geq  0,$ 
the rate relations imposed by rate splitting as specified in \eqref{eqn: Marton R_1} and \eqref{eqn: Marton R_2},
the conditions for a reliable communication to both legitimate receivers, i.e., \eqref{eqn: Marton cond on cover index}, \eqref{eqn: cond Marton Rx1 pe1}, \eqref{eqn: cond Marton Rx1 pe2}, \eqref{eqn: cond Marton Rx2 pe1}, \eqref{eqn: cond Marton Rx2 pe2}, and
the conditions for joint secrecy of the messages at the eavesdropper, i.e., \eqref{eqn: Marton JoS Cond on R_{r}}, \eqref{eqn: Marton IndS Cond on R_{1r}+R_{1c}}, \eqref{eqn: Marton IndS Cond on R_{2r}+R_{2c}}, \eqref{eqn: Marton IndS Cond on R_{1c}+R_{2c}}. 
Eliminating $R_{1k}, R_{2k}, R_{1s}, R_{2s}, R_{1c}, R_{2c}, R_{1r}, R_{2r}, R_r$ by applying Fourier-Motzkin procedure \cite{ElGamal:2012}, we obtain the region of $(R_1, R_2)$ as given in \eqref{eqn: JoS Marton} in Theorem~\ref{thm: JoS Marton}.


\section{Proof of \eqref{eqn: upper bound nR_1^1}} \label{Sec: Appdix proof using Costas EPI}

Recall Costa's EPI as described in the following.
\begin{lemma}\cite[Theorem 1]{src:Costa1985} 
	Let $X$ be an arbitrarily distributed $n$-dimensional random variable. Let $N$ be a $n$-dimensional Gaussian vector, independent of $X,$ and with covariance matrix proportional to the identity matrix, then
		\begin{equation}\label{eqn: Costas EPI}
			e^{\frac{2}{n}h(X+\beta N)} \geq (1-\beta^2) e^{\frac{2}{n}h(X)} + \beta^2 e^{\frac{2}{n}h(X+N)}, 
		\end{equation} 
	where $\beta\in [0,1].$ 
\end{lemma}

Consider the Gaussian DBC under our investigation. Due to the degradedness order $Y_1^n\to Y_2^n\to Z^n,$ we could write
	\begin{align*}
		Y_2^n &= Y_1^n + \beta (N^n_{12}+N^n_{2e}),\\
		Z^n	&= Y_1^n + N^n_{12}+N ^n_{2e},
	\end{align*} 
where $N^n_{12}\sim \mathcal{N} (\mathbf{0}, (\sigma_2^2-\sigma_1^2)\mathbf{I})$ and $N_{2e}\sim \mathcal{N} (\mathbf{0}, (\sigma_2^2-\sigma_1^2)\mathbf{I}),$ and
\begin{equation}\label{eqn: value of beta}
	\beta=\sqrt{\frac{\sigma_2^2-\sigma_1^2}{\sigma_e^2-\sigma_1^2}}.
\end{equation}
Note that $N^n_{12}, N^n_{2e}$ are independent of $Y^n_1$ and $M_2.$ 

Now applying Costa's EPI as described in \eqref{eqn: Costas EPI}, we have
\begin{equation*}
  e^{\frac{2}{n}h(Y_2^n|M_2)} \geq (1-\beta^2) e^{\frac{2}{n}h(Y_1^n|M_2)} + \beta^2 e^{\frac{2}{n}h(Z^n)}.
\end{equation*}
Dividing both sides by $e^{\frac{2}{n}h(Z^n)},$ we obtain
\begin{equation*}
	e^{\frac{2}{n}[h(Y_2^n|M_2)-h(Z^n|M_2)]} \geq (1-\beta^2) e^{\frac{2}{n}[h(Y_1^n|M_2)-h(Z^n|M_2)]} + \beta^2.
\end{equation*}
Replacing $h(Y_2^n|M_2)-h(Z^n|M_2)$ and $\beta$ by their realizations as specified in \eqref{eqn: R_22} and \eqref{eqn: value of beta}, respectively, we obtain
	\begin{equation*}
		\frac{\alpha P+\sigma_2^2}{\alpha P+\sigma_e^2} \geq \frac{\sigma_e^2-\sigma_2^2}{\sigma_e^2-\sigma_1^2}e^{\frac{2}{n}[h(Y_1^n|M_2)-h(Z^n|M_2)]}+\frac{\sigma_2^2-\sigma_1^2}{\sigma_e^2-\sigma_1^2}.
	\end{equation*}
Easy calculation gives
	\begin{equation*}
		h(Y_1^n|M_2)-h(Z^n|M_2)\leq \frac{n}{2}\log \frac{\alpha P+\sigma_1^2}{\alpha P+\sigma_e^2},
	\end{equation*} 
i.e., \eqref{eqn: upper bound nR_1^1}. This concludes our proof.


\section{Proof of \eqref{eqn: upper bound h(Z^n|M_1, M_2)}} \label{Sec: Appdix proof using EPI}

Recall Shannon's EPI as described in the following.
\begin{lemma}\cite{src:Shannon1948}
	For any two independent, $n$-dimensional random variable $X$ and $N,$
		\begin{equation}\label{eqn: EPI}
			e^{\frac{2}{n} h(X+N)} \geq e^{\frac{2}{n} h(X)}+e^{\frac{2}{n} h(N)}.
		\end{equation}
\end{lemma}

Consider the Gaussian DBC under our investigation. We could write
	\begin{equation*}
		Z^n	= Y_1^n + N_{1e}^n,
	\end{equation*} 
where $N_{1e}^n\sim \mathcal{N} (\mathbf{0}, (\sigma_e^2-\sigma_1^2)\mathbf{I})$ and $N_{1e}^n$ is independent of $Y_1^n$ and $(M_1, M_2).$ Therefore, applying EPI as described in \eqref{eqn: EPI}, we have 
\begin{equation*}
	e^{\frac{2}{n}h(Z^n|M_1, M_2)}\geq e^{\frac{2}{n}h(Y_1^n|M_1, M_2)}+e^{\frac{2}{n}h(N_{1e}^n)}.
\end{equation*}
That is, 
\begin{equation*}
	e^{\frac{2}{n}h(Z^n|M_1, M_2)}\geq e^{\frac{2}{n}[h(Y_1^n|M_1, M_2)-h(Z^n|M_1, M_2)]}\cdot e^{\frac{2}{n}h(Z^n|M_1, M_2)}+e^{\frac{2}{n}h(N^n_{1e})}.
\end{equation*}
Replacing $h(Y_1^n|M_1, M_2)-h(Z^n|M_1, M_2)$ by its realization as specified in \eqref{eqn: nR_1^2 value} and $h(N_{1e}^n)$ by 
\begin{equation*}
	h(N_{1e}^n)=\frac{n}{2}\log 2\pi e (\sigma_e^2-\sigma_1^2),
\end{equation*}
we obtain 
\begin{equation*}
	e^{\frac{2}{n}h(Z^n|M_1, M_2)}\geq \frac{\gamma\alpha P+\sigma_1^2}{\gamma\alpha P+\sigma_e^2} e^{\frac{2}{n}h(Z^n|M_1, M_2)}+2\pi e (\sigma_e^2-\sigma_1^2).
\end{equation*}
Easy calculation gives
\begin{equation*}
	h(Z^n|M_1, M_2) \geq \frac{n}{2} \log 2\pi e (\gamma\alpha P+\sigma_e^2).
\end{equation*}
i.e., \eqref{eqn: upper bound h(Z^n|M_1, M_2)}. This concludes our proof.


\section{Proof of Corollary~\ref{Cor: Gaussian looser outer bound}} 
\label{app:CorollaryGaussianOuterBound}

\begin{IEEEproof}
	The upper bound on $R_2$ remains the same to \eqref{eqn: IndS Gaussian Outer bound on R2}; while the upper bounds on $R_1$ and $R_1+R_2$ are obtained by combining \eqref{eqn: IndS Gaussian Outer bound on R1} and \eqref{eqn: IndS Gaussian Outer bound on R2}. In more details, we have 
	\begin{align*}
		R_1 \stackrel{(a)}{\leq} &  C\left(\frac{\alpha(1-\gamma) P}{\gamma\alpha P+\sigma_1^2}\right)- C\left(\frac{\alpha(1-\gamma) P}{\gamma\alpha P+\sigma_e^2}\right)+R_2\\
			\leq &  C\left(\frac{\alpha P}{\sigma_1^2}\right)- C\left(\frac{\alpha P}{\sigma_e^2}\right)+R_2\\
			\stackrel{(b)}{\leq} & 	C\left(\frac{\alpha P}{\sigma_1^2}\right)- C\left(\frac{\alpha P}{\sigma_e^2}\right)+C\left(\frac{(1-\alpha)P}{\alpha P+\sigma_2^2}\right)-C\left(\frac{(1-\alpha) P}{\alpha P+\sigma_e^2}\right)
	\end{align*}
	where $(a)$ is according to \eqref{eqn: IndS Gaussian Outer bound on R1}; 
	$(b)$ is via replacing $R_2$ by its upper bound as given in \eqref{eqn: IndS Gaussian Outer bound on R2}. 
	
	On the other hand, according to \eqref{eqn: IndS Gaussian Outer bound on R1}, we have
	\begin{align*}
		R_1 & \leq 
			C\left(\frac{\alpha(1-\gamma) P}{\gamma\alpha P+\sigma_1^2}\right)- C\left(\frac{\alpha(1-\gamma) P}{\gamma\alpha P+\sigma_e^2}\right)+C\left(\frac{(1-\gamma\alpha)P}{\gamma\alpha P+\sigma_e^2}\right)\\
			& =
			C\left(\frac{\alpha(1-\gamma) P}{\gamma\alpha P+\sigma_1^2}\right)+ C\left(\frac{(1-\alpha) P}{\alpha P+\sigma_e^2}\right)\\ 
			& \leq C\left(\frac{\alpha P}{ \sigma_1^2}\right)+ C\left(\frac{(1-\alpha) P}{\alpha P+\sigma_e^2}\right).
	\end{align*}	
	Summing it up with $R_2$ which is upper bounded by \eqref{eqn: IndS Gaussian Outer bound on R2}, we get the desired upper bound on $R_1+R_2.$ This concludes our proof.
\end{IEEEproof}


\section{Proof of Theorem~\ref{thm:ConstantGap}} 
\label{app:ConstantGap}

\begin{IEEEproof}
	Consider the gap between the inner and outer bounds as specified in \eqref{eqn: IndS Gaussian Inner bound} and \eqref{eqn: IndS Gaussian looser Outer bound}, respectively. If we take the same choice of $\alpha$ in both bounds, the gap may occur only in the $R_1+R_2$ term that is upper bounded by
				\begin{equation*}
					\left[C\left(\frac{\alpha P}{\sigma_1^2}\right)+C\left(\frac{(1-\alpha)P}{\alpha P+\sigma_2^2}\right)\right]- \left[C\left(\frac{\alpha P}{\sigma_1^2}\right)-C\left(\frac{\alpha P}{\sigma_e^2}\right)+C\left(\frac{(1-\alpha)P}{\alpha P+\sigma_2^2}\right)\right]\\
					= C\left(\frac{\alpha P}{\sigma_e^2}\right).
				\end{equation*}
A first observation is that both bounds coincide at $\alpha=0.$ Furthermore, we consider their subregions in the following two cases for comparison.
\begin{itemize}
	\item Consider the case as $R_2\leq C\left(\frac{(1-\alpha) P}{\alpha P+\sigma_e^2}\right).$ The corresponding subregions of $(R_1, R_2)$ in the inner and outer bound are the same, i.e.,
 	\begin{align*}
 		R_{1}\leq & C\left(\frac{\alpha P}{\sigma_1^2}\right)-C\left(\frac{\alpha P}{\sigma_e^2}\right)+C\left(\frac{(1-\alpha)P}{\alpha P+\sigma_2^2}\right)-C\left(\frac{(1-\alpha) P}{\alpha P+\sigma_e^2}\right)\\
 		R_{2}\leq & \min\left\{C\left(\frac{(1-\alpha) P}{\alpha P+\sigma_e^2}\right), C\left(\frac{(1-\alpha)P}{\alpha P+\sigma_2^2}\right)-C\left(\frac{(1-\alpha) P}{\alpha P+\sigma_e^2}\right)\right\}
 	\end{align*}		 
	\item Consider the other case as $C\left(\frac{(1-\alpha) P}{\alpha P+\sigma_e^2}\right) < R_2\leq C\left(\frac{(1-\alpha)P}{\alpha P+\sigma_2^2}\right)-C\left(\frac{(1-\alpha) P}{\alpha P+\sigma_e^2}\right).$ Note that this case is possible only if 		
			\begin{equation}\label{eqn: CS Gaussion Sub condition}
				C\left(\frac{(1-\alpha) P}{\alpha P+\sigma_e^2}\right)<C\left(\frac{(1-\alpha)P}{\alpha P+\sigma_2^2}\right)-C\left(\frac{(1-\alpha) P}{\alpha P+\sigma_e^2}\right). 
			\end{equation}
		The above inequality holds for 
		\begin{equation}\label{eqn: value of alpha sub1}
								0< \alpha < 1 \quad \mbox{as}\quad \sigma_e^2\geq P+2\sigma_2^2; 
		\end{equation}
		or
		\begin{equation}\label{eqn: value of alpha sub2}
			0< \alpha < \frac{(\sigma_e^2-\sigma_2^2)^2}{P(P+\sigma_2^2)}-\frac{\sigma_2^2}{P} \quad \mbox{as}\quad \sigma_e^2\leq P+2\sigma_2^2. 
		\end{equation}
		(The calculation of \eqref{eqn: value of alpha sub1} and \eqref{eqn: value of alpha sub2} is similar to the one given in Appendix \ref{Sec: Appdix calculation alpha}.)
	
	Recall that the gap occurs only in the $R_1+R_2$ term that is upper bounded by $C(\frac{\alpha P}{\sigma_e^2}).$ More specifically, 
	\begin{enumerate}
		\item as $\sigma_e^2\geq P+2\sigma_2^2,$ we have for $0< \alpha<1,$
					\begin{equation*}
					C\left(\frac{\alpha P}{\sigma_e^2}\right) \stackrel{(a)}{<} C\left(\frac{ P}{\sigma_e^2}\right) \stackrel{(b)}{\leq}
									C\left(\frac{P}{P+2\sigma_2^2}\right) \leq C(1) = 0.5
					\end{equation*}
		where $(a)$ is by the fact that $C(x)$ is an increasing function with respect to $x$ and $\alpha$ is upper bounded by 1; $(b)$ is due to the fact that $\sigma_e^2\geq P+2\sigma_2^2.$
		\item as $\sigma_e^2\leq P+2\sigma_2^2,$ we have for $0< \alpha < \frac{(\sigma_e^2-\sigma_2^2)^2}{P(P+\sigma_2^2)}-\frac{\sigma_2^2}{P},$
					\begin{equation*}
					C\left(\frac{\alpha P}{\sigma_e^2}\right) \stackrel{(c)}{<} 
									C\left(\frac{\sigma_e^2-\sigma_2^2}{P+\sigma_2^2}-\frac{\sigma_2^2(P+\sigma_e^2)}{\sigma_e^2(P+\sigma_2^2)}\right)\\
									\stackrel{(d)}{\leq }  	C\left(1-\frac{\sigma_2^2(P+\sigma_e^2)}{\sigma_e^2(P+\sigma_2^2)}\right) \leq C(1) = 0.5
					\end{equation*}
		where $(c)$ is by the fact that $C(x)$ is an increasing function with respect to $x$ and $\alpha$ is upper bounded by $\frac{(\sigma_e^2-\sigma_2^2)^2}{P(P+\sigma_2^2)}-\frac{\sigma_2^2}{P}$; $(d)$ is due to the fact that $(\sigma_e^2-\sigma_2^2)/(P+\sigma_2^2)\leq 1$ since $\sigma_e^2\leq P+2\sigma_2^2.$
	\end{enumerate}
\end{itemize}
This concludes our proof.
\end{IEEEproof}


\section{Calculation for \eqref{eqn: CS Gaussion condition}} \label{Sec: Appdix calculation alpha}
To find $\alpha$ such that \eqref{eqn: CS Gaussion condition} holds, we consider
\begin{equation*}
	C\left(\frac{(1-\alpha)P}{\alpha P+\sigma_2^2}\right)-2C\left(\frac{(1-\alpha) P}{\alpha P+\sigma_e^2}\right) \leq 0
\end{equation*}
which is equivalent to having $$\frac{1}{2}\log \frac{(P+\sigma_2^2)(\alpha P+\sigma_e^2)^2}{(\alpha P+\sigma_2^2)(P+\sigma_e^2)^2} \leq 0,\quad  \mbox{i.e.,} \quad\frac{(P+\sigma_2^2)}{(P+\sigma_e^2)^2} (\alpha P+\sigma_e^2)^2 \leq (\alpha P+\sigma_e^2)-(\sigma_e^2-\sigma_2^2).$$
Note that this inequality can be formulated as a quadratic inequality with respect to $\alpha P+\sigma_e^2$. Accordingly, denoting 
$x	=\alpha P+\sigma_e^2$,
$A=\frac{(P+\sigma_2^2)}{(P+\sigma_e^2)^2}$, and
$C=\sigma_e^2-\sigma_2^2$,
we represent the inequality above by $f(x)=Ax^2-x+C\leq 0.$
Here, as $A\geq 0$, $f(x)$ is convex and this inequality holds when 
$$\frac{1-T}{2A}\leq x \leq \frac{1+T}{2A}, \quad \mbox{where}\quad T=\sqrt{1-4AC}=\frac{P+2\sigma_2^2-\sigma_e^2}{P+\sigma_e^2}.$$ 
Here, $T\geq 0$ as the assumptions in the theorem implies that $P\geq \frac{\sigma_e^2}{\sigma_2^2}(\sigma_e^2-2\sigma_2^2) \geq (\sigma_e^2-2\sigma_2^2)$, where the last inequality is due to $\sigma_e^2\geq \sigma_2^2$. Using the values of $T$, $A$, and $x$ in this last condition, we obtain that $f(x)\geq 0$ if and only if
\begin{equation*}
\frac{(\sigma_e^2-\sigma_2^2)^2}{P(P+\sigma_2^2)}-\frac{\sigma_2^2}{P} \leq\alpha \leq 1.
\end{equation*}

\bibliographystyle{IEEEtran}
\bibliography{IEEEabrv,CitedReferences}
\end{document}